\documentclass[sigconf]{acmart}

\usepackage{booktabs}
\usepackage{algorithm}
\usepackage{algpseudocode}
\usepackage{mdframed}
\usepackage{array}
\usepackage{tabularx} 
\usepackage{nicefrac}
\usepackage{balance}
\usepackage{lipsum}
\usepackage{xcolor}
\usepackage{amsthm}
\usepackage{pifont}
\usepackage{xspace}
\usepackage{graphics}
\usepackage{graphicx}
\usepackage{multirow}
\usepackage{outlines}
\usepackage{enumerate}
\usepackage{enumitem}
\usepackage{hyperxmp}
\usepackage{subcaption}
\usepackage{float}
\usepackage{stfloats}
\usepackage{xurl}

\newcommand{\ie}{{\it i.e.,}\xspace}
\newcommand{\eg}{{\it e.g.,}\xspace}

\newcommand{\greybox}[1]{\vspace{0.1in}\begin{mdframed}[backgroundcolor=gray!15]{\vspace{0.05in}#1\vspace{0.05in}}\end{mdframed}}

\newcommand{\parab}[1]{\vspace{0.06in}\noindent{\bf #1}}

\newtheorem{definition}{Definition}

\newcommand{\sysName}{\textsc{ReFlex}\xspace}

\setcopyright{none}
\renewcommand\footnotetextcopyrightpermission[1]{}

\begin{document}

% Title
\title{New primitives for bounded degradation in network service}

% Authors
\author{Simon Kassing$^1$, Vojislav Dukic$^1$, Ce Zhang$^1$, Ankit Singla$^1$}
\affiliation{%
  \institution{$^1$ETH Zurich}
  \country{}
  \vspace{0.5cm}
}

\begin{abstract}
  Certain new ascendant data center workloads can absorb \textit{some} degradation in network service, not needing fully reliable data transport and/or their fair-share of network bandwidth. This opens up opportunities for superior network and infrastructure multiplexing by having this \textit{flexible} traffic cede capacity under congestion to regular traffic with stricter needs. We posit there is opportunity in network service primitives which permit degradation within certain bounds, such that flexible traffic still receives an acceptable level of service, while benefiting from its weaker requirements. We propose two primitives, namely guaranteed partial delivery and bounded deprioritization. We design a budgeting algorithm to provide guarantees relative to their fair share, which is measured via probing. The requirement of budgeting and probing limits the algorithm's applicability to large flexible flows.

We evaluate our algorithm with large flexible flows and for three workloads of regular flows of small size, large size and a distribution of sizes. Across the workloads, our algorithm achieves less speed-up of regular flows than fixed prioritization, especially for the small flows workload (1.25$\times$ vs. 6.82$\times$ in the 99th \%-tile). Our algorithm provides better guarantees in the workload with large regular flows (with 14.5\% vs. 32.5\% of flexible flows being slowed down beyond their guarantee). However, it provides not much better or even slightly worse guarantees for the other two workloads. The ability to enforce guarantees is influenced by flow fair share interdependence, measurement inaccuracies and dependency on convergence. We observe that priority changes to probe or to deprioritize causes queue shifts which deteriorate guarantees and limit possible speed-up, especially of small flows. We find that mechanisms to both prioritize traffic and track guarantees should be as non-disruptive as possible.

%%%%%%%%%%%%%%%%%%%%%%%%%%%%%%%%%%%%%%%%%%%%%
%%%%%%%%%%%%%%%%%%%%%%%%%%%%%%%%%%%%%%%%%%%%%
%%%%%%%%%%%%%%%%%%%%%%%%%%%%%%%%%%%%%%%%%%%%%
%%%%%%%%%%%%%%%%%%%%%%%%%%%%%%%%%%%%%%%%%%%%%
%%%%%%%%%%%%%%%%%%%%%%%%%%%%%%%%%%%%%%%%%%%%%

\end{abstract}

\pagestyle{plain}
\maketitle

\section{Introduction}

While the diversity of data center network applications and their service requirements continues to grow, network service primitives have lagged these changes. Most deployed data center networks expose to applications only reliable (TCP) or unreliable (UDP) service, and fair-share or reserved bandwidth or some simple form of prioritization. 

For many applications, these primitives are either insufficient, or too rigid. While there is ample work on stricter service models that provide a notion of quality of service~\cite{ballani2014offering, wang2014sdn, rygielski2013network}, we argue that there is under-explored value in more \textit{relaxed} models of network service, whereby applications with weaker service requirements cede the network to those with stricter requirements, as long as their own needs are met.

Specifically, some applications can tolerate limited data loss without significant app-level deterioration, and thus do not need fully reliable data transfer, while others may accept limited deprioritization in favor of applications with stricter requirements. One high-value class of loss-tolerant applications is distributed machine learning training. Recent work in the ML research community has shown~\cite{icml-unreliable} that if network loss is bounded by some acceptable upper limit, this causes minimal or no performance impact for many types of ML training tasks: delivering, \eg $80\%$ of each model update between workers may be enough to maintain convergence rate and achieve the same training loss, with fully reliable delivery being unnecessary. Given that ML is the fastest growing cloud workload~\cite{MITsmr}, offering an appropriate bounded-loss primitive that bridges that TCP-UDP binary would be attractive.

Similarly, some applications can tolerate being deprioritized such that they receive less than their fair share of network bandwidth. Instances of such applications are long-running backup or administrative tasks, and model updates in asynchronous ML training for tasks where real-time model freshness is not always needed.

Note that in the above ``flexible'' applications, \emph{some} loss or deprioritization is acceptable, but it is desired that it be \textit{bounded}: if loss exceeds a specified limit, ML training jobs may experience large training loss; if deprioritized tasks are continually starved, application expectations may be violated.

Being able to offer network service that accounts for such flexibility in applications would naturally allow superior multiplexing: one could colocate on the same infrastructure, regular workloads that expect standard network service (\eg Web search or ML inference queries), with flexible workloads, such that bursts in regular traffic benefit from the flexible traffic ceding network capacity. Improving infrastructure multiplexing in this manner requires: (a) ensuring that regular traffic is serviced preferentially; and (b) the extent of this preferential treatment is bounded such that the corresponding deterioration in network service for flexible traffic is controlled, thus causing only negligible or acceptable application-level performance impact. In particular, we explore the co-design of both loss and deprioritization guarantees, and when data should be considered lost.

The most commonly available networking primitives are unfortunately incapable of exploiting this opportunity by simultaneously meeting both the above requirements. At the transport level, we only have abstractions for fully reliable (TCP) or entirely best-effort (UDP) transmission, instead of \emph{guaranteed partial delivery}. At the network level, we have a variety of prioritization and scheduling mechanisms, but none provide a notion of \textit{bounded deprioritization} for certain classes of traffic. The concept of applications with more flexible network requirements ceding to more demanding applications has been partially explored. LEDBAT~\cite{ledbat} makes use of delay measurements to enforce it does not cause congestion in the network and makes use of a congestion control procedure less aggressive than TCP. The concept of loss flexibility similarly has been explored by Approximate Transport Protocol (ATP)~\cite{atp}. Under-explored however is the co-design of both loss and deprioritization guarantees, and when data should be considered lost.

We take the first steps in this direction by: (a) introducing primitives describing partial delivery and bounded degradation goals; and (b) designing mechanisms for the network that implement these primitives. We propose a simple extension to the application-network interface: an application sending data can specify what fraction of it must be reliably delivered, and/or what degradation in performance is acceptable compared to the default fair-sharing of the network. Using this interface, different applications (\eg different ML training approaches and workloads) can specify arbitrarily different degrees of acceptable degradation in the same network. We also design and implement algorithms for the network to exploit the headroom these relaxations of the network service objectives provide, by dropping or deprioritizing flexible traffic in favor of regular traffic, while ensuring that the specified (relaxed) guarantees for flexible traffic are indeed met.

We have evaluated our approach, \textbf{\sysName}, using packet-level simulation. We show that by allowing (guaranteed) partial delivery and bounded deprioritization for flexible traffic, the performance of regular traffic can be improved. We compare our approach with standard prioritization methods, and investigate the extent to which both are able to bound degradation as well as speed-up workloads.

%%%%%%%%%%%%%%%%%%%%%%%%%%%%%%%%%%%%%%%%%%%%%
%%%%%%%%%%%%%%%%%%%%%%%%%%%%%%%%%%%%%%%%%%%%%
%%%%%%%%%%%%%%%%%%%%%%%%%%%%%%%%%%%%%%%%%%%%%
%%%%%%%%%%%%%%%%%%%%%%%%%%%%%%%%%%%%%%%%%%%%%
%%%%%%%%%%%%%%%%%%%%%%%%%%%%%%%%%%%%%%%%%%%%%

\section{Why new networking primitives?}
\label{sec:why-new-primitives}

Network traffic for a growing class of applications offers two degrees of flexibility that cannot be exploited by today's networks: (a) such traffic does not need the predominant reliable delivery of TCP; and (b) such traffic can be deprioritized in favor of more other time-critical traffic. However, providing \emph{no} network service assurance is also unacceptable, as that can lead to substantial application-level performance degradation. We next discuss each of these potential relaxations to the network service model in greater detail.

%%%%%%%%%%%%%%%%%%%%%%%%%%%%%%%%%%%%%%%%%%%%%
%%%%%%%%%%%%%%%%%%%%%%%%%%%%%%%%%%%%%%%%%%%%%
%%%%%%%%%%%%%%%%%%%%%%%%%%%%%%%%%%%%%%%%%%%%%
%%%%%%%%%%%%%%%%%%%%%%%%%%%%%%%%%%%%%%%%%%%%%
%%%%%%%%%%%%%%%%%%%%%%%%%%%%%%%%%%%%%%%%%%%%%

\subsection{Guaranteed partial delivery}

Traditionally, networking has provided only two extremes: (1) a fully reliable in-order data delivery primitive, using TCP or its derivatives; and (2) entirely best-effort data delivery using UDP, with no reliability guarantees. While many existing applications depend on one of these service models, new emerging workloads can benefit from guaranteed partial reliability. For instance, in certain types of distributed machine learning training, losing some fraction, \eg 20\% of each model update transferred between workers, or between workers and a parameter server, has no or negligible impact on the ML training task~\cite{icml-unreliable}. Note that this is different from adaptive applications like video streaming, where there is first a compromise in application performance by picking lower video quality, and then data for the lower quality still needs to be delivered reliably using TCP, or is entirely best-effort, again with additional observable degradation.

To characterize flexible traffic that only requires guaranteed partial delivery, we define a reliability factor as follows:
\begin{definition}
\label{def:r}
Each flow $f$ has a \textbf{reliability factor $r\in[0,1]$}, which is a lower bound on the fraction of its data that must be delivered reliably by the transport protocol. For regular, inflexible traffic, and more broadly, for any traffic for which $r$ is not explicitly specified, $r=1$ by default.
\end{definition}

A naive way of exploiting the flexibility specified using $r<1$ would be to simply trim the payload before each transfer, for instance, transfer the first $80\%$ bytes for each model update reliably using TCP, and then drop the last $20\%$ bytes. While this is indeed ``partially reliable'' delivery, this is a poor strategy: what if the network was congested earlier in the transfer, but is under-utilized later? In this case, sending less data does not have much upside in terms of improving network multiplexing. Thus, we would like to use this flexibility adaptively, dropping data when the network is busy with regular, inflexible traffic, while delivering it when the network is less congested. The knowledge of payload size in advance is fundamental to partial delivery: if the control mechanism only knows the final payload size once the transfer is complete, it is unable to finish early before delivering the entire payload.

Unfortunately, today's binaries of reliable and unreliable transfer do not accommodate such sophisticated, adaptive switching between reliable and unreliable transmission. Interpreted in light of the above definition, using UDP effectively implies $r=0$, while using TCP enforces $r=1$. Note that our definition does not enforce that only $r$ fraction of traffic is delivered, rather $r$ is a lower bound, with at least $r$ fraction of traffic sought to be reliably delivered. The need for a new partially-reliable transport protocol has been recognized by a parallel work available as an online manuscript in which the Approximate Transmission Protocol (ATP) has been proposed~\cite{atp}. ATP defines a similar reliability factor and provides guarantees for the data delivered. The key distinction is that ATP decides partial delivery based on actual packets being lost, which differs from our approach which defines partial delivery based on bandwidth loss relative to fair share (\S\ref{sec:abstract-design}). More distinctions to ATP are further explained in \S\ref{sec:fdt-related-work}.

%%%%%%%%%%%%%%%%%%%%%%%%%%%%%%%%%%%%%%%%%%%%%
%%%%%%%%%%%%%%%%%%%%%%%%%%%%%%%%%%%%%%%%%%%%%
%%%%%%%%%%%%%%%%%%%%%%%%%%%%%%%%%%%%%%%%%%%%%
%%%%%%%%%%%%%%%%%%%%%%%%%%%%%%%%%%%%%%%%%%%%%
%%%%%%%%%%%%%%%%%%%%%%%%%%%%%%%%%%%%%%%%%%%%%

\subsection{Bounded deprioritization}
\label{sec:bounded-deprioritization}

Many applications, such as backup or administrative tasks, non-interactive analytics, and ML training workloads where real-time model freshness is not critical, can be bandwidth-hungry, but do not necessarily need to compete fairly with more time-sensitive traffic such as user-facing Web search or ML inference queries. Networking does offer three broad types of primitives to benefit regular traffic by exploiting such flexibility, but as we discuss in the following, none of these provides any bounds on how much worse the performance of the flexible traffic can be.

\parab{Background transport:} LEDBAT~\cite{ledbat} attempts to capture unused network bandwidth, and is commonly used over the Internet today for transfers like software updates. However, LEDBAT does not compete fairly with TCP traffic at all, and thus, if used for flexible traffic, cannot bound the degree to which flexible traffic is deprioritized: if the network is seeing a large volume of regular traffic, flexible traffic using LEDBAT would be completely starved.

\parab{Prioritization:} Superficially, running flexible traffic at low priority using multiple queues available in most commodity switches today is a natural way of favoring regular traffic over flexible traffic. A variety of more sophisticated scheduling schemes are also known that allow multiple degrees of low priority~\cite{pfabric} for different degrees of flexibility in traffic, or weighted prioritization (instead of absolute) with configurable weights~\cite{cisco-priority}. However, regardless of the details of how prioritization is implemented, it offers no guarantees: the performance of deprioritized traffic can be arbitrarily worse than its fair-share of network bandwidth. If there is a surge in high priority traffic, low priority traffic suffers starvation. While starvation can be somewhat reduced using ``aging'', whereby a flow's priority increases over time~\cite{karuna}, this is a heuristic with no guarantees. A broader issue is that priorities are meaningful only relative to other network-wide traffic's priorities and volume, which an individual sender does not know. Thus, even weighted prioritization results in no guarantees: if traffic with many different degrees of flexibility co-exists in a network, each class of flexible traffic has no knowledge of what share of bandwidth its weight will translate to, given that other weights are unknown to it.

\parab{Deadline-aware networking:} In a similar vein to priorities, one can also specify explicit deadlines for different traffic flows to finish~\cite{d2tcp}. Deadlines exhibit some of the same problems as noted for priorities above. In addition, setting deadlines is especially difficult for flows that have soft deadlines, where the performance degrades gradually with time, and it is unclear what to do when a deadline is not met. The flexible workloads we describe are of this type, \eg even if there is no explicit deadline of when an ML model must finish an update, delaying updates gradually increases staleness, and degrades performance. While one can potentially set deadlines for other traffic and use leftover bandwidth for flexible traffic~\cite{karuna}, this provides no guarantees to flexible traffic. As work on deadline-aware scheduling notes, such workloads must be managed using heuristics~\cite{d3}, with no guarantees.

Thus, none of the existing networking primitives allow the specification of a bounded degree of deprioritization, where a guarantee can be specified such that it holds independent of other traffic. Further, primitives to bound starvation can actually cause flexible traffic to not back off appropriately (\eg with weighted prioritization) or worse, increase in aggressiveness (\eg with aging) at unfortunate times when the network is most congested.

Given the above discussion on the limitations of existing primitives, one may even ponder whether any guarantee on bounded deprioritization is possible to frame and achieve at all without explicit reservations, which are unsuitable for our purposes of degraded service. We find that while absolute guarantees are unachievable simply because other traffic is variable and can congest the network, a guarantee can be framed relative a flow's fair-share bandwidth. To do so, we define how aggressive a flexible flow is as follows.

\begin{definition}
\label{def:alpha}
Each flow $f$ has an \textbf{aggressiveness factor $\alpha\in[0,\infty)$}, which specifies the fraction of $f$'s max-min fair-share bandwidth that will be guaranteed for $f$ over its transmission. For regular, inflexible traffic, and more broadly, for any traffic for which $\alpha$ is not explicitly specified, $\alpha=\infty$ by default.
\end{definition}

Like our discussion of the reliability factor above, $\alpha$ is only a lower bound, with a flow seeking at least that fraction of its fair-share bandwidth. TCP, by targeting (but not always achieving) max-min fair-share bandwidth, effectively enforces $\alpha=1$. Flexible flows with $\alpha <1$ thus cede capacity to regular flows, as well as flexible flows with higher $\alpha$ values. In certain cases with other flexible flows present, a flow might actually achieve a rate higher than their max-min fair share. An $\alpha$ value greater than $1$ specifies how much rate speed-up a flexible flow permits. Regular flows want the maximum amount of rate speed-up fairly available, as such for them by default $\alpha$ is set to $\infty$. In practice an $\alpha>1$ is generally unable to be guaranteed as it depends on other flows ceding their fair share (\ie having their $\alpha$ less than 1).

Note that our above framing allows flexible flows to specify a bounded degree of deprioritization compared to their default TCP-driven fair-share state. This is particularly attractive for applications with soft deadlines --- there is no explicit notion of a hard deadline, but performance degrades with time, so specifying an $\alpha <1$ allows specifying how much degradation compared to the default case is acceptable. Indeed, the fair-share depends on other network traffic and varies over time, but it can be independently estimated by regularly probing the network (such a mechanism is described \S\ref{sec:probing-mechanism}).

%%%%%%%%%%%%%%%%%%%%%%%%%%%%%%%%%%%%%%%%%%%%%
%%%%%%%%%%%%%%%%%%%%%%%%%%%%%%%%%%%%%%%%%%%%%
%%%%%%%%%%%%%%%%%%%%%%%%%%%%%%%%%%%%%%%%%%%%%
%%%%%%%%%%%%%%%%%%%%%%%%%%%%%%%%%%%%%%%%%%%%%
%%%%%%%%%%%%%%%%%%%%%%%%%%%%%%%%%%%%%%%%%%%%%

\section{Abstract algorithm design}
\label{sec:abstract-design}

%%%%%%%%%%%%%%%%%%%%%%%%%%%%%%%%%%%%%%%%%%%%%
%%%%%%%%%%%%%%%%%%%%%%%%%%%%%%%%%%%%%%%%%%%%%
%%%%%%%%%%%%%%%%%%%%%%%%%%%%%%%%%%%%%%%%%%%%%
%%%%%%%%%%%%%%%%%%%%%%%%%%%%%%%%%%%%%%%%%%%%%
%%%%%%%%%%%%%%%%%%%%%%%%%%%%%%%%%%%%%%%%%%%%%

\subsection{Budgeting algorithm for bounded degradation}
\label{sec:abstract-algorithm}

The key challenges in enforcing bounded degradation stem from the lack of information: 

\begin{enumerate}
    \item A flow's size may not always be known at its start~\cite{flow-size-in-advance}, and as it may finish anytime, we must ensure that it has delivered at least $r$ fraction of its data at \textit{every instant}.
    \item As flows arrive and finish, a flexible flow's fair share bandwidth continually evolves, and it is unclear when to incur deprioritization such that the $\alpha$-bound holds.
\end{enumerate}

\noindent\sysName addresses these challenge by building up a budget for degradation of flexible flows and then exploiting it. Flexible flows receive normal service for a period such that they run ahead of their performance goals (growing the budget), and then are degraded opportunistically when needed to speed up regular flows (depleting the budget). For now, we assume that the fair share for each flow is known to it at all times. We shall later incorporate probing to estimate the fair share.

\begin{figure}[t]
    \centering
    \includegraphics[width=0.48\textwidth]{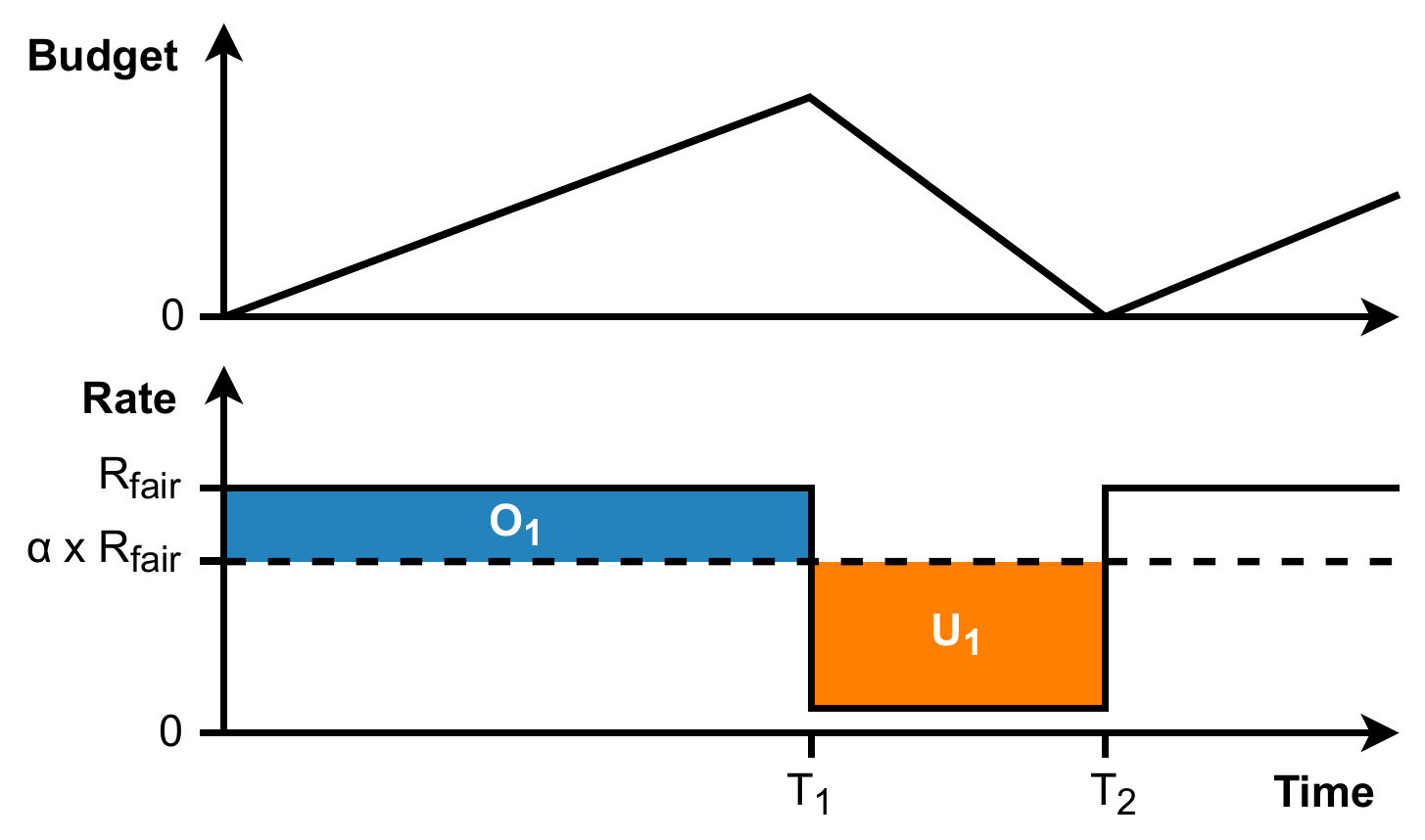} 
    \caption{The algorithm is based upon collecting budget and once it has been built up ($O_1$), having the budget be spent to prioritize regular (more time-critical) flows ($U_1$). Budget can be built up through $\alpha$ by having a rate $>\alpha\times R_{\text{fair}}$ or through $r$ in a similar way.}
    \label{fig:budget_intro}
\end{figure}

In the following, we shall illustrate this budget buildup and exploitation in the context of flexible flows that seek fully reliable delivery ($r=1$), but can tolerate bounded deprioritization ($\alpha <1$). The logic for flows that allow partial delivery but not deprioritization ($r<1$, $\alpha=1$) is similar. For flows that allow both, deprioritization is used first because in any case, low-priority flows cede capacity to regular flows; loss is invoked after the deprioritization budget is fully depleted. 

Regardless of the number of different flows with different degrees, $\alpha_i$, of deprioritization permissible, \sysName uses two priority queues at each switch. Commodity switches used today typically provide at least $8$ priority queues~\cite{cisco8queues}.

We explain the algorithm for switching between high (default, used for regular flows) and low priorities visually using Fig.~\ref{fig:budget_intro}. A flexible flow starts by fairly competing with other flows at high priority, and receives its fair-share rate $R_{fair}$. Until time $T_1$, it sends $S_1$ bytes at this rate. Since we need only guarantee $\alpha\cdot R_{fair}$ bandwidth, this implies that $(1-\alpha)\cdot S_1$ more bytes than needed to meet the guarantee are delivered by $T_1$. This is the area marked $O_1$ in Fig.~\ref{fig:budget_intro}. These ``overdelivered'' bytes represent the accumulated budget.

Once sufficient budget is accumulated, the flow switches to low priority, ceding capacity to other flows, until time $T_2$. At low priority, the flow may potentially receive less than its fair share of bandwidth, thus sending fewer bytes than it could have at the fair share. These ``underdelivered'' bytes, shown in area $U_1$ in Fig.~\ref{fig:budget_intro}, are a depletion of the budget. Given our assumption of knowing the fair share at any time, we can always calculate the depletion in the budget and ensure that the budget is always non-negative. When the budget is completely drained, the flow is switched back to high priority for rebuilding its budget.

Note that in the above, the time when the flow switches back to high priority, $T_2$, is obvious: whenever the budget reaches zero. However, $T_1$ is a design parameter: how much budget to accumulate before switching to low priority? If $T_1$ is small, a flow will switch more frequently, as it builds up and expends buffer. In our abstract description thus far, this is immaterial, but in a practical system, there is a convergence period in which a flow arrives at its new bandwidth allocation each time such a switch takes place, and $T_1$ must be set such that it is several times the convergence time. We provide a detailed evaluation of this parameter in \S\ref{sec:challenge-probing}.

\vspace{0.1in}
\noindent We make several remarks about this simple algorithm:
\begin{enumerate}[topsep=3pt,itemsep=0pt]
    \item Budget building must only use past history, as any assumptions on future bandwidth can be mistaken, and violate the guarantee of $\alpha$.
    \item For partial delivery, a flow can finish anytime, and must have sent $r$ fraction of its data \textit{at every instant}.
    As flow size being known in advance is fundamental to partial delivery, the loss budget is known too, and can be used to improve performance (\S\ref{sec:decision-partial-delivery}).
    \item Low priority can, but does not always drain the budget; the budget is drained only while a flow receives bandwidth smaller than $\alpha\cdot R_{fair}$.
\end{enumerate}

%%%%%%%%%%%%%%%%%%%%%%%%%%%%%%%%%%%%%%%%%%%%%
%%%%%%%%%%%%%%%%%%%%%%%%%%%%%%%%%%%%%%%%%%%%%
%%%%%%%%%%%%%%%%%%%%%%%%%%%%%%%%%%%%%%%%%%%%%
%%%%%%%%%%%%%%%%%%%%%%%%%%%%%%%%%%%%%%%%%%%%%
%%%%%%%%%%%%%%%%%%%%%%%%%%%%%%%%%%%%%%%%%%%%%

\subsection{Max-min guarantee in multi-hop}
\label{sec:achieving-max-min}

Up until now, we have primarily considered the scenario of flows sharing a single link, in which max-min fairness is straight-forward to calculate and reason about. However, in a max-min fair allocation system, downstream changes have upstream effects. More specifically, removing some traffic (or moving it to low priority) on one path can cause performance \textit{degradation} on another.

\begin{figure}
    \centering
    \includegraphics[width=0.48\textwidth]{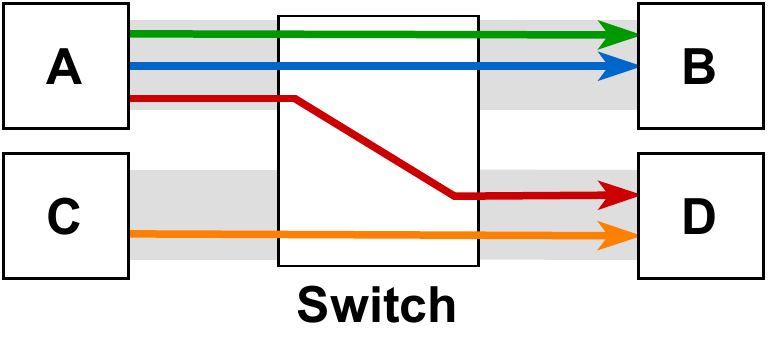}
    \caption{A max-min fair allocation system in which removing one flow between A and B will degrade the performance of the orange flow between C and D. 
    }
    \label{fig:max-min}
\end{figure}

Consider an example with 4 hosts attached to a switch, as shown in Fig.~\ref{fig:max-min}. Two flows are going from A to B, one from A to D and one from C to D. If all the flows are at the same priority (default), flows that originate from A have 1/3 units of bandwidth, while the orange C-D flow utilizes 2/3 units of bandwidth. If one flow from A to B switches to low priority, e.g. the top (green) flow, it will receive 0 units, while all other flows would get 1/2, including the (orange) one between C and D. This further implies that, the regular C-D flow experiences a slowdown of 0.25 compared to the default, just because a flexible flow on another path switches to low priority.

Note however, that the above is not a violation of our bounded deprioritization guarantee: for any flow, it gets $\alpha$ fraction of its fair-share bandwidth, but due to the nature of max-min fair allocation, the fair share can itself change as other flows shift their priorities. This is precisely why we do not state (and cannot fulfill) guarantees framed in terms of flow completion time. The above discussion is meant to highlight that the fair-share-degradation guarantee is weaker than an FCT-degradation guarantee, and \sysName only provides the former. The fair-share-degradation guarantee of a flow is only as strong as the fair share it can achieve given the priorities of other flows, which thus can be less than its fair share when all flows are of equal priority. We refer to this phenomena as flow fair share interdependence.

%%%%%%%%%%%%%%%%%%%%%%%%%%%%%%%%%%%%%%%%%%%%%
%%%%%%%%%%%%%%%%%%%%%%%%%%%%%%%%%%%%%%%%%%%%%
%%%%%%%%%%%%%%%%%%%%%%%%%%%%%%%%%%%%%%%%%%%%%
%%%%%%%%%%%%%%%%%%%%%%%%%%%%%%%%%%%%%%%%%%%%%
%%%%%%%%%%%%%%%%%%%%%%%%%%%%%%%%%%%%%%%%%%%%%

\section{Practical algorithm design}
\label{sec:from-theory-to-practice}

The abstract budgeting algorithm as described in \S\ref{sec:abstract-design} suffices to achieve guaranteed partial delivery regardless of a real transport implementation's quirks, as long as fully reliable delivery is correctly implemented using acknowledgments and retries. However, the bounded deprioritization guarantee relies on two assumptions: (a) each flow knowing its fair share at all times; and (b) the transport protocol converging to fair share instantly after flow arrivals, departures, and priority switches. In practice, these assumptions do not hold. For (b), the transport protocol's degree of divergence from fair share is, of course, inherited by \sysName. For (a), in our approach we choose to have each flow probe the network periodically to approximately estimate its fair share, as we describe next. Having only approximate estimates for fair-share bandwidth, and incurring some time for convergence to the fair share, both lead to imprecision in the $\alpha$-bound, which we later explore experimentally.

%%%%%%%%%%%%%%%%%%%%%%%%%%%%%%%%%%%%%%%%%%%%%
%%%%%%%%%%%%%%%%%%%%%%%%%%%%%%%%%%%%%%%%%%%%%
%%%%%%%%%%%%%%%%%%%%%%%%%%%%%%%%%%%%%%%%%%%%%
%%%%%%%%%%%%%%%%%%%%%%%%%%%%%%%%%%%%%%%%%%%%%
%%%%%%%%%%%%%%%%%%%%%%%%%%%%%%%%%%%%%%%%%%%%%

\subsection{Probing mechanism}
\label{sec:probing-mechanism}

We propose the usage of regular probing for each flexible flow to determine its fair share. The key challenge is estimating a flow's fair share through probing is that all flows must operate approximately in synchronous fashion at high priority in the probing period, such that each flow can independently arrive at its fair share estimate. We rely on a data center wide synchronization primitive for this purpose. While recent work has claimed data center wide synchronization within tens of nanoseconds~\cite{syncGoogle, syncHakim}, for our purposes, microsecond-scale synchronization using PTP~\cite{ptpimplementation} is sufficient. Note that while such synchronized probing may sound aggressive, it is merely a brief return to today's default behavior, with more operational time spent using the results of probing.

Each flexible flow operates in a (synchronized) cyclic fashion with three phases: (a) \textit{warmup}, with all flexible flows at high priority; (b) \textit{measure}, with all flexible flows at high priority; and (c) \textit{exploit}, where each flow individually decides whether to switch to low priority. The warmup phase allows time for convergence to fair share, before a fair-share estimate is made from the measure phase. We define the following parameters:

\begin{itemize}
    \item $T_{int}$: interval duration (ms)
    \item $D_{warmup}$: warmup phase duration (no. intervals)
    \item $D_{measure}$: measure phase duration (no. intervals)
    \item $D_{exploit}$: exploit phase duration (no. intervals)
\end{itemize}

Once the fair share estimate is available after the measure phase, each flow calculates the amount of potential expense (\ie budget loss) of going low priority in the upcoming phase. This expense amounts to $\alpha\times R_{fair}\times D_{exploit}\times T_{int}$. Only if the current budget is larger than the expense, does the flexible switch to low priority for the exploit phase.

The fair share estimate which is established in the measure phase is used throughout all subsequent intervals to accumulate or drain budget. As such, although a flexible flow can only switch to low priority during the exploit phase, it collects budget in all phases (with the exclusion of the first partial cycle it needs to synchronize up).

Note that explicit coordination is not required across flows: the lengths of the warmup, measure, and exploit phases are set network-wide. A newly started flexible flow starts at high priority until the next synchronized cycle starts, for which its uses its own (network-wide synchronized) clock. It can calculate at any time which part of the cycle is currently active using one simple modulo operation. If the timesteps are on the order of hundreds of microseconds or more, microsecond-scale synchronization is clearly sufficient. Additionally note that if transport protocol convergence were faster, \eg by approximating RCP in modern programmable switches~\cite{power-of-flexible-packet-processing} or implementing PERC~\cite{jose2015high}, we could eschew the warmup phase entirely, thus increasing the fraction of time flows can operate in the exploit phase.

%%%%%%%%%%%%%%%%%%%%%%%%%%%%%%%%%%%%%%%%%%%%%
%%%%%%%%%%%%%%%%%%%%%%%%%%%%%%%%%%%%%%%%%%%%%
%%%%%%%%%%%%%%%%%%%%%%%%%%%%%%%%%%%%%%%%%%%%%
%%%%%%%%%%%%%%%%%%%%%%%%%%%%%%%%%%%%%%%%%%%%%
%%%%%%%%%%%%%%%%%%%%%%%%%%%%%%%%%%%%%%%%%%%%%

\subsection{Algorithm}

\parab{Budget adjustment algorithm.} The core of the algorithm is the manner in which it accounts budget, which is abstractly described in \S\ref{sec:abstract-design}. The algorithm requires a fair share estimate $R_{fair}$ to account for the budget accumulation or drain. The budget must be updated at a time granularity fine enough to (a) incorporate the frequency at which the fair share estimate is renewed, and (b) before a priority decision is made. In the probing mechanism, this is performed at $T_{int}$ interval granularity, which meets both criteria. In the adjustment algorithm, two budgets are maintained:

\begin{itemize}

    \item Rate degradation budget $B_{\alpha}$. It increases each update when the actual rate $R_{actual}$ is higher than $\alpha\times R_{fair}$.
    
    \item Reliable delivery degradation budget $B_{r}$. With the assumption that the flow size $F$ is known in advance, $B_{r}$ starts at the $(1-r)\times F$. Only if $r<1$ is the flow size required in the algorithm. We leave exploration of other (later) signaling of flow size to future work.
    
\end{itemize}

The budget is decreased by first draining the rate degradation budget $B_{\alpha}$, and only if that is empty, draining $B_{r}$. The reasoning behind this is that draining $B_{r}$ results in less flow data being sent out, which is undesirable if the targeted fair share portion is still being achieved. The $B_{\alpha}$ can become negative if consistently we achieve lower rate than intended -- which in turn will trigger going at high priority to remedy this (see next section). The budget algorithm pseudo-code is depicted in Alg.~\ref{alg:budget}.

\begin{algorithm}[ht!]
\caption{Budget adjustment logic for each elapsed time period of the flow in which the fair share rate $R_{fair}$ is known.}
\label{alg:budget}
\begin{algorithmic}[1]
\Statex \textbf{Initial state:}
\Statex $F$: flow size (known if $r<1$, else $\infty$), $S_{sent}=0$, $B_{\alpha}=0$, $B_r = (1-r)\times F$ % $B_{r}=0$ \textbf{if} flow size is not known \textbf{else} $B_r = (1-r)\times F$
\Statex ~
\Function{ADJUST\_BUDGET}{$ACK_{byte}$, $T_{elapsed}$}
    \State $R_{actual} = ACK_{byte}\;/\;T_{elapsed}$
    \State $S_{sent} = S_{sent} + ACK_{byte}$
    \State $B_\alpha = B_\alpha - T_{elapsed} * (\alpha\times R_{fair} - R_{actual})$
    % \If{flow size is not known}
    %     \State $B_r = B_r + (1 - r) * ACK_{byte}$
    % \EndIf
    \If{$B_\alpha < 0$}
        \State $B_{r,before} = B_r$
        \State $B_r = B_r + B_\alpha$
        \If{$B_r >= 0$}
            \State $B_\alpha = 0$
        \Else
            \State $B_\alpha = B_r$ 
            \State $B_r = 0$
        \EndIf
        \State $S_{sent} = S_{sent} + (B_{r,before} - B_r)$\Comment{$B_r$ drainage reduces how much to send}
    \EndIf
\EndFunction
\end{algorithmic}
\end{algorithm}

%%%%%%%%%%%%%%%%%%%%%%%%%%%%%%%%%%%%%%%%%%%%%
%%%%%%%%%%%%%%%%%%%%%%%%%%%%%%%%%%%%%%%%%%%%%
%%%%%%%%%%%%%%%%%%%%%%%%%%%%%%%%%%%%%%%%%%%%%
%%%%%%%%%%%%%%%%%%%%%%%%%%%%%%%%%%%%%%%%%%%%%
%%%%%%%%%%%%%%%%%%%%%%%%%%%%%%%%%%%%%%%%%%%%%

\parab{Control algorithm with probing.} The control algorithm has four tasks: (1) it determines whether the flow is finished earlier due to reduced reliability, (2) it awaits to be synchronized with the other flexible flows, (3) after a complete warmup and measure phase, it calculates the fair share which is used in the next cycle, and (4) it decides whether to go at low or high priority during the exploit phase. In each update, the current interval is identified with $ID_{int}$, which is determined using a modulo operation of the clock and the interval duration $T_{int}$. The control algorithm pseudo-code is depicted in Alg.~\ref{alg:control}.

\begin{algorithm}[ht!]
\caption{Control algorithm with probing. The update is called at the $T_{int}$ interval with an incremented $ID_{int}$ each time. The $ID_{int}$ is determined using the global synchronized clock each flexible-capable host maintains.}
\label{alg:control}
\begin{algorithmic}[1]
\Statex \textbf{Initial state:}
\Statex $R_{fair} = $\textbf{ not set}, $P_{prev}=\;$NONE, SYNC = \textbf{false}, $D_{probe}=0$, $T_{probe}=0$, $I_{exploit\;low}$ = \textbf{false}
\Statex ~
\Function{UPDATE}{$ID_{int}$, $ACK_{byte}$, $T_{elapsed}$}
    \If{$R_{fair}$ \textbf{is set}}
        \State \texttt{ADJUST\_BUDGET}($ACK_{byte}$, $T_{elapsed}$)
    \EndIf
    \If{$S_{sent} \geq F$}
        \State \textbf{return} FINISHED
    \EndIf
    \If{$P_{prev}$ == MEASURE \textbf{and} SYNC}
        \State $D_{probe} = D_{probe} + ACK_{byte}$ 
        \State $T_{probe} = T_{probe} + T_{elapsed}$
    \EndIf
    \State $P_{upcoming}$ = \texttt{DETERMINE\_PHASE}($ID_{int}$, $D_{warmup}$, $D_{measure}$, $D_{exploit}$)
    \If{$P_{prev}$ == EXPLOIT \textbf{and} $P_{upcoming}$ == WARMUP}
        \State SYNC = \textbf{true}
        \State $I_{exploit\;low}$ = \textbf{false}
    \EndIf
    \If{$P_{prev}$ == MEASURE \textbf{and} $P_{upcoming}$ == EXPLOIT \textbf{and} SYNC}
        \State $R_{fair} = D_{probe}  / T_{probe}$
        \State $D_{probe} = 0,\;\;T_{probe} = 0$
    \EndIf
    \State $L_{potential}$ = $\alpha \times R_{fair} \times D_{exploit} \times T_{int}$
    \If{$P_{prev}$ == MEASURE \textbf{and} $P_{upcoming}$ == EXPLOIT \textbf{and} $R_{fair}$ \textbf{is set}}
        \State $I_{exploit\;low}$ = \textbf{true} \textbf{if} $B_{\alpha} + B_{r} > L_{potential}$ \textbf{else} \textbf{false}
    \EndIf
    \State $P_{prev} = P_{upcoming}$
    \State \textbf{return} LOW \textbf{if} $I_{exploit\;low}$ \textbf{else} HIGH
\EndFunction
\end{algorithmic}
\end{algorithm}

%%%%%%%%%%%%%%%%%%%%%%%%%%%%%%%%%%%%%%%%%%%%%
%%%%%%%%%%%%%%%%%%%%%%%%%%%%%%%%%%%%%%%%%%%%%
%%%%%%%%%%%%%%%%%%%%%%%%%%%%%%%%%%%%%%%%%%%%%
%%%%%%%%%%%%%%%%%%%%%%%%%%%%%%%%%%%%%%%%%%%%%
%%%%%%%%%%%%%%%%%%%%%%%%%%%%%%%%%%%%%%%%%%%%%

\subsection{Probing consequences and limitations}
\label{sec:probing-consequences-limitations}

The introduction of probing to periodically determine the fair share has certain consequences and limitations.

\parab{What is the effect of an inaccurate fair share estimate?} An inaccurate fair share estimate can either be too low or too high. In the former (too low estimate), the budget will increase too much as the gap $R_{actual} - \alpha\times R_{fair}$ is larger than in actuality, leading potentially to an inflated budget and thus a rate which can deteriorate beyond the theoretical guarantee. In the latter (too high estimate), the budget will not build up as much, thus resulting in less flexibility. An inaccurate estimate can be due to several factors. Firstly, it takes time for a congestion control protocol to converge, which can lead to inaccuracies. This can be ameliorated by having large enough phases, as well as the choice of a quickly converging congestion control protocol. Secondly, the measured fair share can be out of date because new flows can arrive or existing ones depart. This is especially the case when there are flows present whose completion time last shorter than a full cycle, which would lead to a reduced rate during whichever phase it is active.

\parab{Flexibility limited to relatively large flows.} We acknowledge that our service degradation primitives are only meaningful for relatively large flows, which run for at least several milliseconds. We believe this to be reasonable: there is, after all, little advantage from degrading service for short flows that don't consume most of the network's capacity. However, by degrading service for a few such large flows, we can benefit a large number of short flows, which typically have more stringent service requirements.

\parab{Speed-up limited to exploit phase.} In the warmup and measure phases the flexible flow is unable to be at low priority. As a consequence, the maximum fraction of time a flexible flow can be at low priority is $D_{exploit} / (D_{warmup} + D_{measure} + D_{exploit})$. This is especially impactful if there are short regular flows, as if they occur in the first two phases they will receive only fair share rather than be sped up. This can be ameliorated with a relatively longer exploit phase, which in turn however would reduce the fair share estimate accuracy.

\parab{Time synchronization required among flexible hosts.} Participating hosts which wish to participate in the starting of flexible flows must be time synchronized. As such, the deployability of \sysName is limited to those settings.

%%%%%%%%%%%%%%%%%%%%%%%%%%%%%%%%%%%%%%%%%%%%%
%%%%%%%%%%%%%%%%%%%%%%%%%%%%%%%%%%%%%%%%%%%%%
%%%%%%%%%%%%%%%%%%%%%%%%%%%%%%%%%%%%%%%%%%%%%
%%%%%%%%%%%%%%%%%%%%%%%%%%%%%%%%%%%%%%%%%%%%%
%%%%%%%%%%%%%%%%%%%%%%%%%%%%%%%%%%%%%%%%%%%%%

\section{Implementation}
\label{sec:fdt-implementation}

For ease of understanding, we have thus far described \sysName at an abstract, algorithmic level. Implementing it requires small changes at multiple levels of the networking stack, which we discuss in a top-down manner.

\parab{Application changes:} Applications that wish to send flexible traffic must correctly use an extended sockets API to specify the degradation parameters for their traffic. An application using only deprioritization does not see any change in terms of transport behavior, except performance. However, applications allowing partial delivery might see data loss, and thus must prepare data correctly such that partial deliveries are usable. 

\sysName must be able to discard data if reduced reliability is permitted. When \sysName decides to discard some data segments, it will remove these data segments from the send buffer. As such, application have to ensure that \textit{any} data segment transmitted over a flexible socket can be removed from the flow without hindering the receiver from being able to make use of the rest of the data segments. \sysName must similarly be informed of the application data segmentation strategy, which could range from a static constant (\eg every $B$ bytes is an individual data segment) to dynamically informed strategies. This would incur a header overhead for each data segment. In this work, we assume that data is segmentable at the byte level. We leave investigation of data segmentation and discard decision strategies to future work.

\parab{Application-transport interface changes:} The sockets API has to be extended to support the creation of flexible flows. For every flexible flow, the application must be able to specify the parameters $\alpha$ and $r$ through the socket API. By default, $\alpha$ is set to $\infty$ and $r$ is set to $1$, such that legacy applications do not need to change, and can continue operating as regular traffic, while benefiting from flexible traffic's behavior. 

\parab{Transport level changes at hosts:} Our design requires that the transport implementation be extended with the \sysName state machine. For each flexible flow, it must track the budget and the time slots (warmup, probe, exploit) using a system time synchronization protocol like PTP, perform fair-share measurements, decide on whether to operate in low or high priority or / and discard or retransmit data, and tag packets with the appropriate priority levels.

\sysName benefits if the transport protocol converges fast to fair-share bandwidth. Thus, in our implementation, we use DCTCP instead of TCP. DCTCP, by making use of congestion markings (ECN), can make more accurate rate adjustment decisions than TCP. We use DCTCP because it is already commonly used in data centers, but if RCP or PERC were implemented to achieve even faster convergence, \sysName's performance would improve further.

\parab{Network-level changes at switches:} \sysName's bounded deprioritization uses $2$ priority queues, while commodity data center switches commonly support $8$ queues. The remaining available priority levels can still be utilized for explicit prioritization outside the domain of \sysName, \eg to prioritize network control messages. 

An alternative to such in-network prioritization is to simply have ``wimpier'' congestion control logic kick in when flows are operating in degraded mode, \eg back off by a larger multiplicative decrease factor on packet loss than regular flows. We have not explored such a strategy yet.

%%%%%%%%%%%%%%%%%%%%%%%%%%%%%%%%%%%%%%%%%%%%%
%%%%%%%%%%%%%%%%%%%%%%%%%%%%%%%%%%%%%%%%%%%%%
%%%%%%%%%%%%%%%%%%%%%%%%%%%%%%%%%%%%%%%%%%%%%
%%%%%%%%%%%%%%%%%%%%%%%%%%%%%%%%%%%%%%%%%%%%%
%%%%%%%%%%%%%%%%%%%%%%%%%%%%%%%%%%%%%%%%%%%%%

\section{Evaluation}
\label{sec:fdt-evaluation}

We evaluate \sysName using packet-level simulations. We first explain the experimental setup including network devices, congestion control, and default parameterization in \S\ref{sec:fdt-experimental-setup}. Next, we show the motivation of \sysName by comparing it in a simple scenario against other prioritization schemes in \S\ref{sec:prior-scheme-comparison}, and showcase how partial delivery works in \S\ref{sec:decision-partial-delivery}. We explore the parameterization of the probing mechanism in \S\ref{sec:challenge-probing}. Finally, we run \sysName for larger workloads to examine the utility that flexible flows offer and compare it to the performance of fixed weighted prioritization in \S\ref{sec:utility-of-flexibility}.

%%%%%%%%%%%%%%%%%%%%%%%%%%%%%%%%%%%%%%%%%%%%%
%%%%%%%%%%%%%%%%%%%%%%%%%%%%%%%%%%%%%%%%%%%%%
%%%%%%%%%%%%%%%%%%%%%%%%%%%%%%%%%%%%%%%%%%%%%
%%%%%%%%%%%%%%%%%%%%%%%%%%%%%%%%%%%%%%%%%%%%%
%%%%%%%%%%%%%%%%%%%%%%%%%%%%%%%%%%%%%%%%%%%%%

\subsection{Experimental setup}
\label{sec:fdt-experimental-setup}

For the packet-level simulations we make use of ns-3~\cite{ns3} with the basic-sim module~\cite{basic-sim, abbas2020securing, hanjing2020making, hypatia}. The default setup is as follows (unless otherwise specified in the upcoming sections). The network consists of 10~Gbit/s links with a delay of 1~$\mu s$. Each network device has two traffic control queue discipline queues, with a weighted scheduling mechanism which it draws from using deficit round robin. The high priority queue has 90\% of bandwidth, and the low priority 10\% of bandwidth. Both traffic control queues are configured with Random Early Detection (RED) as its queue discipline, with a hard marking threshold of 65 packets and a maximum size of 466 packets (both following recommendation from \cite{dctcp}). Each network device finally has a simple tail drop queue of 20 packets. Network devices use an MTU of 1500 byte.

We use DCTCP as the congestion control protocol. The various timeout values have been adjusted to be effective in a low latency (2-4$~\mu s$ base RTT), high throughput (10~Gbit/s) environment, in particular the minimum RTO is set to 1~ms, the initial RTT measurement to 2~ms, the connection timeout to 2~ms, the delayed ACK timeout to 1~ms, the persist timeout to 8~ms and the maximum segment lifetime to 8~ms. The segment size is set to 1380~byte. The clock granularity is set to 1~ms. It is not possible to have the throughput match the theoretical line rate due to the overhead of the point-to-point, IP and TCP (with TS option) headers. As a result, approximately 96\% of line rate is the maximum throughput. The send and receive buffers are set to 1~MiB.

\sysName is by default configured with $T_{int}=5$~ms, and with $D_{warmup} : D_{measure} : D_{exploit}$ set to 1~:~1~:~3. This matches the low latency and relatively high throughput setting. Parameterization is further explored in \S\ref{sec:challenge-probing}.

%%%%%%%%%%%%%%%%%%%%%%%%%%%%%%%%%%%%%%%%%%%%%
%%%%%%%%%%%%%%%%%%%%%%%%%%%%%%%%%%%%%%%%%%%%%
%%%%%%%%%%%%%%%%%%%%%%%%%%%%%%%%%%%%%%%%%%%%%
%%%%%%%%%%%%%%%%%%%%%%%%%%%%%%%%%%%%%%%%%%%%%
%%%%%%%%%%%%%%%%%%%%%%%%%%%%%%%%%%%%%%%%%%%%%

\subsection{Prioritization scheme comparison}
\label{sec:prior-scheme-comparison}

\textbf{Use cases.} The difference of \sysName to the prioritization alternatives described in \S\ref{sec:bounded-deprioritization} is best illustrated by an example. Consider a single link with a bandwidth of 10~Gbit/s over which two flows are started. We bring forward two scenarios, the first to showcase speed-up and the second to showcase starvation. In both cases, the flexible flow wants its performance to be degraded by at most 10\% of its fair share ($\alpha=0.9$). The regular flow expects usual performance and thus does not want its performance degraded.

\begin{itemize}
    \item Case 1: 40~Gbit flexible flow starts at T=0~s, and a 2~Gbit regular flow starts at T=2~s
    \item Case 2: 40~Gbit regular flow starts at T=0~s, and a 2~Gbit flexible flow starts at T=2~s
\end{itemize}

\begin{figure}[t]
    \centering
    \includegraphics[width=0.48\textwidth]{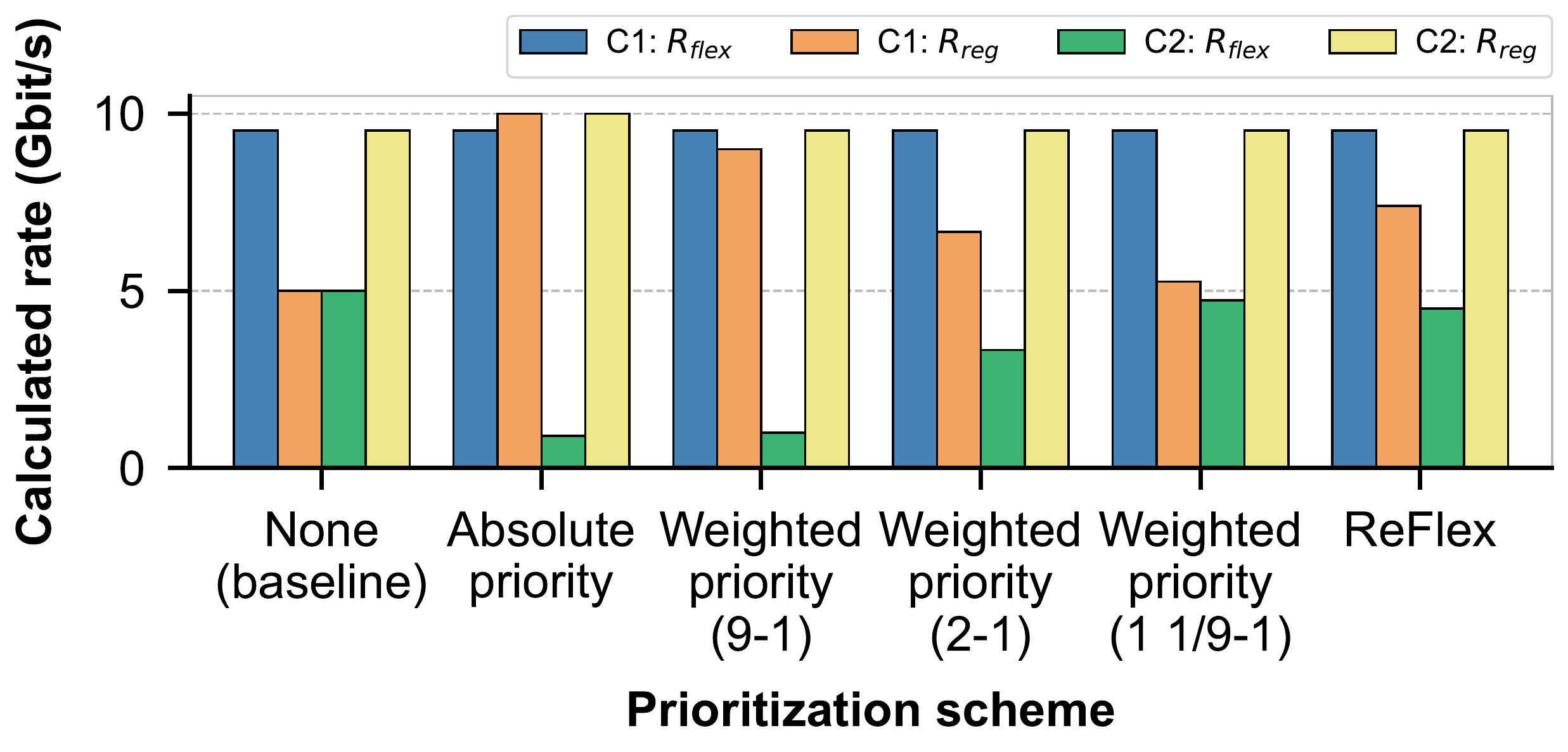} 
    \caption{Calculation of different prioritization schemes for the two cases.}
    \label{fig:fdt-priority-calculation}
\end{figure}

\begin{figure}[t]
    \centering
    \includegraphics[width=0.48\textwidth]{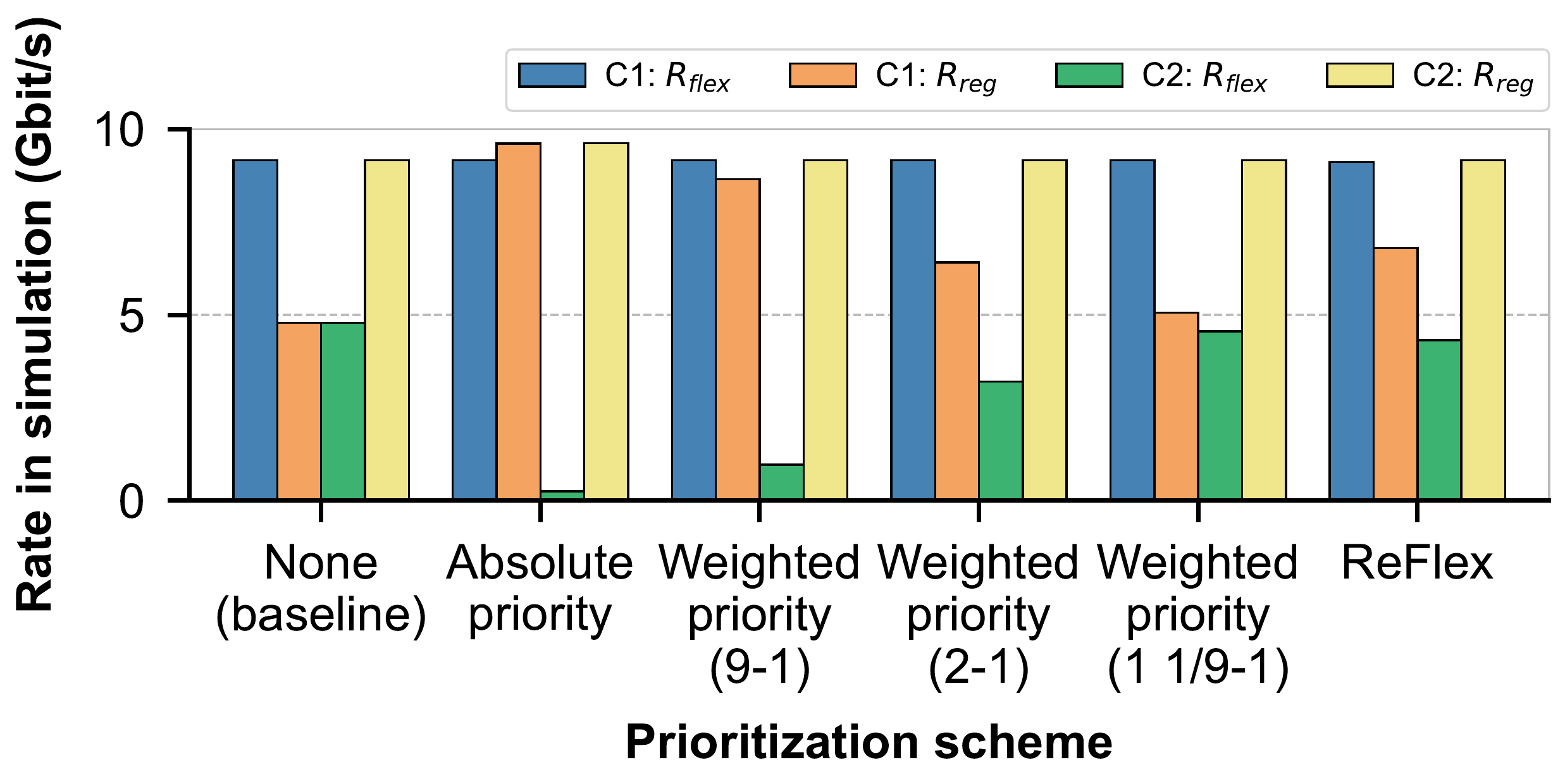} 
    \caption{Packet-level simulation results of different prioritization schemes for the two cases.}
    \label{fig:fdt-priority-simulation}
\end{figure}

\begin{figure*}[t]
	\centering
    \begin{subfigure}[b]{0.46\textwidth}
        \centering
        \includegraphics[width=\textwidth]{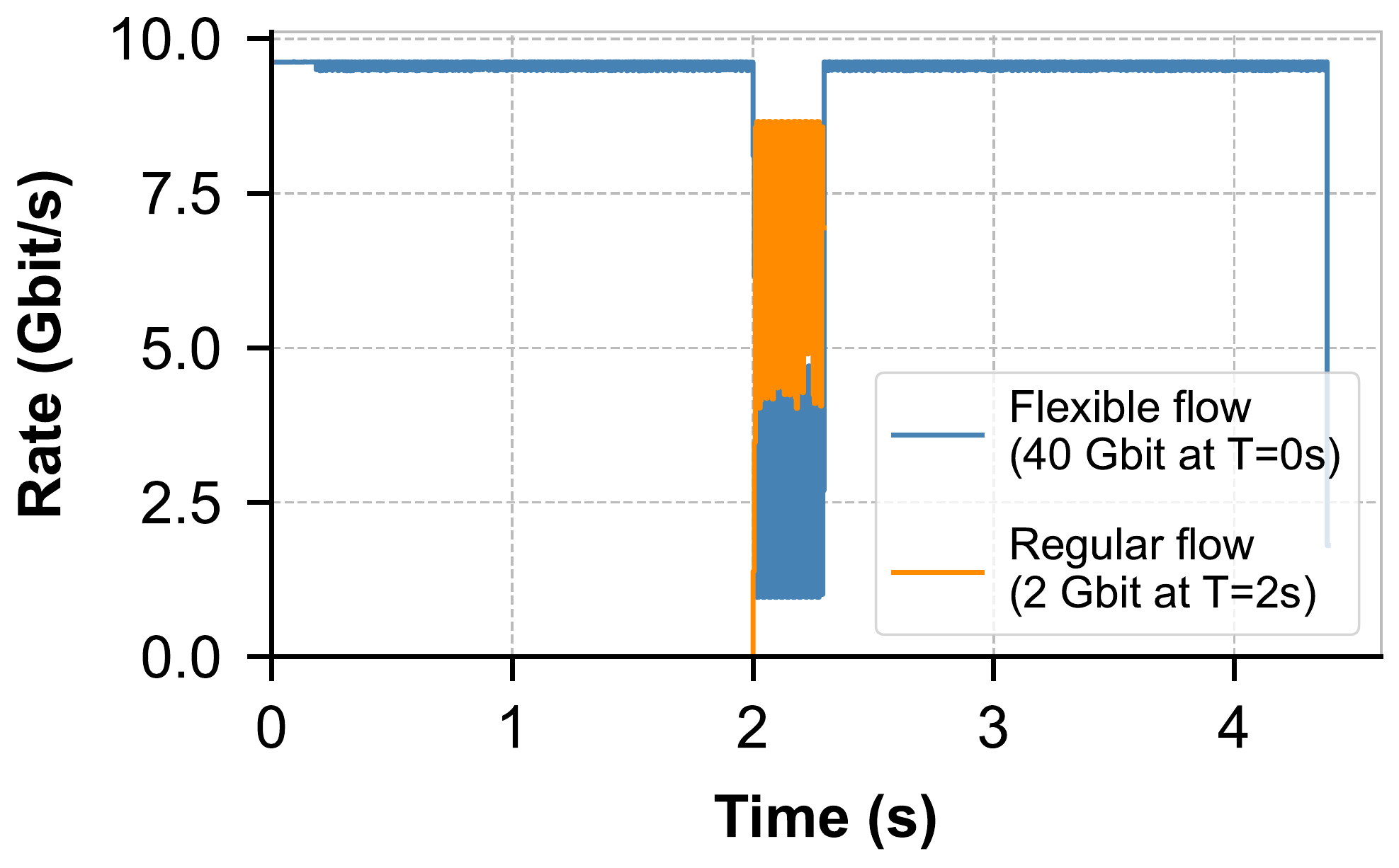}
        \caption{Entire timespan}
    \end{subfigure}
    \hfill
    \begin{subfigure}[b]{0.46\textwidth}
        \centering
        \includegraphics[width=\textwidth]{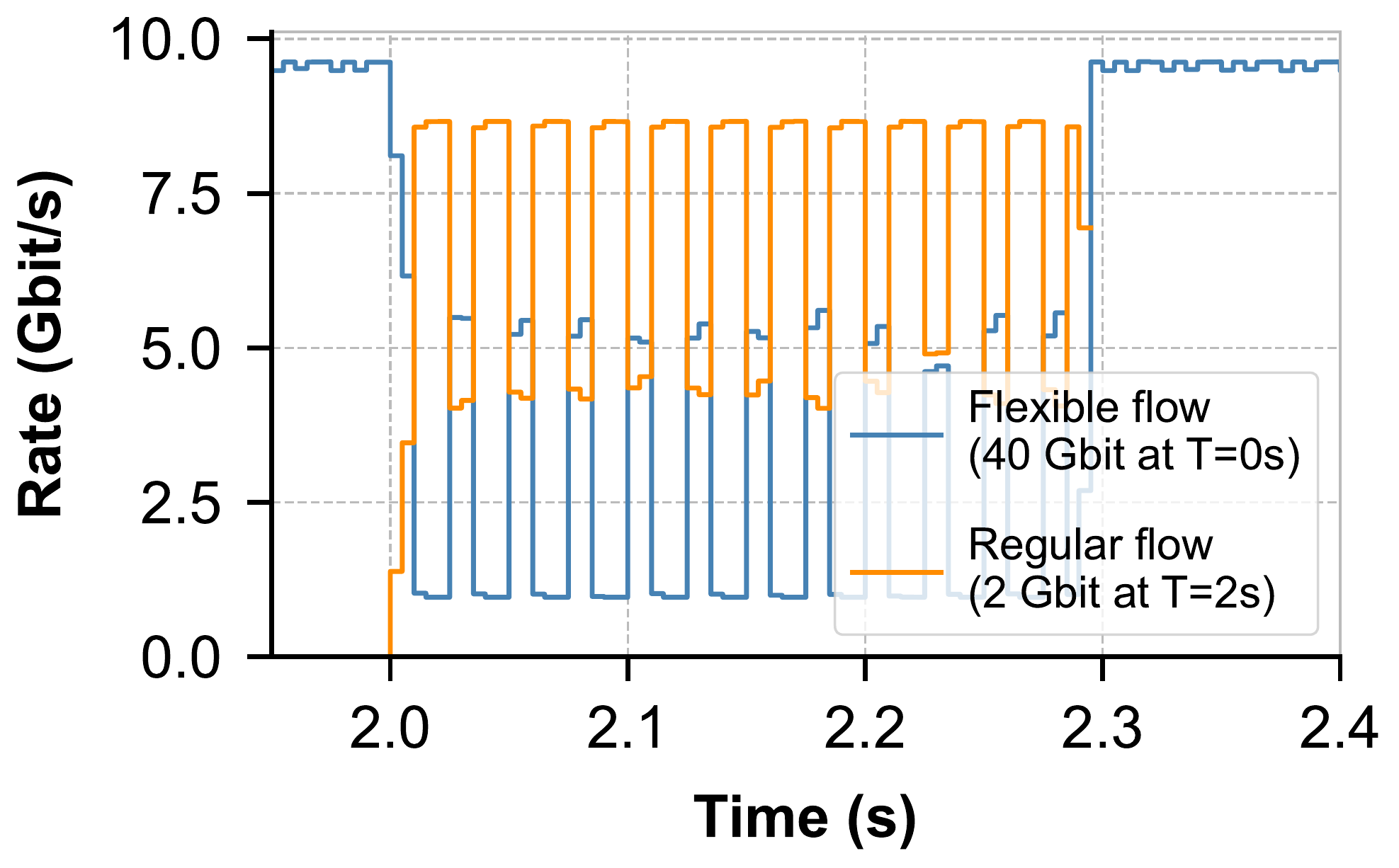}%
        \caption{Period of competition: 2.0s--2.4s}
    \end{subfigure}
    \caption{For Case 1, \sysName yields considerable speed-up for the regular flow, although it its effect is reduced due to the warmup and measure phases.}
    \label{fig:fdt-priority-reflex-rates}
    \vspace{-6pt}
\end{figure*}

\textbf{Results.} In the baseline setting (\ie without any prioritization scheme), the flows compete fairly and thus in both cases the flexible and regular flow achieve 5~Gbit/s when both are active. When using absolute prioritization, the flexible flow achieves 0~Gbit/s when both are active, and the regular flow 10~Gbit/s. Thus, in Case 2 this result in the flexible flow being finished only after the regular flow is done. For weighted prioritization, we consider three variants for comparison: regular flows having 9x more weight, 2x more weight, and 1$\frac{1}{9}$x more weight. Among these three variants, either there is little speed-up of the regular flow in Case 1, or starvation of the flexible flow in Case 2. In Case 1, \sysName builds up budget in the first 20~s which means that it can be on low priority when the regular flow arrives, giving it all the high priority bandwidth. However, because of the presence of probing with 1:1:3 phases, the high priority can only be consumed 60\% of the time. This is shown in Fig.~\ref{fig:fdt-priority-reflex-rates} in the changing between the fair share at high priority and the reduced rate at low priority. Thus, the regular flow rate calculated rate is 7.4~Gbit/s. In Case 2, \sysName make sure the flexible flow to maintain the $\alpha=90\%$ guarantee, therefore not permitting the regular flow to degrade the flexible flow's rate too much. This yields a rate of 4.5~Gbit/s for the flexible flow. An overview of the average rates achieved is shown in Fig.~\ref{fig:fdt-priority-calculation} for calculation. Barring the header overhead, the calculation results match those of the simulation in Fig.~\ref{fig:fdt-priority-simulation}. Note that in Fig.~\ref{fig:fdt-priority-reflex-rates} there is a small rate fluctuation for the flexible flow, which is caused by retransmission due to reordering. Among the compared prioritization schemes, \sysName is able to (a) achieve a significant speed-up of the regular flow in Case 1, and (b) prevent starvation of the flexible flow in Case 2 comparable to the baseline. This is evidently workload dependent: if flexible flows are not starved, other fixed (weighted) prioritization schemes perform better.

It is possible to achieve better speed-up in Case 1 through either increasing weight of the high priority queue, or, more effectively, by changing the ratio of the probing phases to have a relatively longer exploit phases. The former has as a side effect that congestion control protocols do not respond well to complete starvation, as they will be deprived of any network signals. The effects of the latter are further explored in \S\ref{sec:challenge-probing}.

\textbf{Different from  deadlines.} The most distinguishing characteristic of \sysName is that it defines its guarantees in relation to the network state, in particular the fair share each flow should achieve. This is distinctly different from explicit deadlines, which are set agnostic to network state. Indirectly, an absolute time deadline corresponds to a certain desired rate provided the data transfer size. A simple approach would be to set these absolute time deadlines based on the line rate. With this approach, for Case 1 one would set the deadline of the flexible to T=4.4~s (4~s + 10\% for flexibility) and of the regular to T=2.2~s, and in Case 2 of the flexible to T=2.2~s (0.2~s + 10\% for flexibility) and of the regular to T=4~s. Case 1 would be possible for the deadlines to be met, however in Case 2 it is not feasible. Of course, one can also have or set more relaxed deadlines, which still nevertheless is based on an assumption of the number of present flows and their target rate. The deadline-aware mechanism, both centralized scheduler such as FastPass~\cite{fastpass} and decentralized such as D2TCP~\cite{d2tcp}, must decide how to handle infeasible deadlines as well as how to accommodate the non-deadline-aware traffic.

\greybox{\textbf{Priority scheme comparison takeaways:} \sysName makes use of probing to prevent a flexible flow from being starved. It is in low priority as long as it cumulatively receives the portion of fair share it requires. The potential benefit of \sysName is (a) \textit{workload dependent} because it requires flexible flows to be starved to outperform other prioritization schemes, and (b) \textit{objective dependent} as it is useful only if it is desirable in moment of high utilization to not fully prioritize regular flows.}

% Partial delivery
\begin{figure*}[t]
	\centering
    \begin{subfigure}[b]{0.46\textwidth}
        \centering
        \includegraphics[width=\textwidth]{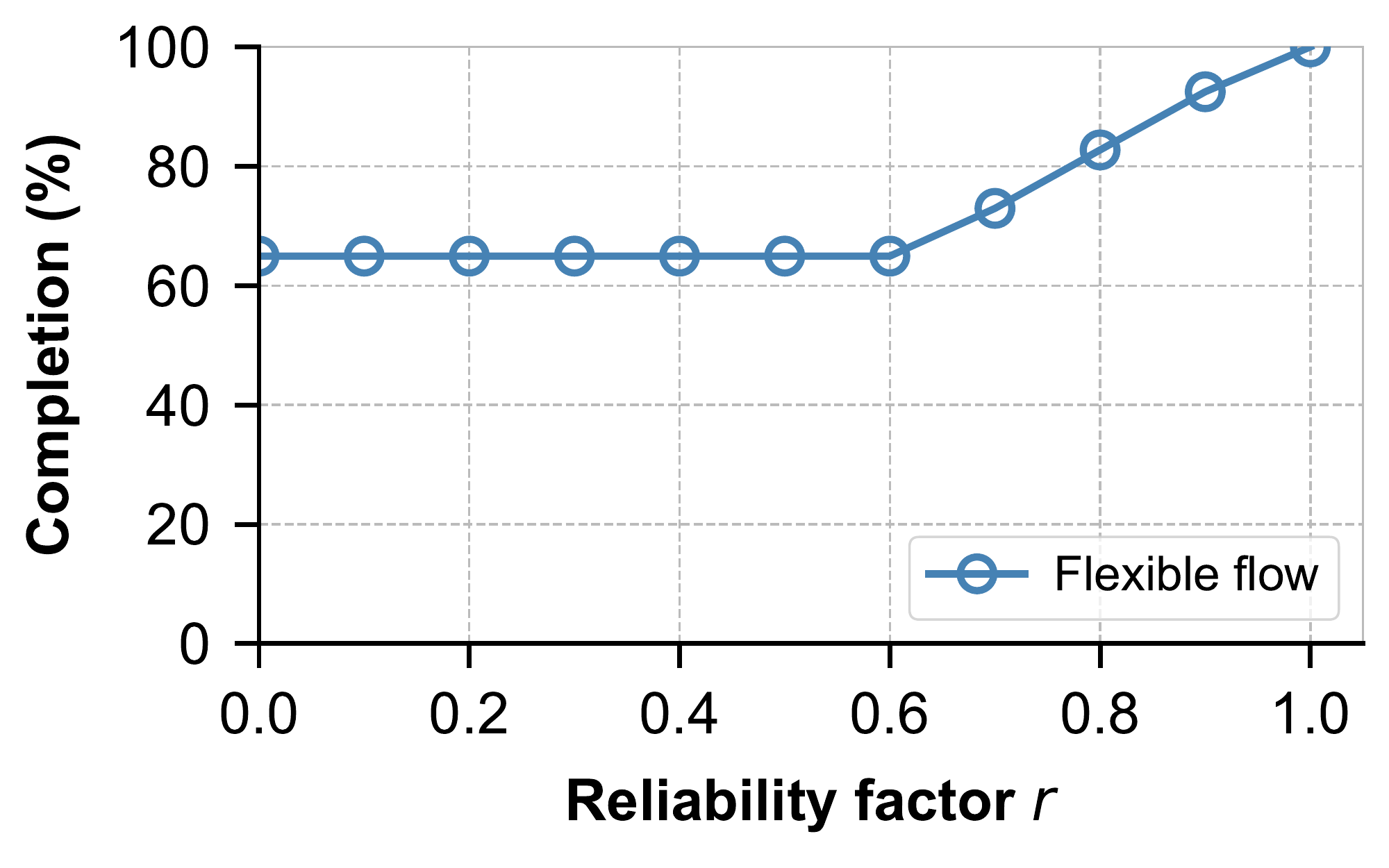}%
        \caption{Partial delivery}
        \label{fig:partial-delivery:reliability-achieved}
    \end{subfigure}
 	\hfill
    \begin{subfigure}[b]{0.46\textwidth}
        \centering
        \includegraphics[width=\textwidth]{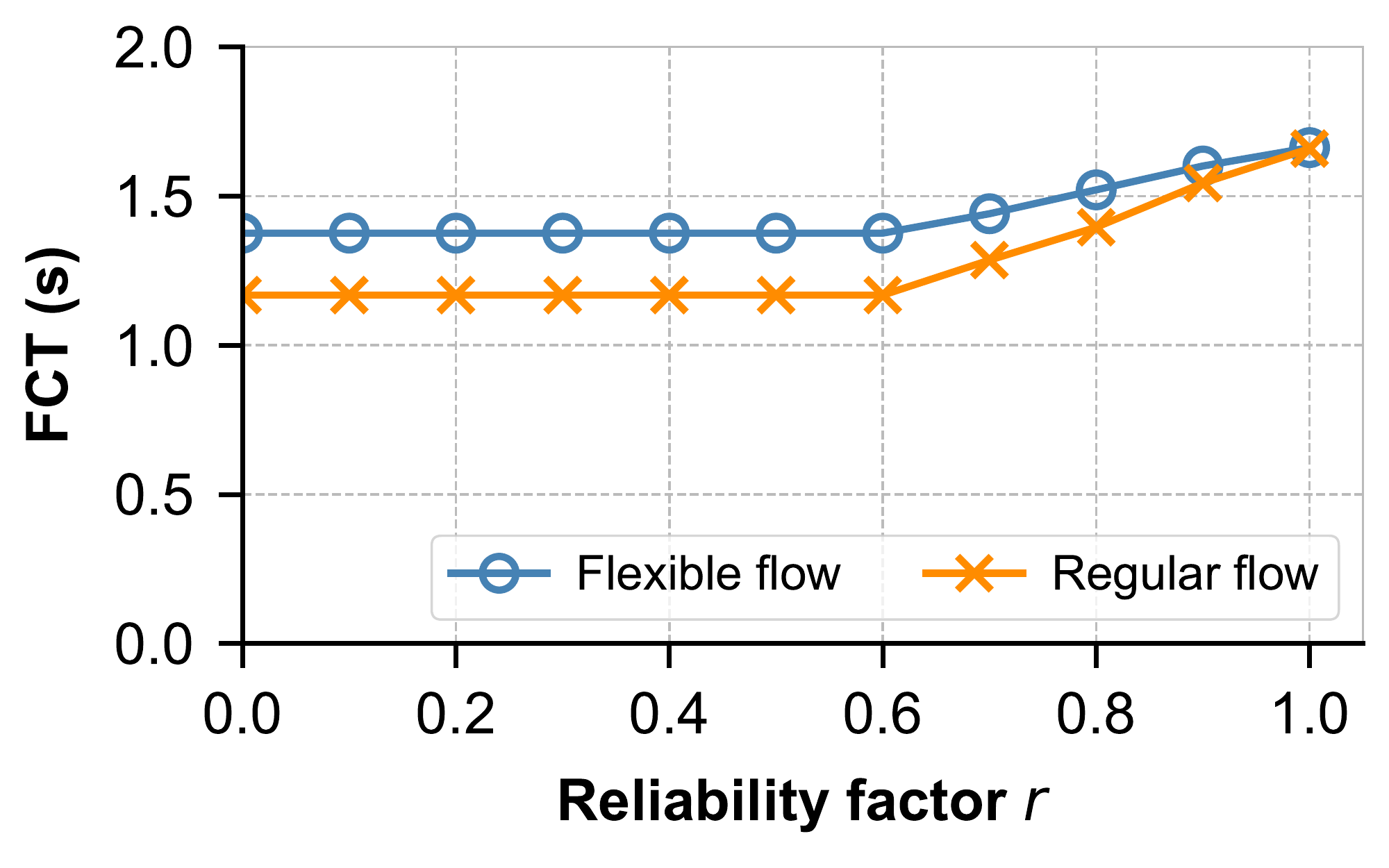}%
        \caption{Flow completion time}
        \label{fig:partial-delivery:flow-fct}
    \end{subfigure}
    \caption{Results for a single flexible flow of different reliability factor $r$ competing with a single regular flow. The flexible flow discards data in order to speed-up the regular flow.}
    \label{fig:partial-delivery}
    \vspace{-6pt}
\end{figure*}

%%%%%%%%%%%%%%%%%%%%%%%%%%%%%%%%%%%%%%%%%%%%%
%%%%%%%%%%%%%%%%%%%%%%%%%%%%%%%%%%%%%%%%%%%%%
%%%%%%%%%%%%%%%%%%%%%%%%%%%%%%%%%%%%%%%%%%%%%
%%%%%%%%%%%%%%%%%%%%%%%%%%%%%%%%%%%%%%%%%%%%%
%%%%%%%%%%%%%%%%%%%%%%%%%%%%%%%%%%%%%%%%%%%%%

\subsection{Decision of partial delivery}
\label{sec:decision-partial-delivery}

In this section, we turn our focus to partial delivery. For flexible flows which permit partial delivery, \sysName makes the decision whether to not deliver part of data in the same manner it drains the budget: through the difference between the actual rate and the fair share rate multiplied with aggressiveness factor $\alpha$. Through going at low priority, the algorithm permits other (regular) flows if present to reduce the budget and in turn the portion of data delivered by the flexible flow. Unlike $\alpha$, a flow actually immediately starts with a budget with the flow size known in advance.

We showcase with a single flexible flow of 8~Gbit competing against a regular flow of the same size. We vary the reliability factor $r$ from 0 (no reliability guarantee) to 1 (complete reliability guarantee). The results are shown in Fig.~\ref{fig:partial-delivery}. The competitive period lasts until the regular flow finishes. Up until around $r=0.6$, the lower the reliability factor $r$ the less data of it is delivered (Fig.~\ref{fig:partial-delivery:reliability-achieved}), and thus simultaneously the faster the flexible completes itself (blue line in Fig.~\ref{fig:partial-delivery:flow-fct}). Similarly, the regular flow is significantly sped up as the flexible yields the bandwidth to it (orange line in Fig.~\ref{fig:partial-delivery:flow-fct}). The ability of the flexible flow to degrade is limited by the bandwidth it achieves at low priority, which is at most 10\% of bandwidth in the current setting if there is competition, as well as the bandwidth it achieves at high priority during the warmup and measure phases, which is 50\%. Once the regular flow finished, there is no competition, as such the flexible at that moment will perform full delivery. For these reasons, at $r<0.6$ there is no further degradation of the flexible flow. At $r=0.0$, the flexible flow completes 65\%. The fastest rate of the regular flow is similarly limited, to around 7.1~Gbit/s on average, which corresponds to an FCT of 1.12~s -- which we approximately observe in Fig.~\ref{fig:partial-delivery:flow-fct} with 6.9~Gbit/s and an FCT of 1.17~ms.

Another interesting case is for when the regular flow would continue for the entire duration of the flexible flow. In that case, it would complete at a rate of 2.5~Gbit/s and drain budget at a rate of 2.3~Gbit/s. As such it would finish after 1.66~s with a completion percentage of around 52\%, which means that reducing $r$ beyond 0.52 would not yield any speed-up or further partial delivery. 

\greybox{\textbf{Partial delivery takeaways:} \sysName trades off data loss proportional to its reduced rate over time. Only if there is competition will it complete partially. The amount it can discard is determined by parameter $r$ as well as the probing parameterization: the longer the exploit phase, the more data it can potentially discard.}

%%%%%%%%%%%%%%%%%%%%%%%%%%%%%%%%%%%%%%%%%%%%%
%%%%%%%%%%%%%%%%%%%%%%%%%%%%%%%%%%%%%%%%%%%%%
%%%%%%%%%%%%%%%%%%%%%%%%%%%%%%%%%%%%%%%%%%%%%
%%%%%%%%%%%%%%%%%%%%%%%%%%%%%%%%%%%%%%%%%%%%%
%%%%%%%%%%%%%%%%%%%%%%%%%%%%%%%%%%%%%%%%%%%%%

\subsection{Challenge of probing}
\label{sec:challenge-probing}

The probing mechanism is parameterized by the interval duration $T_{int}$ and the balance between phases ( $D_{warmup} : D_{measure} : D_{exploit}$ ). These probing parameters impact (a) the maximum possible speed-up \sysName can fundamentally offer, and (b) the ability to fulfill its guarantees.

% Different interval duration T_int
\begin{figure*}[t]
	\centering
    \begin{subfigure}[b]{0.48\textwidth}
        \centering
        \includegraphics[width=\textwidth]{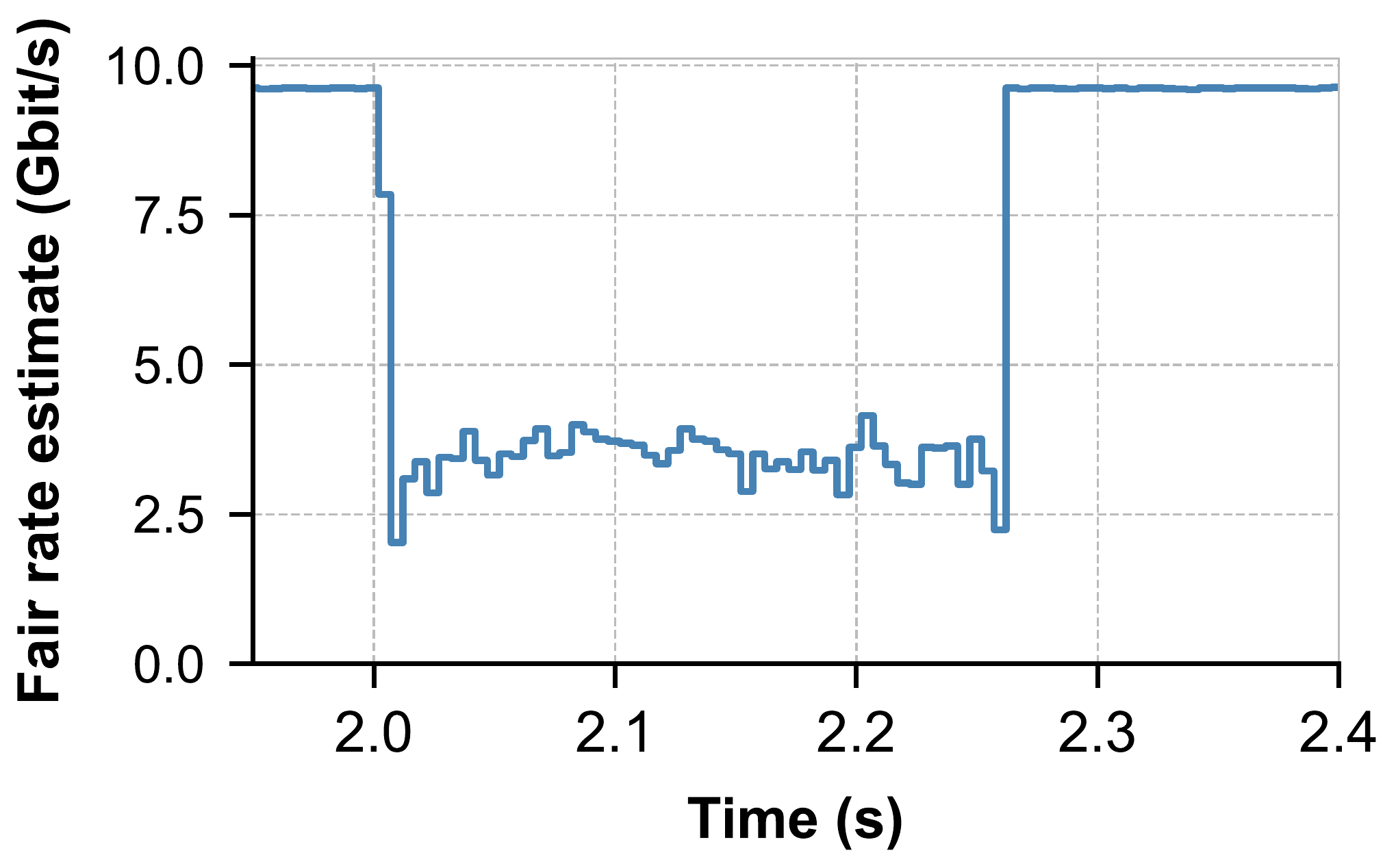}%
        \caption{$T_{int}=$ 1ms}
        \vspace{0.3cm}
    \end{subfigure}
 	\hfill
    \begin{subfigure}[b]{0.48\textwidth}
        \centering
        \includegraphics[width=\textwidth]{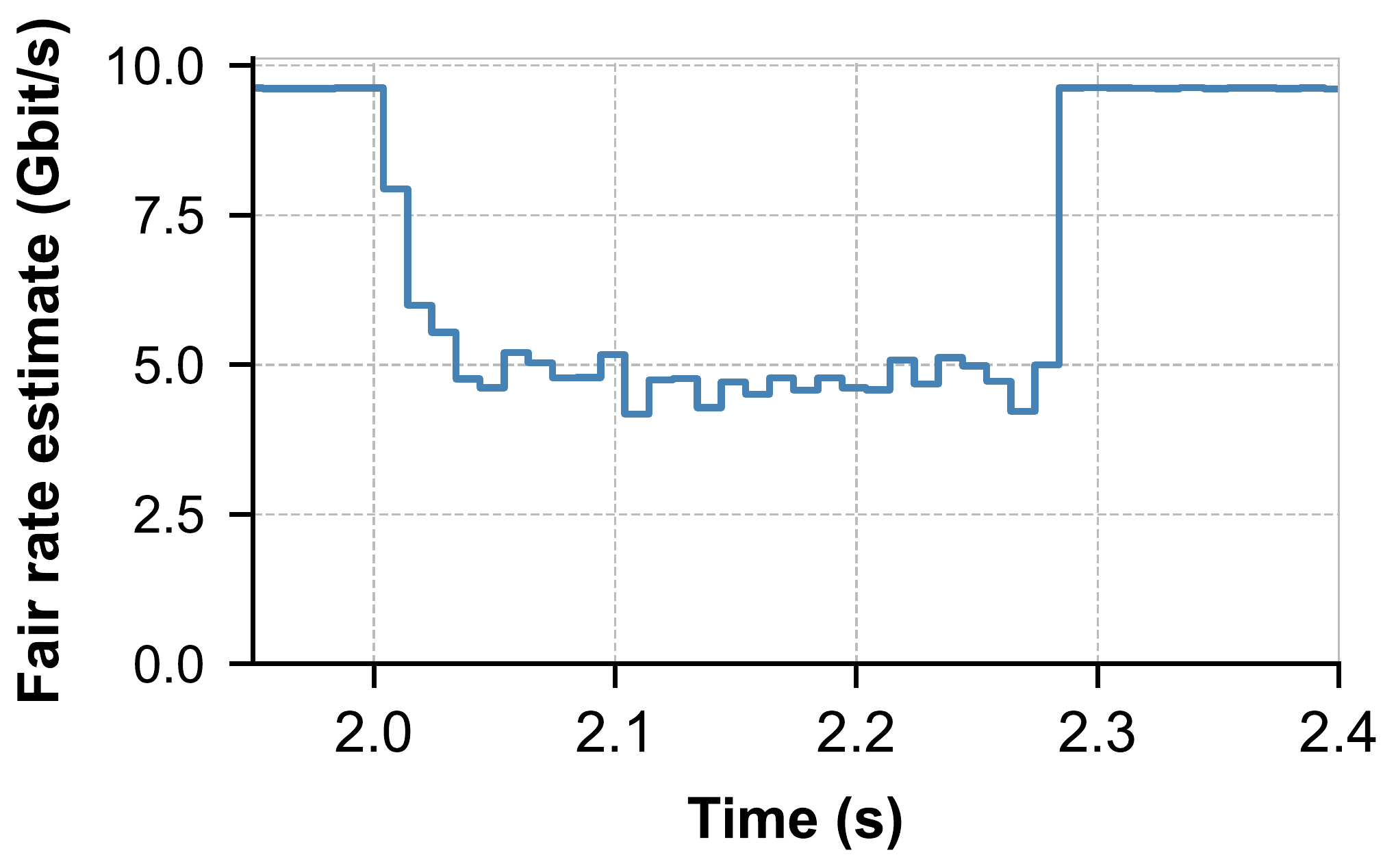}%
        \caption{$T_{int}=$ 2ms}
        \vspace{0.3cm}
    \end{subfigure}
    \begin{subfigure}[b]{0.48\textwidth}
        \centering
        \includegraphics[width=\textwidth]{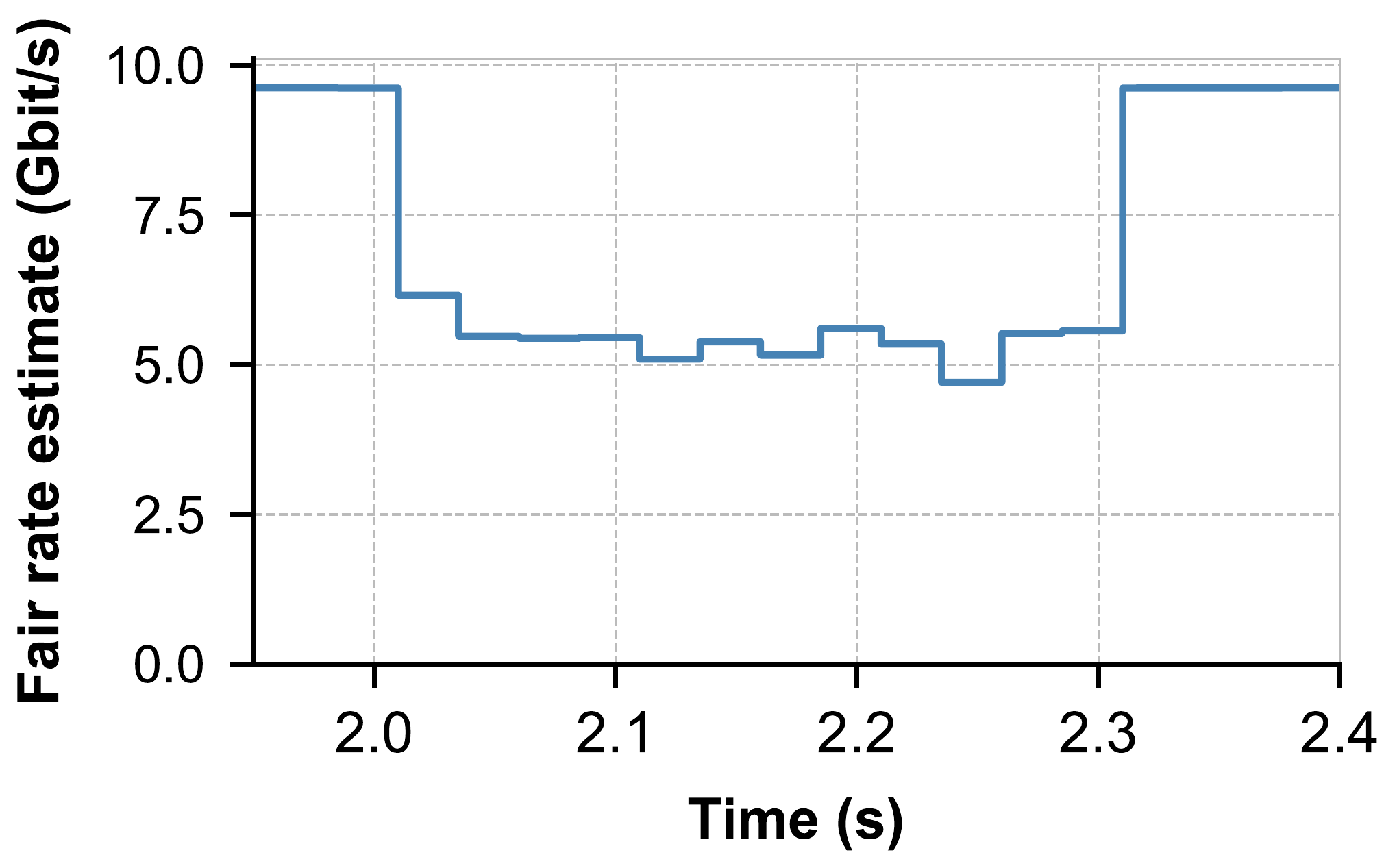}%
        \caption{$T_{int}=$ 5ms}
    \end{subfigure}
    \hfill
    \begin{subfigure}[b]{0.48\textwidth}
        \centering
        \includegraphics[width=\textwidth]{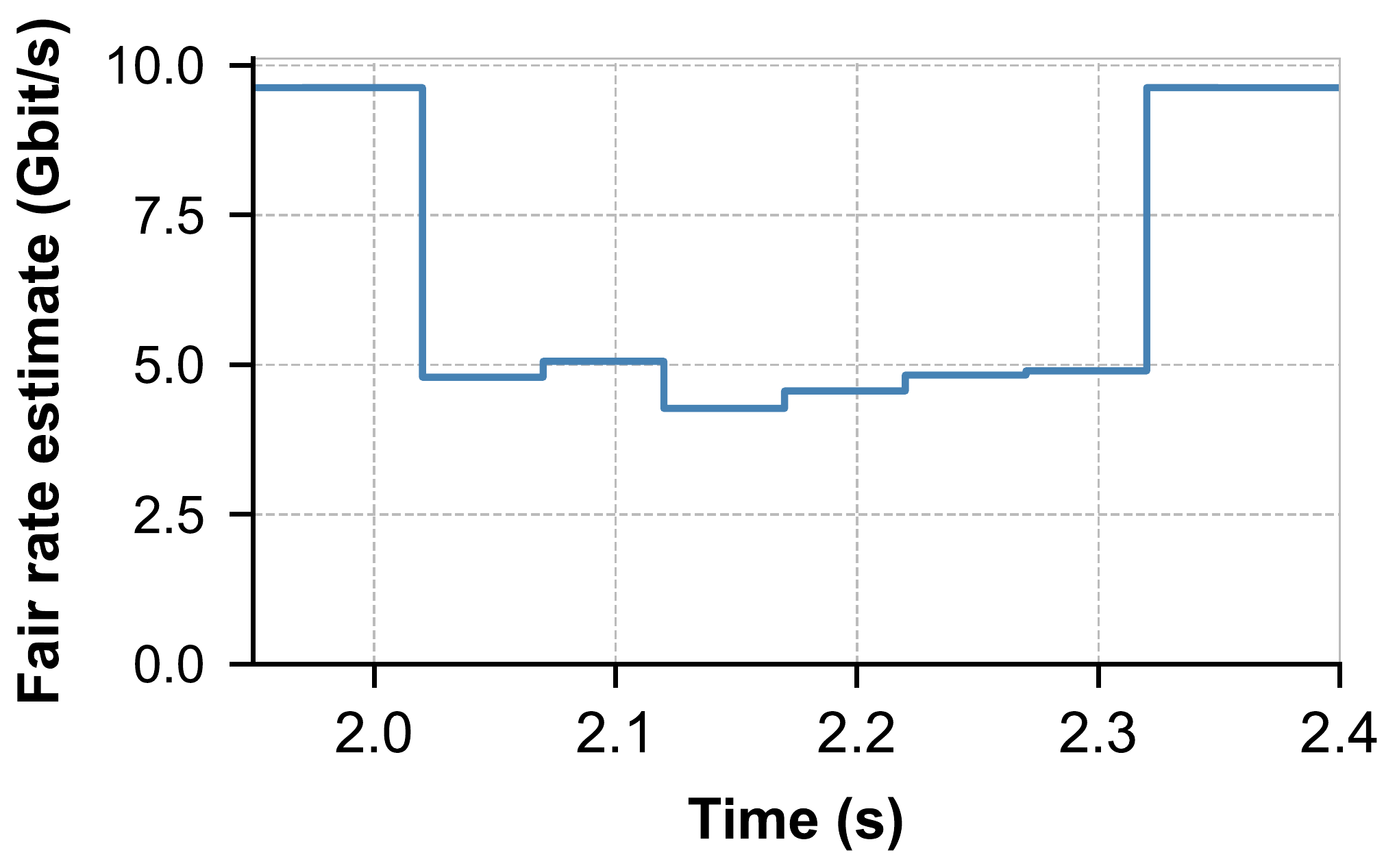}%
        \caption{$T_{int}=$ 10ms}
    \end{subfigure}
    \caption{Fair share estimated rates for Case 1 for different interval duration. A too low interval duration leads to too low (or incorrect) fair share estimates.}
    \label{fig:interval-duration}
    \vspace{-6pt}
\end{figure*}

\parab{Interval granularity ( $T_{int}$ ).} We set out to have $T_{int}$ be the minimum interval needed to acquire a measurement of the fair share rate. As such, it is desirable for it to be at least one full RTT (including queueing delay), such that its measurement captures at least a full in-flight congestion window. Likely, several RTTs are required for a consistently accurate fair share estimate. In our simulated network, the RTT over a single link is in the order of several microseconds when empty, and 100s of microseconds when queues are filled. To investigate this granularity, we perform Case 1 from \S\ref{sec:prior-scheme-comparison} for three different $T_{int}$: 1~ms, 2~ms, 5~ms, and 10~ms. For each parameterization, we plot the  $R_{fair}$ which was found during the period of competition between approximately the 0.95~s and 1.2~s timestamps in Fig.~\ref{fig:interval-duration}. At $T_{int}=$~1~ms, the estimate is considerably lower than the actual fair share rate of half. This is caused by the congestion control protocol taking time to converge. At a $T_{int}$ of 2~ms, 5~ms and 10~ms, the fair share rate is correctly maintained around 50\% of the line rate. As a new flow starts exactly at $T=2~s$, the initial estimate is a bit higher as the flows take time to balance their congestion window especially with slow start. There is a fine balance between the interval duration and the responsiveness of \sysName: the lower the interval duration, the quicker a flow can decide if it can become flexible, as well the more recent and thus accurate it is. However, if it fails to converge and obtain an accurate measurement, the fair share estimate will be inaccurate and lead to guarantee violations. As a balance between these two effects, we set the interval duration to $T_{int}=$~5~ms as default. It is important to note that in settings where RTTs are higher, convergence rates will similarly be slower, as such in such networks the interval would need to be higher. Conversely, it could be set lower in networks with even lower latency (and higher throughput).

\begin{figure*}[t]
	\centering
    \begin{subfigure}[b]{0.48\textwidth}
        \centering
        \includegraphics[width=\textwidth]{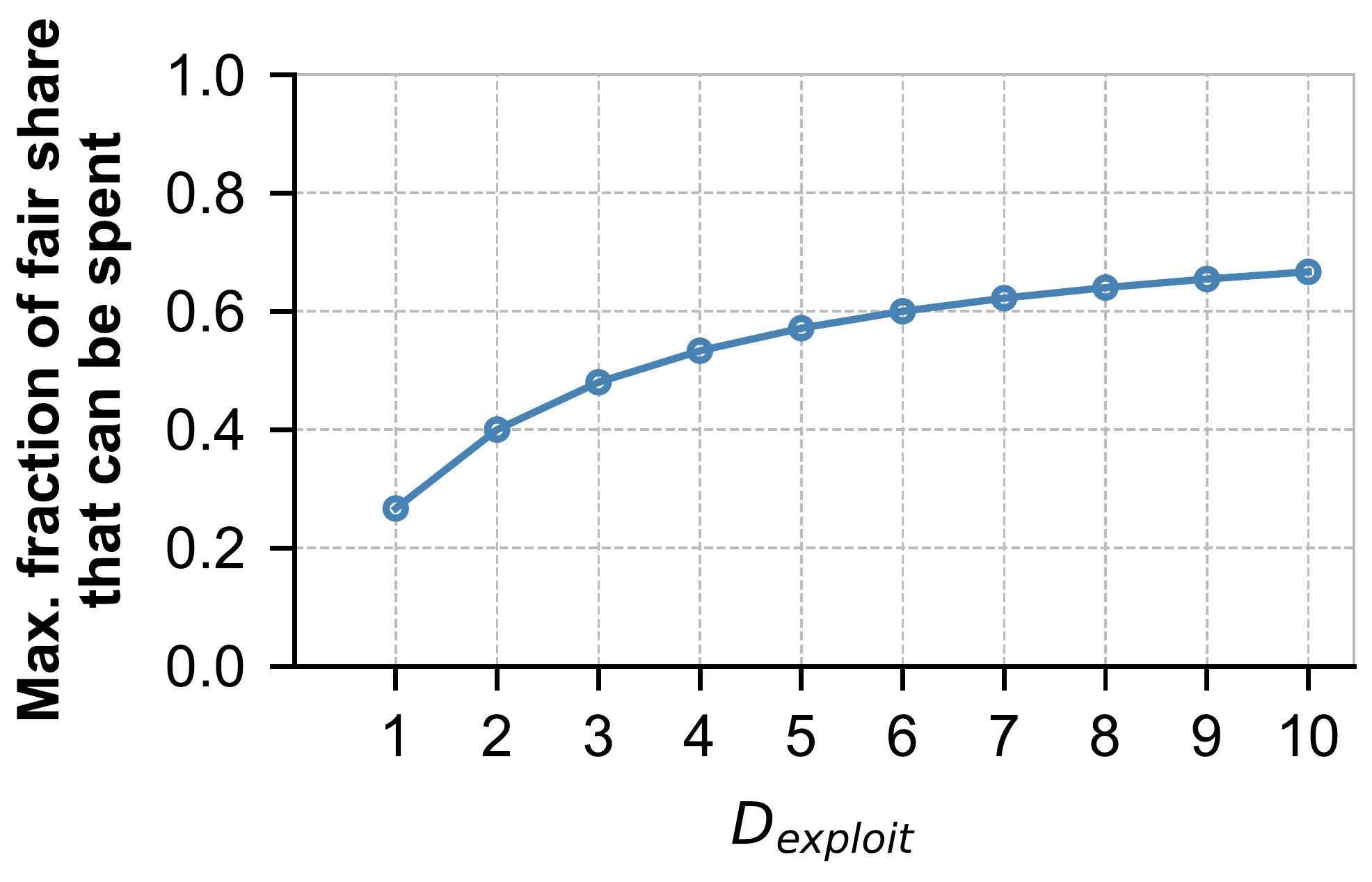}
        \caption{The maximum fraction of the fair share a flexible flow can spent.}
        \label{fig:spentfraction}
    \end{subfigure}
    \hfill
    \begin{subfigure}[b]{0.48\textwidth}
        \centering
        \includegraphics[width=\textwidth]{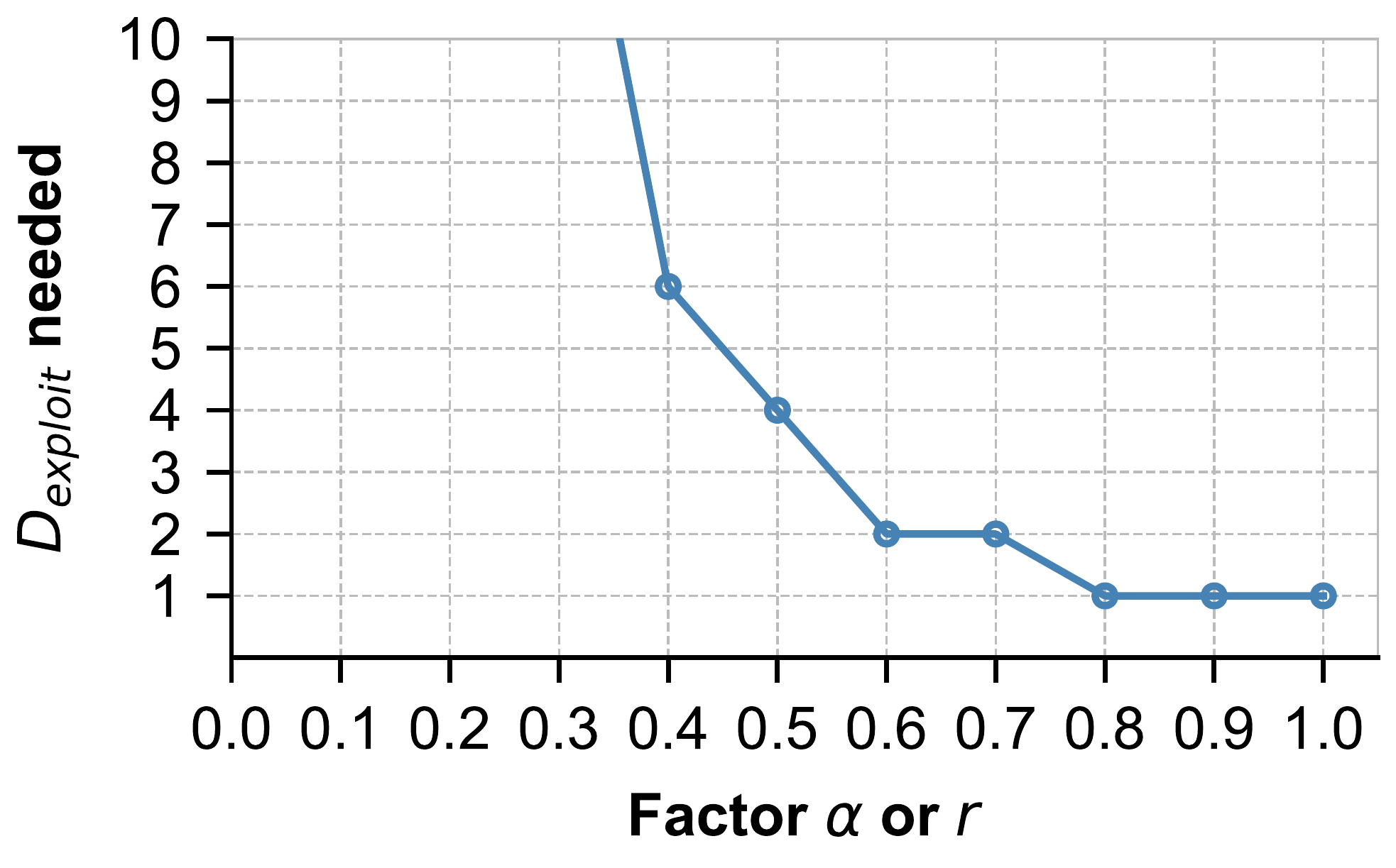}%
        \caption{The exploit phase size needed to be able to spend as much as the factor freedom permits.}
        \label{fig:Dexploitneeded}
    \end{subfigure}
    \caption{Calculation for two flow competing, one regular and one flexible. The longer the exploit phase, the larger portion if it time it can use to speed up other flows. The limits are determined by the time spent in warmup and measure phases (1:1), as well as the underlying priority scheme (which is 9-1 weighted priority in our experiments).}
    %\label{fig:fdt-priority-reflex-rates}
    \vspace{-6pt}
\end{figure*}

\parab{Phase balance ( $D_{warmup} : D_{measure} : D_{exploit}$ ).} The ratio between the phases has a profound impact on the speed-up that \sysName can deliver, as only during the exploit phase can other flows be sped up. We have earlier set the interval duration $T_{int}$ to be high enough to achieve a good enough estimate, and low enough to not cause outdated estimates by itself. With this in consideration, we set both the warmup $D_{warmup}$ and $D_{measure}$ to be 1 in the phase balance. Thus, only the configuration of $D_{exploit}$ remains: the lower, the fresher the fair share estimate is, and the higher, the more speed-up a flexible flow can provide. The speed-up is due to the base proportion of exploit phase (as is depicted in Fig.~\ref{fig:spentfraction}), as well as the probability of a (short) flow starting and completing wholly in the exploit phase. To showcase the latter, we run 1 long flexible flow with $\alpha=0.9$ (which is active the entire duration) and short regular flows of 100~kB at a Poisson arrival rate of 1000 flows/s for 5~s with 1~s experiment warmup and 1~s cooldown. We vary $D_{exploit}\in[1, 2, ..., 10]$.

% Phase balance I
\begin{figure*}[t]
\centering%
\begin{minipage}[t]{.48\textwidth}%
    %\centering
    \includegraphics[width=\textwidth]{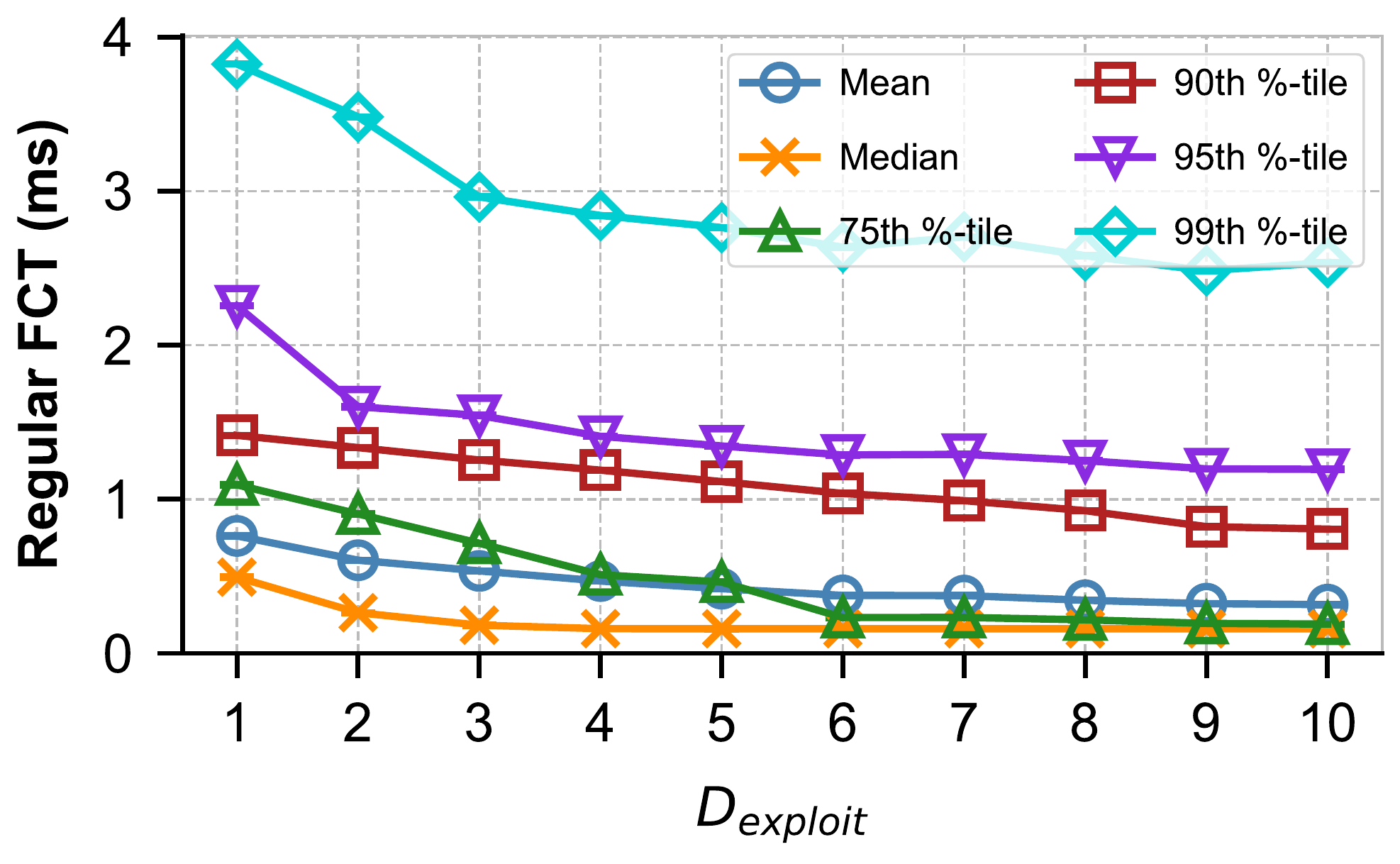} 
    \captionof{figure}{Scenario of a single large flexible flow with many short regular flows arriving. The relatively larger the exploit phase is, the higher the percentiles of the short regular flows that are sped up. However, this only holds if the flexible flow has time to accumulate budget first.}
    \label{fig:Dexploit}
\end{minipage}%
\hfill%
\begin{minipage}[t]{.48\textwidth}%
    %\centering
    \includegraphics[width=\textwidth]{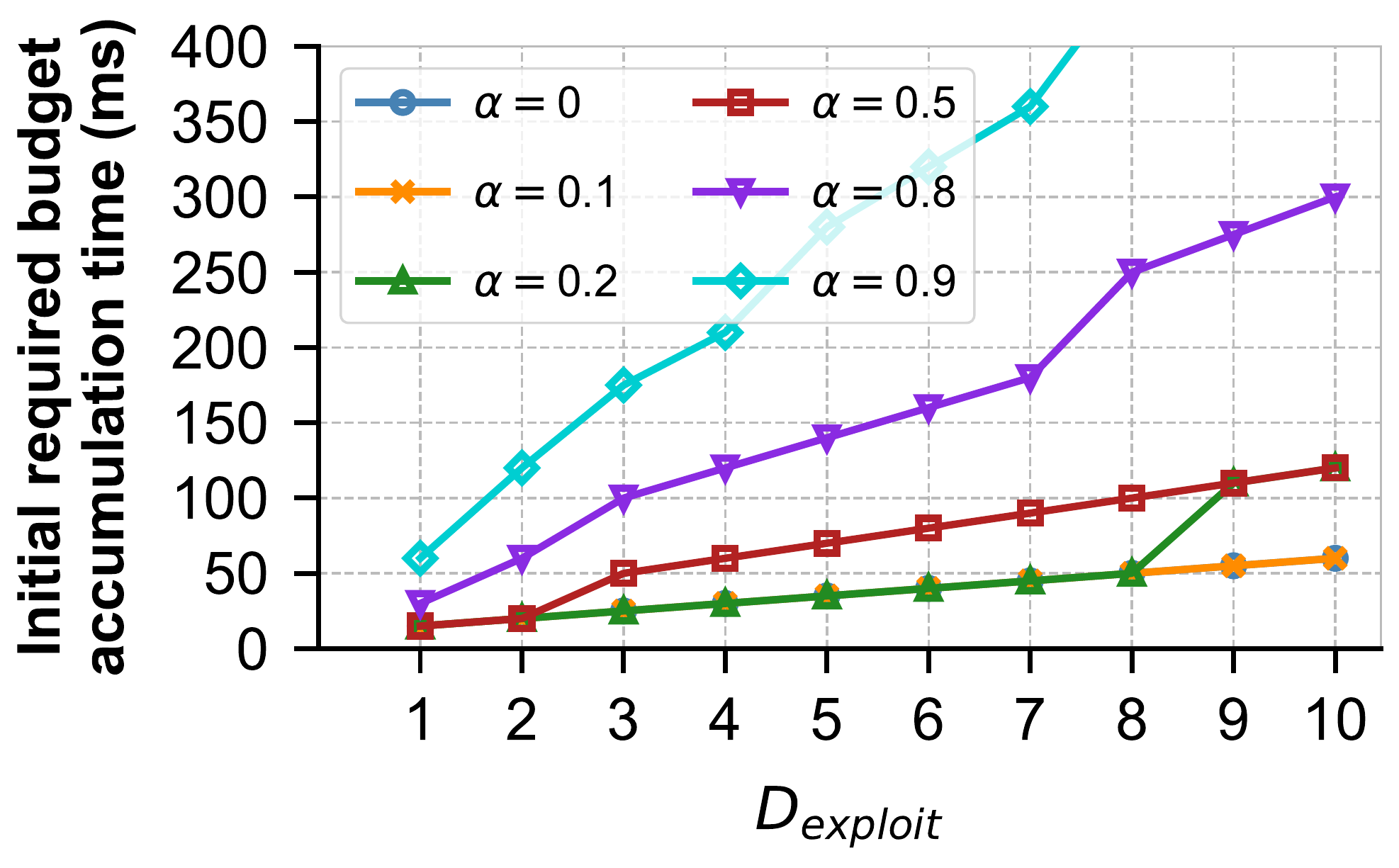} 
    \captionof{figure}{Calculation of the required time it takes to collect sufficient budget to go at low priority during the exploit phase with $T_{int}=5$~ms. At $D_{exploit}=3$ with $\alpha=0.9$ it takes 175~ms to collect initial required budget.}
    \label{fig:timetillexploit}
\end{minipage}%
\end{figure*}

% Phase balance II

We show in Fig.~\ref{fig:Dexploit} the FCT of the regular flows as we increase the length of the exploit phase. The longer the exploit phase, the better higher percentiles perform. This is as expected: for example at $D_{exploit}=3$, approximately 60\% of the time the flexible flow can operate at low priority as the remainder it spends on (preparing) estimation of the fair share. As such, it is especially effective for the median. A longer exploit phase in general improves even the percentiles which fall out of it, as flows finish quicker thus lessening chance of competition, as well as it reduces the frequency at which the queue size changes suddenly due to the flexible flow switching priorities. However, conversely, it increases how much budget must be accumulated as well as increases the lower flow size bound of flows that can react. We plot this need for accumulation in Fig.~\ref{fig:timetillexploit}. The exploit phase duration influences the amount of fraction of fair share that can be spent as well, which is shown in Fig.~\ref{fig:spentfraction} and \ref{fig:Dexploitneeded}. Based on these various tradeoff, we set $D_{exploit}$ to 3 by default. This enables \sysName in our experiments to spend 48\% of its fair share in a two-flow competition (Fig.~\ref{fig:spentfraction}) thus providing benefit for $\alpha$ and $r$ that are greater than or equal to 0.6 (Fig.~\ref{fig:Dexploitneeded}). This parameter value provides a positive impact on the mean, median and the higher percentiles as is shown in Fig.~\ref{fig:Dexploit}. \sysName with $D_{exploit}=3$ is able to accumulate enough budget for $\alpha=0.9$ in 175~ms (Fig.\ref{fig:timetillexploit}). By adding on a cycle (25~ms) for accumulated budget to be spent, we set our flexible flow size in the larger experiments to last 200~ms at perfect line rate (10~Gbit/s) which equates to 250~MB.

% Phase balance effect of many short flows
\begin{figure*}[t]
    \centering
    \includegraphics[width=0.48\textwidth]{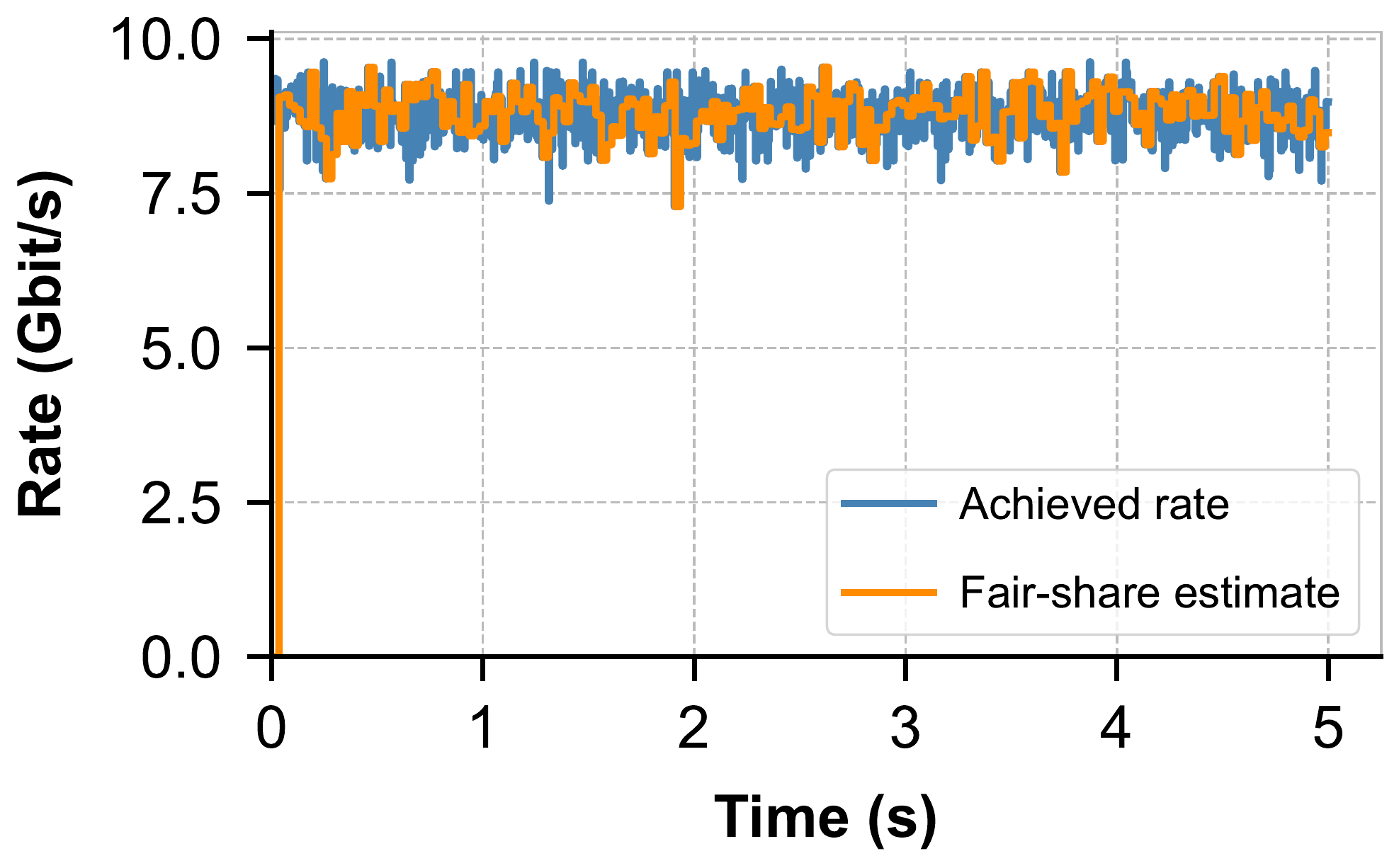} 
    \caption{Achieved rate and fair share estimate for $D_{exploit}=3$. Due to the arrival of many short flows, the achieved rate at low priority is the same as at high priority, and as such the estimate is as well.}
    \label{fig:shortflowseffect}
\end{figure*}

% Convergence dependency
\begin{figure*}[t]
	\centering
    \begin{subfigure}[b]{0.48\textwidth}
        \centering
        \includegraphics[width=\textwidth]{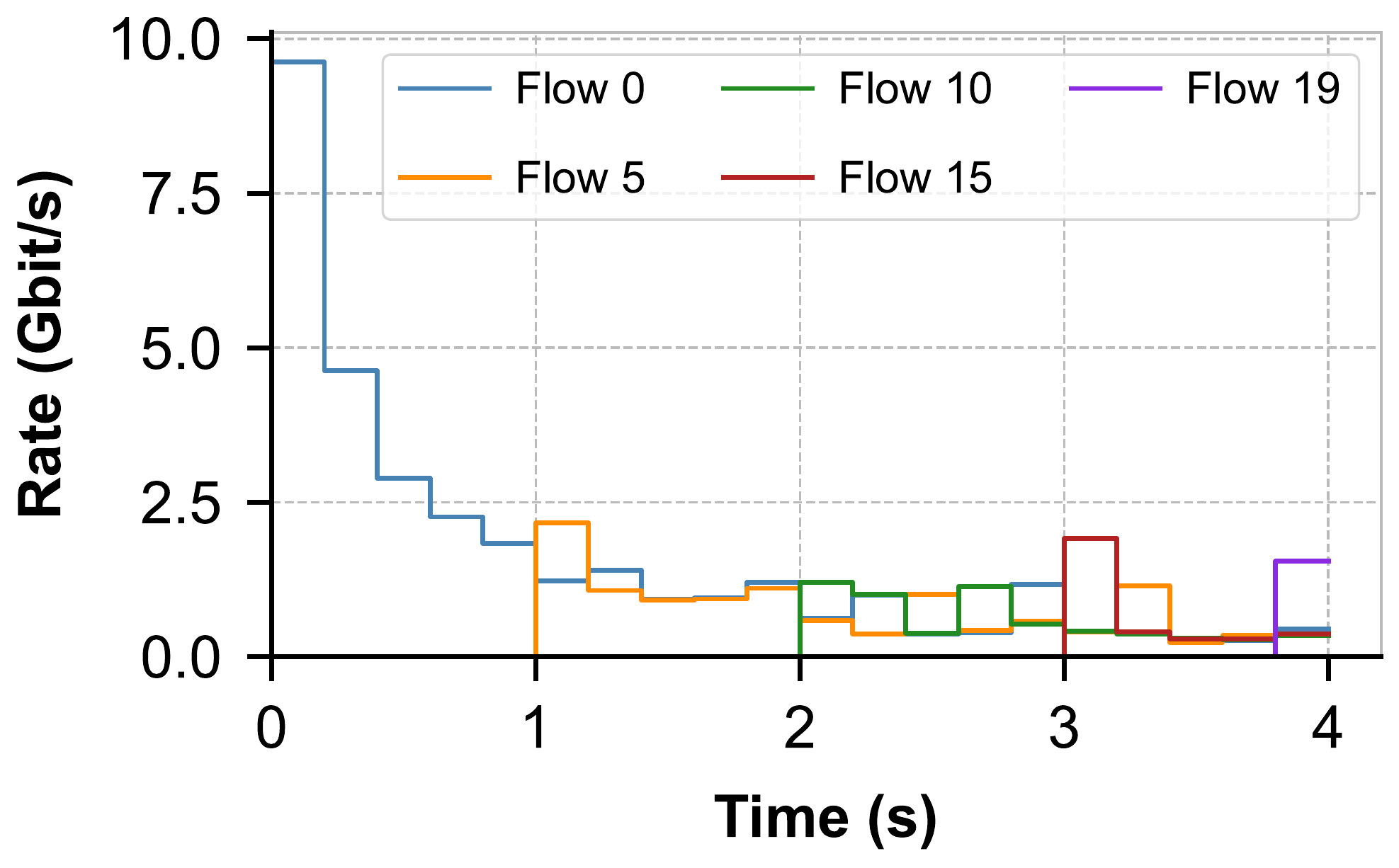}%
        \caption{Actual rates}
        \label{fig:convergence:actual}
    \end{subfigure}
 	\hfill
    \begin{subfigure}[b]{0.48\textwidth}
        \centering
        \includegraphics[width=\textwidth]{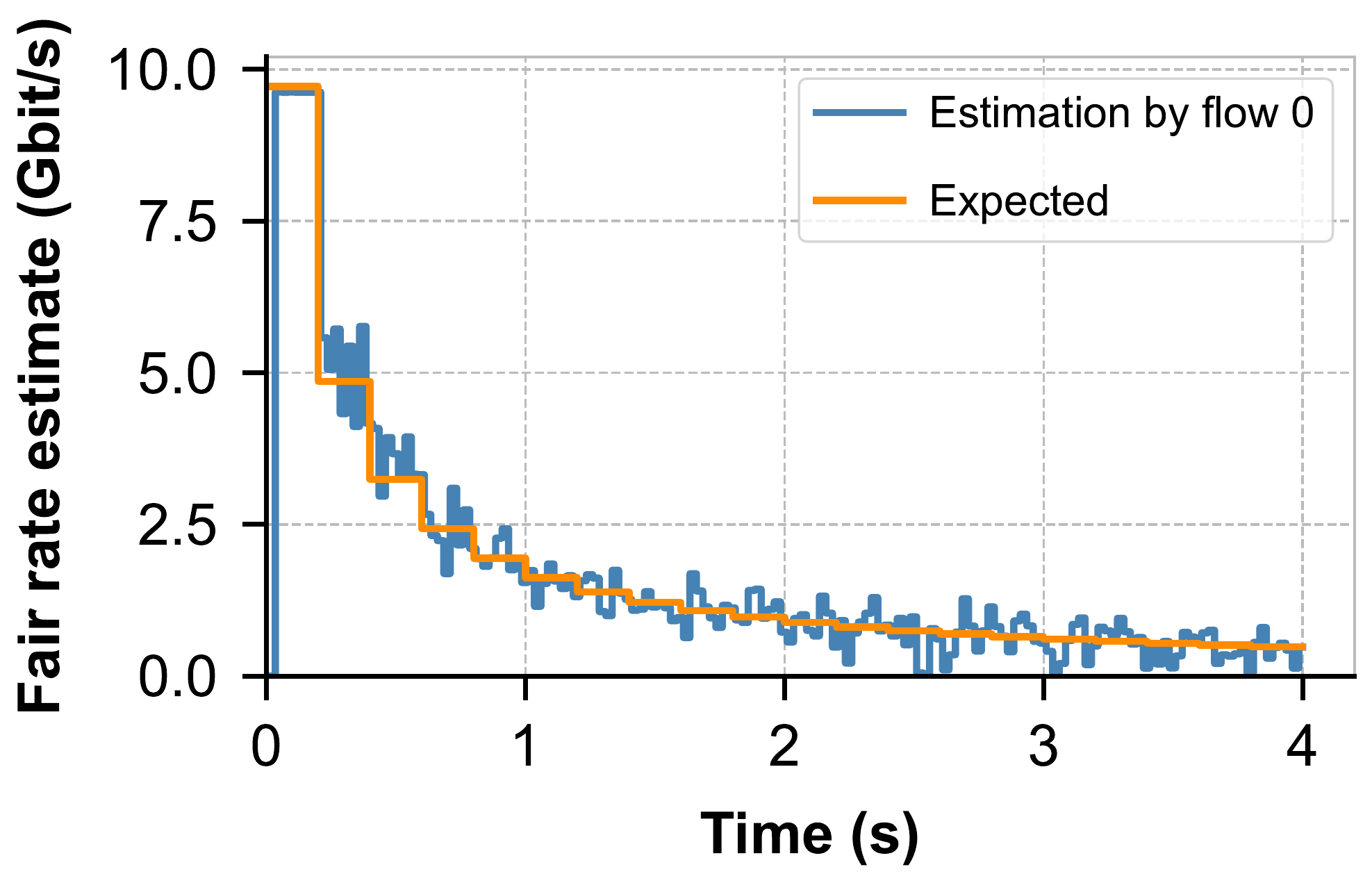}%
        \caption{Fair rate estimates}
        \label{fig:convergence:fair}
    \end{subfigure}
    \caption{Actual rate and fair share estimate for the scenario over a single link in which flexible flows arrive every 200~ms for 4 seconds. As the number of flows increases, the variation of the fair rate estimate increases.}
    \label{fig:convergence}
    \vspace{-6pt}
\end{figure*}

The fair share estimate operates at a set granularity, which is based on the convergence rate of the congestion control protocol given the network conditions (in particular, bandwidth and latency). As a first consequence, flows that are potentially flexible but are too short are unable to achieve estimates before finishing. A second consequence is that frequent short flows, which by definition do not have time to fully compete or convergence before completing, will proportionally reduce the fair share estimated by a long flexible flow. This is demonstrated most clearly in Fig.~\ref{fig:shortflowseffect}, which plots for the $D_{exploit}=3$ the achieved rate and the fair share estimate -- in the periods where it is low priority, it still experiences the fair share estimate, as such not resulting in budget reduction

\parab{Convergence dependency.} We show the convergence behavior of \sysName by starting a long flexible flow every 200~ms for 4 seconds. The \sysName convergence ability is dependent on ability of the underlying transport protocol (in this case, DCTCP) to converge when changing queue. We observe the achieved rates as well as their estimation of the fair rate. In the presence of many flows, arriving flows are able to achieve a rate similar to their fair share rate. However, at larger number of flows (in particular, beyond 10 flows at $T=2$~s), there is increased variance and fairness is more difficult to achieve (depicted in Fig.~\ref{fig:convergence:actual}). This also reflects in the fair share rate estimate, which exhibits similarly exhibits variation. Although the estimate hovers around the expectation (see Fig.~\ref{fig:convergence:fair}), this will result in excessive draining (in case of a too high estimate) or build-up (in case of a too low estimate) of the budget. The ability of \sysName to upholds its guarantees is tied to the ability to estimate fair share and thus the ability to quickly converge.

\greybox{\textbf{Probing parameterization takeaways:} The interval duration must be long enough for the flows to converge and achieve an accurate fair share estimate. This depends on both the throughput as well as latency of the network. The finer grained the interval duration and the shorter the exploit phase, the less time is required for a flexible flow to accumulate enough budget to go at low priority. The longer the exploit phase, the more budget can be spent and as such the more regular flows can be sped up, however the less recent its fair share estimate is.}

%%%%%%%%%%%%%%%%%%%%%%%%%%%%%%%%%%%%%%%%%%%%%
%%%%%%%%%%%%%%%%%%%%%%%%%%%%%%%%%%%%%%%%%%%%%
%%%%%%%%%%%%%%%%%%%%%%%%%%%%%%%%%%%%%%%%%%%%%
%%%%%%%%%%%%%%%%%%%%%%%%%%%%%%%%%%%%%%%%%%%%%
%%%%%%%%%%%%%%%%%%%%%%%%%%%%%%%%%%%%%%%%%%%%%

\subsection{Utility of increased flexibility}
\label{sec:utility-of-flexibility}

Beyond the configuration of probing, the potential improvement in regular flow performance that is achievable with \sysName is determined by its parameters, namely $\alpha$ and $r$. We consider a single ToR network with 20 servers underneath, connected by 10~Gbit/s links. The large flexible flows are set to 250~MB, which is the smallest flow size for the chosen \sysName parameterization for flexibility to provide benefit, as explained in \S\ref{sec:challenge-probing}. This size is both considered a large flow for data centers~\cite{hull,dctcp} as well as in the order of magnitude of medium-sized ML models such as ResNet~\cite{zagoruyko2016wide,he2016deep}. As~\cite{pfabric}, we make use of Poisson arrival of flows which achieves a target average utilization while also providing varying low and high utilization of the network (\eg also described as microbursts~\cite{microbursts}). In the workloads, we aim for 40\% average utilization, which is higher than the 25\% indicated by \cite{jupiter-rising} as yielding increased drop rate. Half of the target utilization we set to come from regular flows, and the other half from flexible flows. It is thus representing a period of higher than usual utilization in which there is significant probability of flow competition. The 250~MB flexible flows arrive at an average Poisson arrival rate of 20 flows per second, thus providing on average utilization of 20\%. 

The experiment is run for 16~s (1~s warm-up, 10~s measure, 5~s cool-down) and is repeated three times with different random seeds for its workload generation. In the figures the mean values are plotted with errors bars representing minimum and maximum across the three repetitions. We run \sysName with full reliability ($r=1$) but a varying aggressiveness factor $\alpha\in[0, 0.1, ..., 1]$, as well as with reduced reliability ($r\in[0, 0.1, ..., 1]$) but fair share aggressiveness ($\alpha=1$). For comparison, we additionally run the baseline strategy without prioritization and the fixed 9-1 weighted prioritization. We consider three workloads for the regular flows.

% Large flows
\begin{figure*}[t]
	\centering
    \begin{subfigure}[b]{0.3\textwidth}
        \centering
        \includegraphics[width=\textwidth]{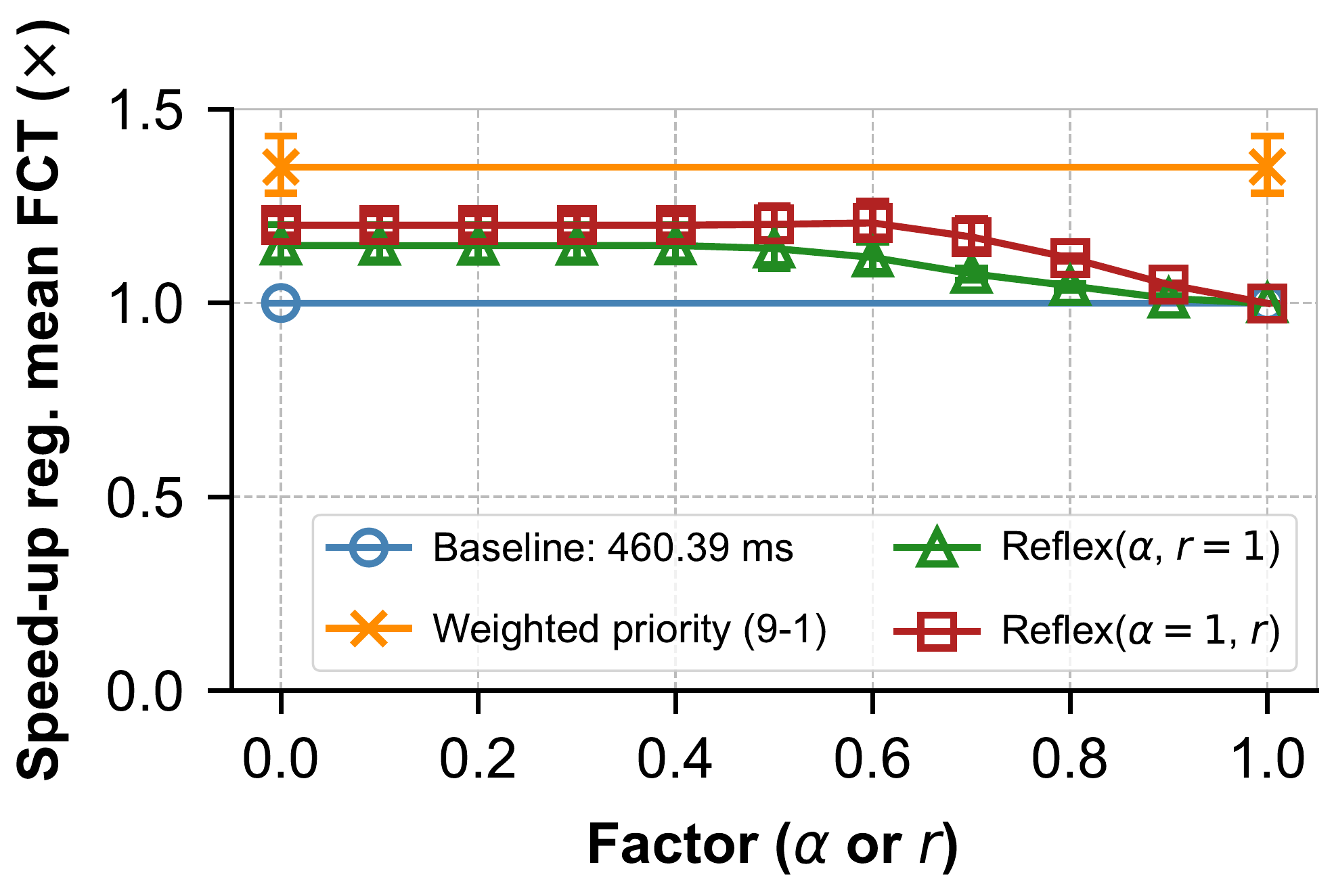}%
        \caption{Regular: mean FCT}
        \label{fig:largeflows:regular:mean}
    \end{subfigure}
 	\hfill
    \begin{subfigure}[b]{0.3\textwidth}
        \centering
        \includegraphics[width=\textwidth]{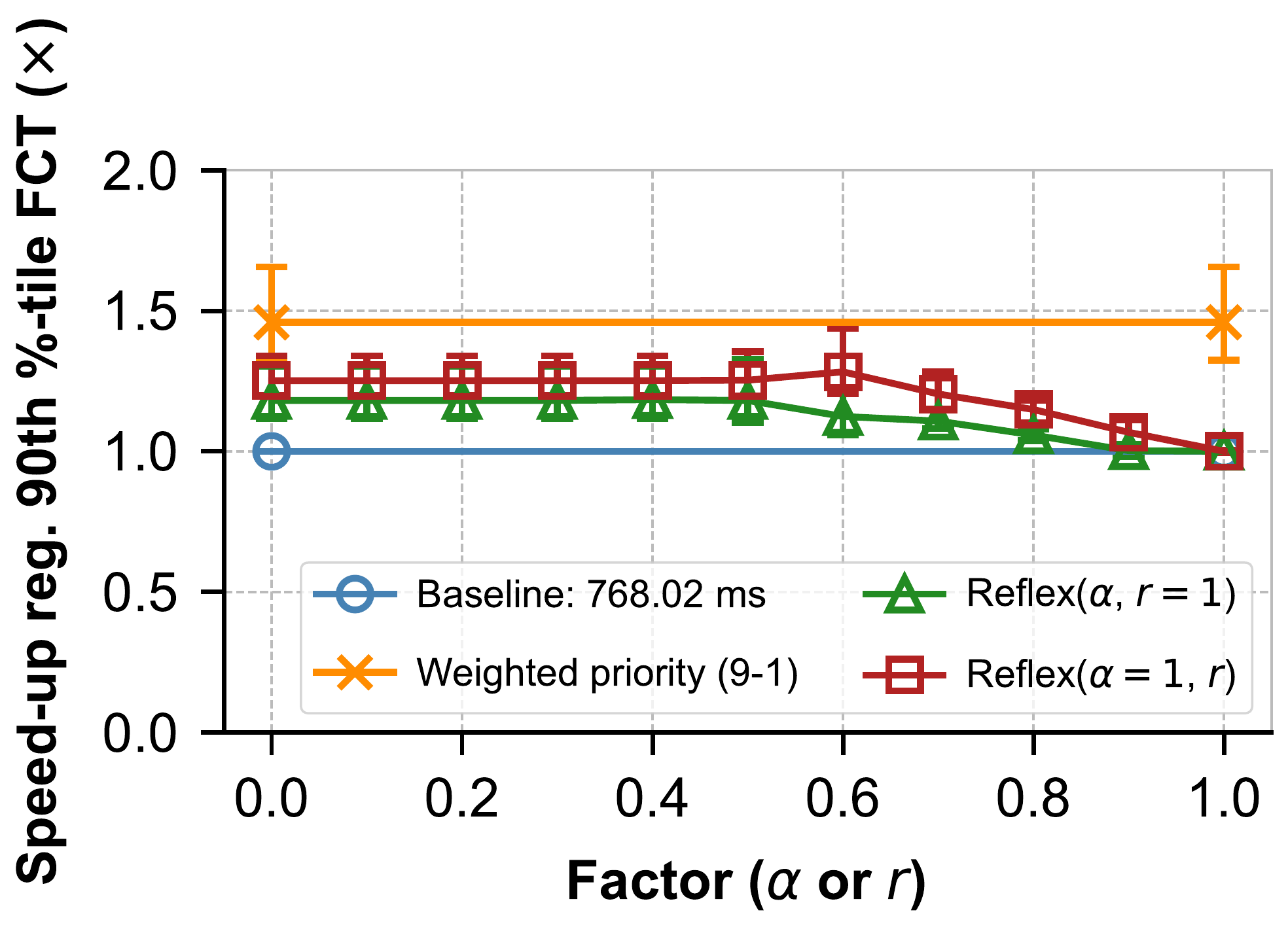}%
        \caption{Regular: 90th \%-tile FCT}
        \label{fig:largeflows:regular:90th}
    \end{subfigure}
 	\hfill
    \begin{subfigure}[b]{0.3\textwidth}
        \centering
        \includegraphics[width=\textwidth]{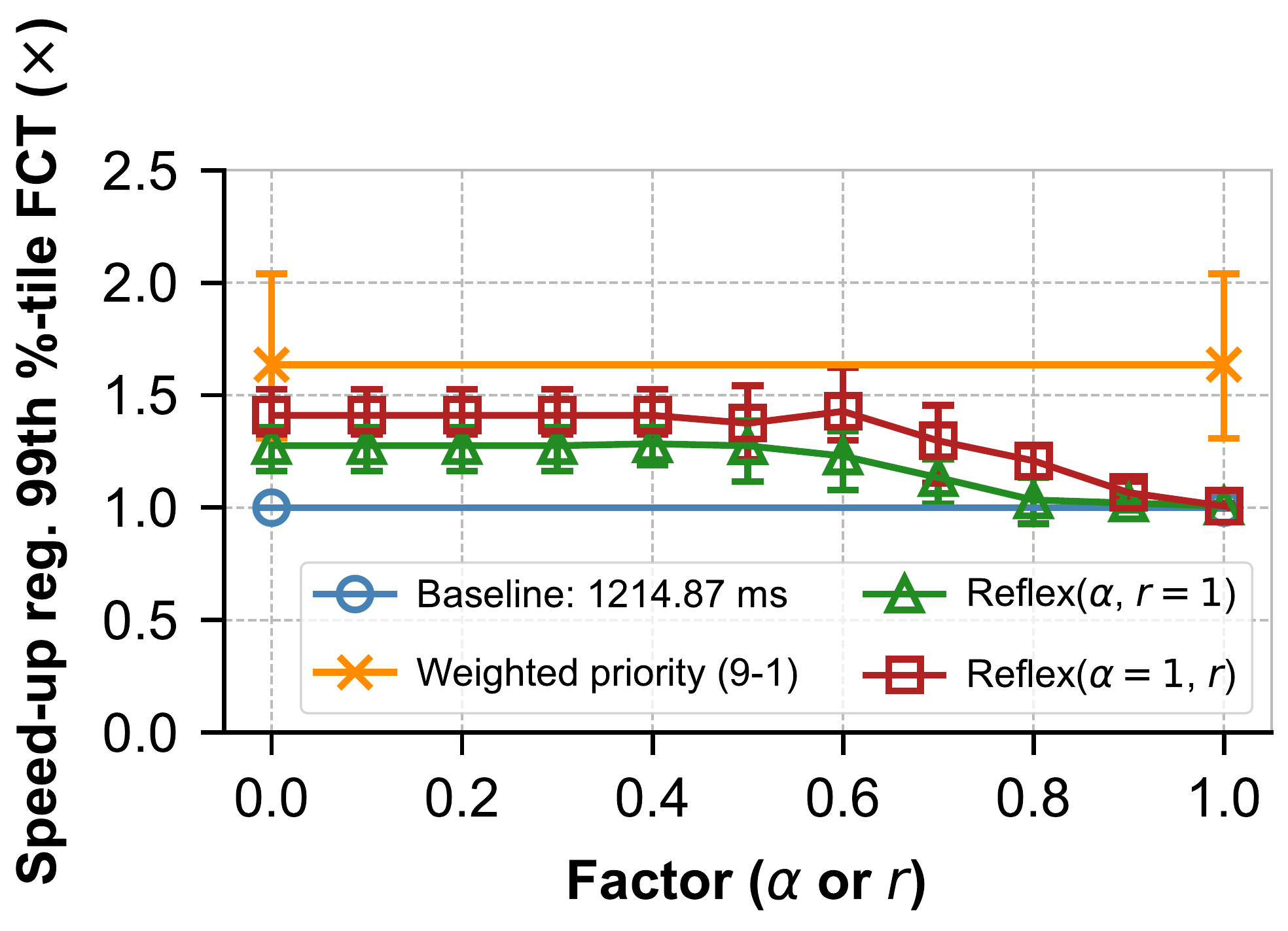}%
        \caption{Regular: 99th \%-tile FCT}
        \label{fig:largeflows:regular:99th}
    \end{subfigure}
    \hfill
    \begin{subfigure}[b]{0.3\textwidth}
        \centering
        \includegraphics[width=\textwidth]{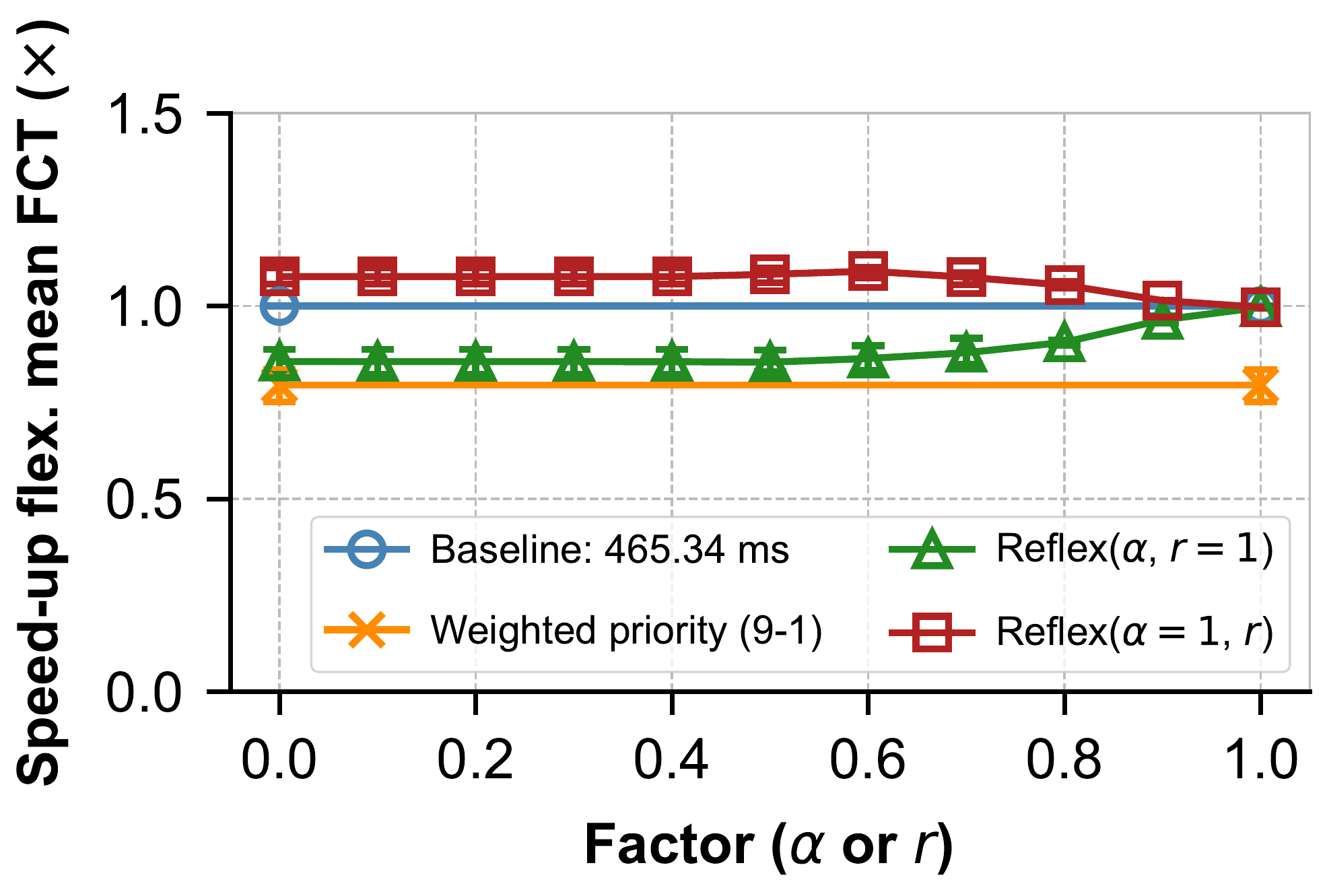}%
        \caption{Flexible: mean FCT}
        \label{fig:largeflows:flexible:mean}
    \end{subfigure}
    \hfill
    \begin{subfigure}[b]{0.3\textwidth}
        \centering
        \includegraphics[width=\textwidth]{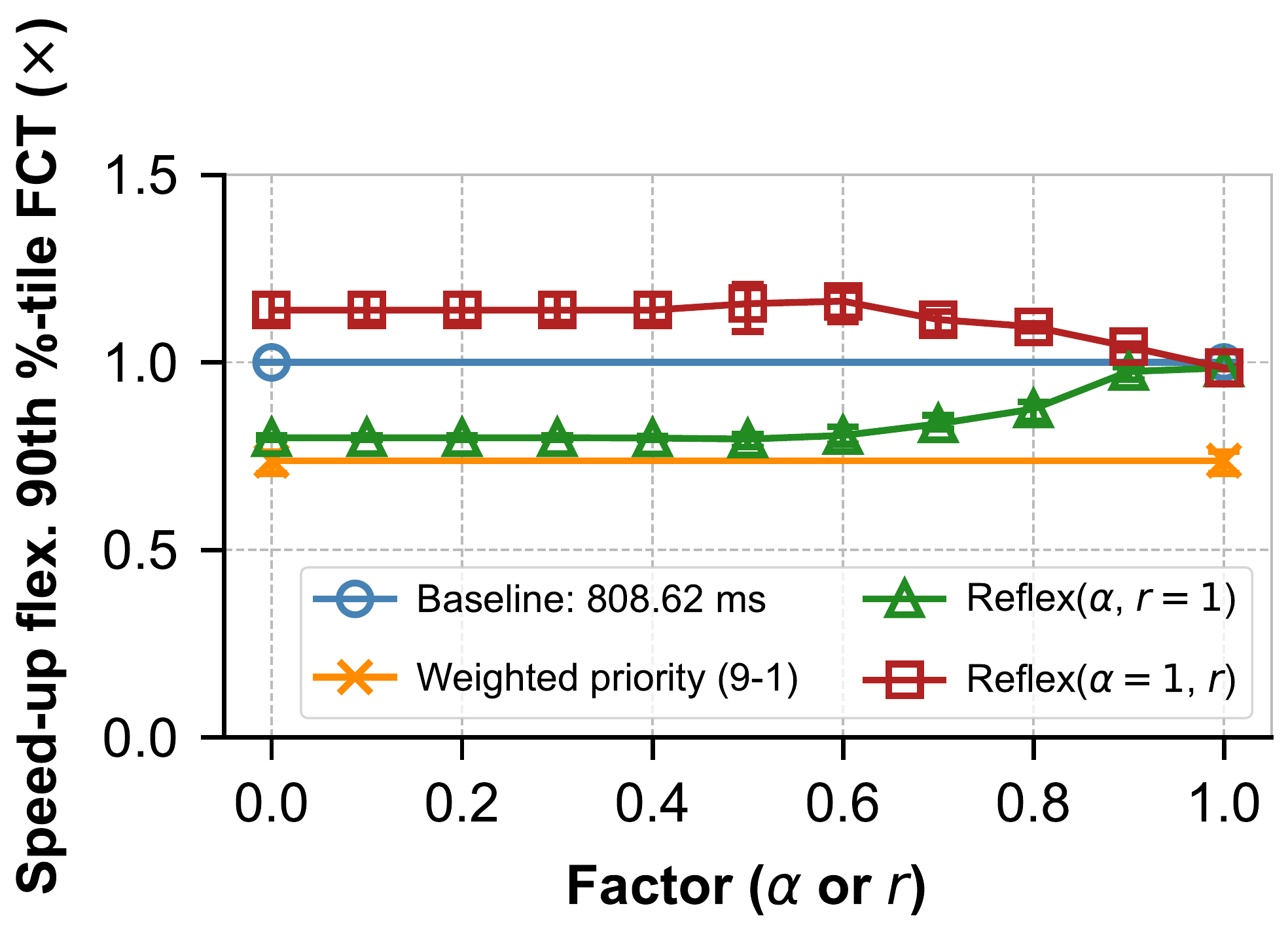}%
        \caption{Flexible: 90th \%-tile FCT}
        \label{fig:largeflows:flexible:90th}
    \end{subfigure}
    \hfill
    \begin{subfigure}[b]{0.3\textwidth}
        \centering
        \includegraphics[width=\textwidth]{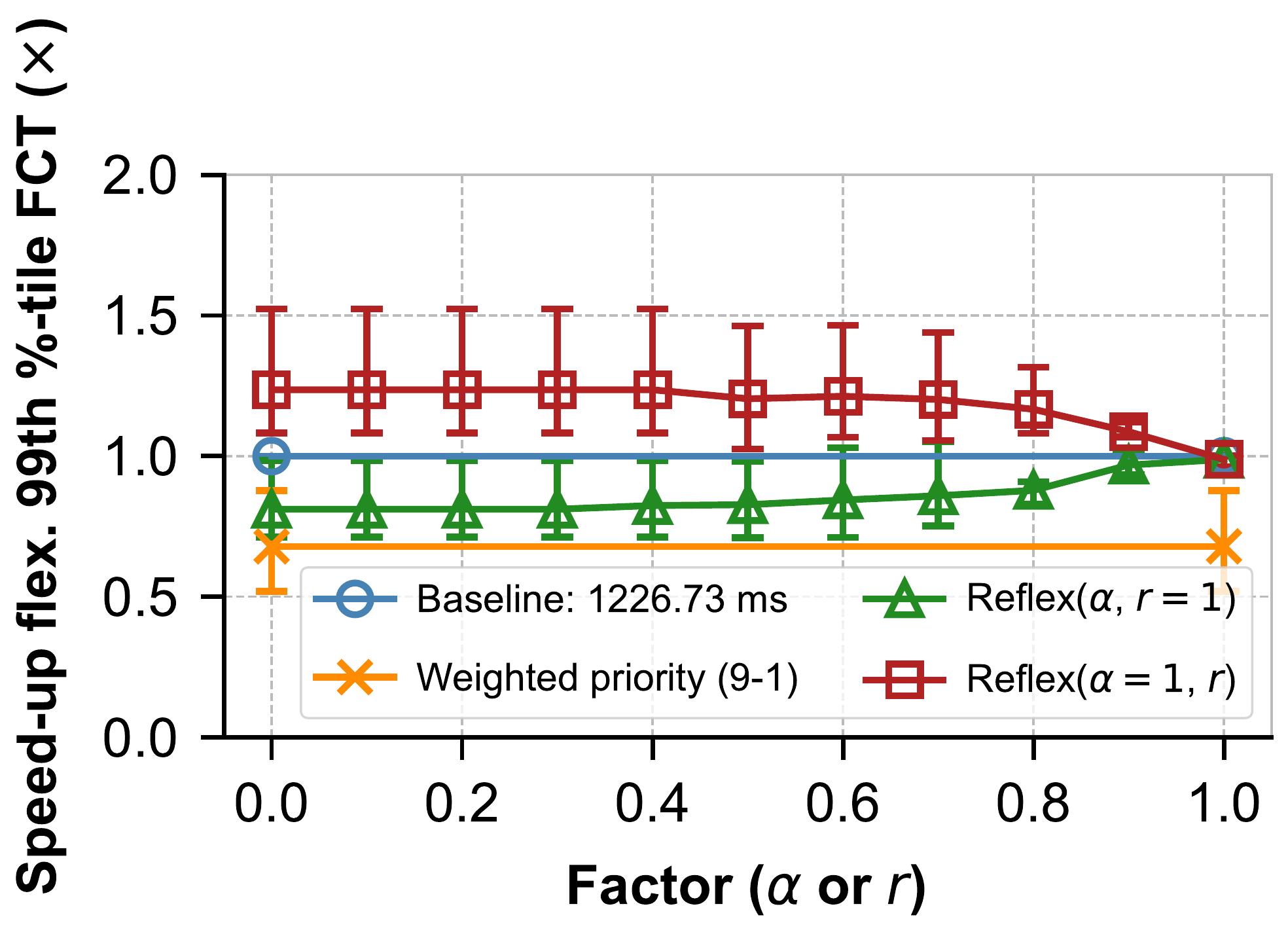}%
        \caption{Flexible: 99th \%-tile FCT}
        \label{fig:largeflows:flexible:99th}
    \end{subfigure}
    \hfill
    \caption{Workload 1: large flow competition. Speed-up for the mean and 90th/99th \%-tile (higher speed-up is better). \sysName provides a tradeoff between no and full prioritization, configurable by $\alpha$ and $r$.}
    \label{fig:largeflows}
    \vspace{-2pt}
\end{figure*}

% CDFs
\begin{figure*}[t]
	\centering
    \begin{subfigure}[b]{0.24\textwidth}
        \centering
        \includegraphics[width=\textwidth]{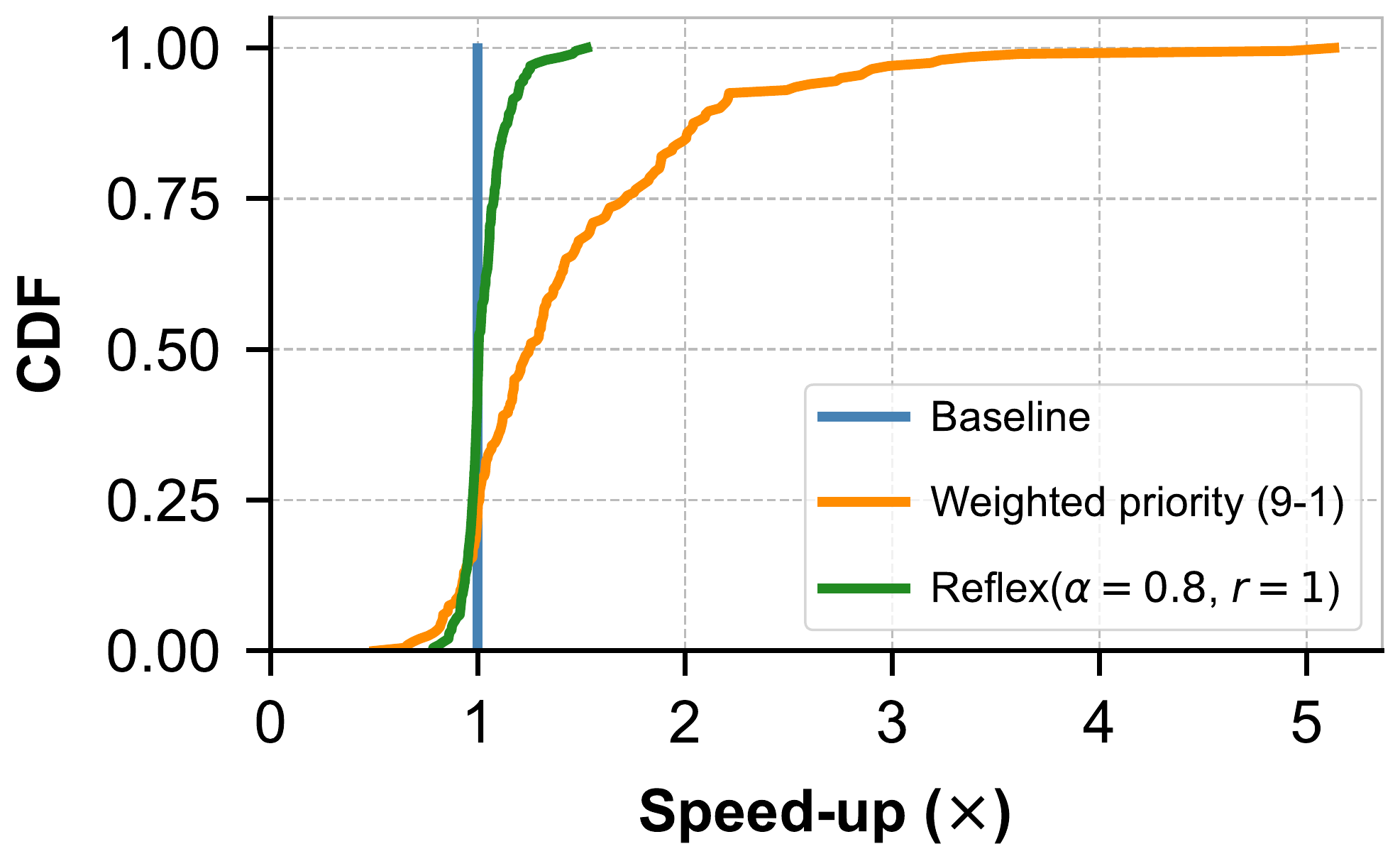}%
        \caption{Regular: speed-up}
        \label{fig:cdf-speedup:regular}
    \end{subfigure}
    \begin{subfigure}[b]{0.24\textwidth}
        \centering
        \includegraphics[width=\textwidth]{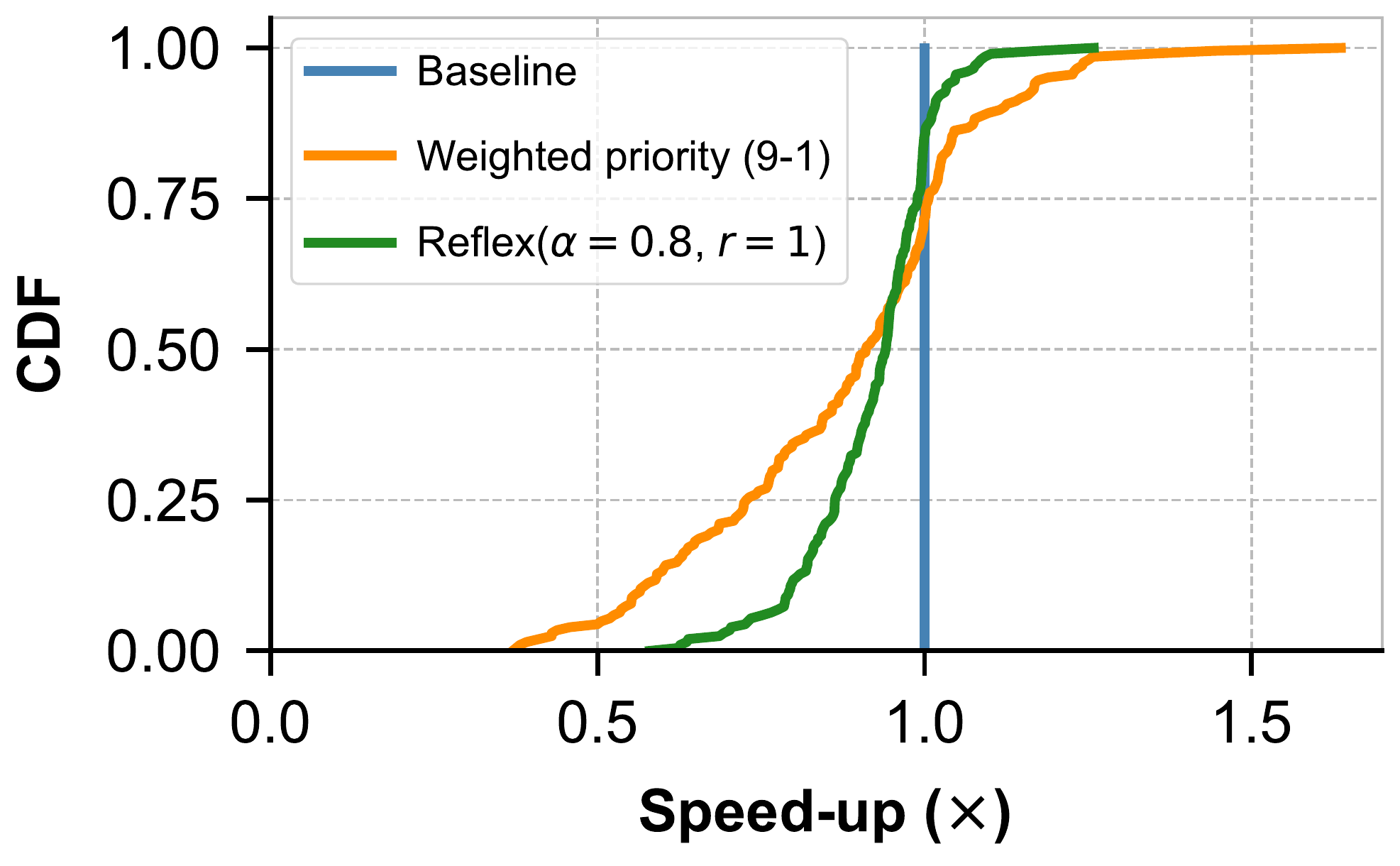}%
        \caption{Flexible: speed-up}
        \label{fig:cdf-speedup:flexible}
    \end{subfigure}
    \begin{subfigure}[b]{0.24\textwidth}
        \centering
        \includegraphics[width=\textwidth]{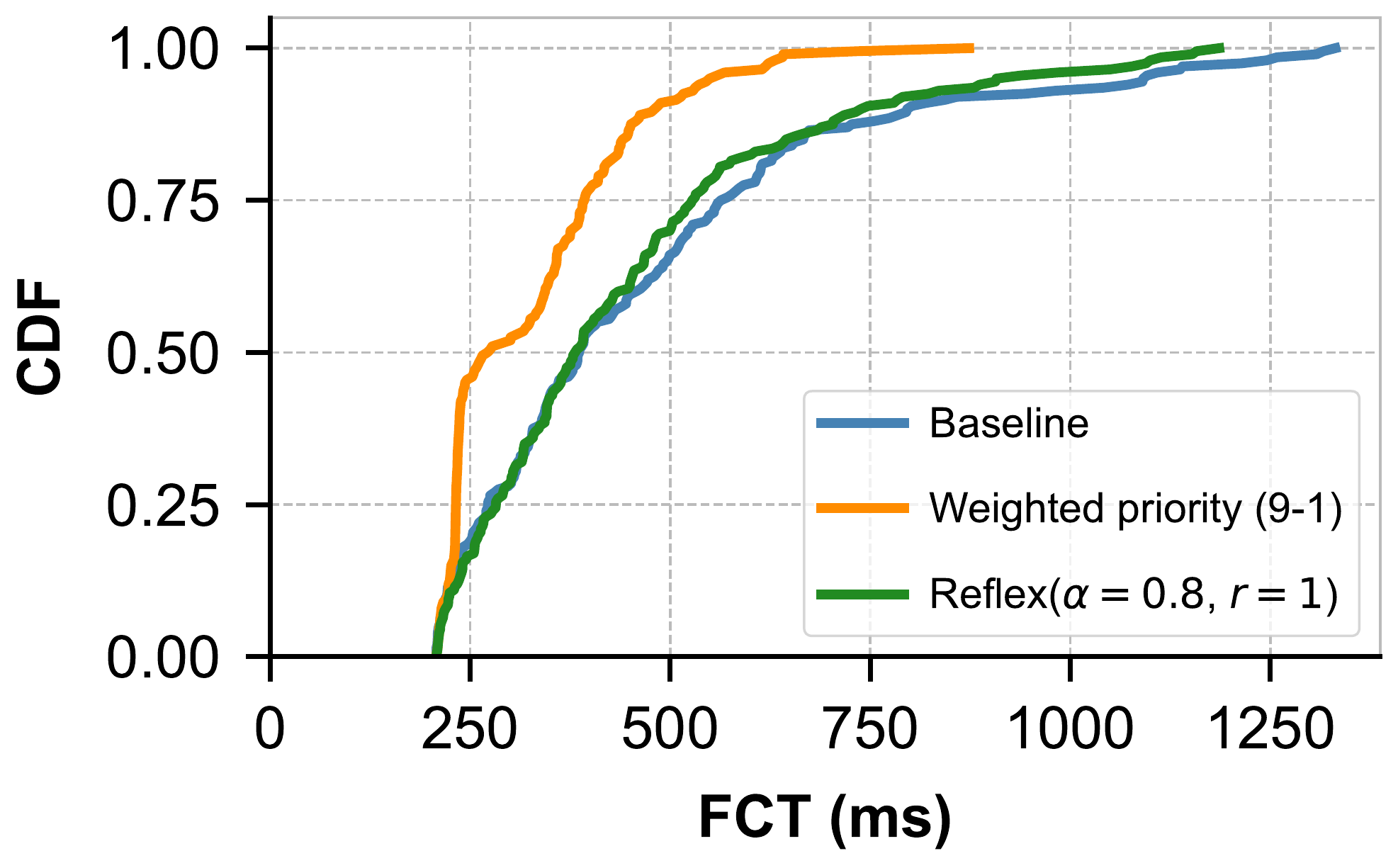}%
        \caption{Regular: FCT}
        \label{fig:cdf-fct:regular}
    \end{subfigure}
    \begin{subfigure}[b]{0.24\textwidth}
        \centering
        \includegraphics[width=\textwidth]{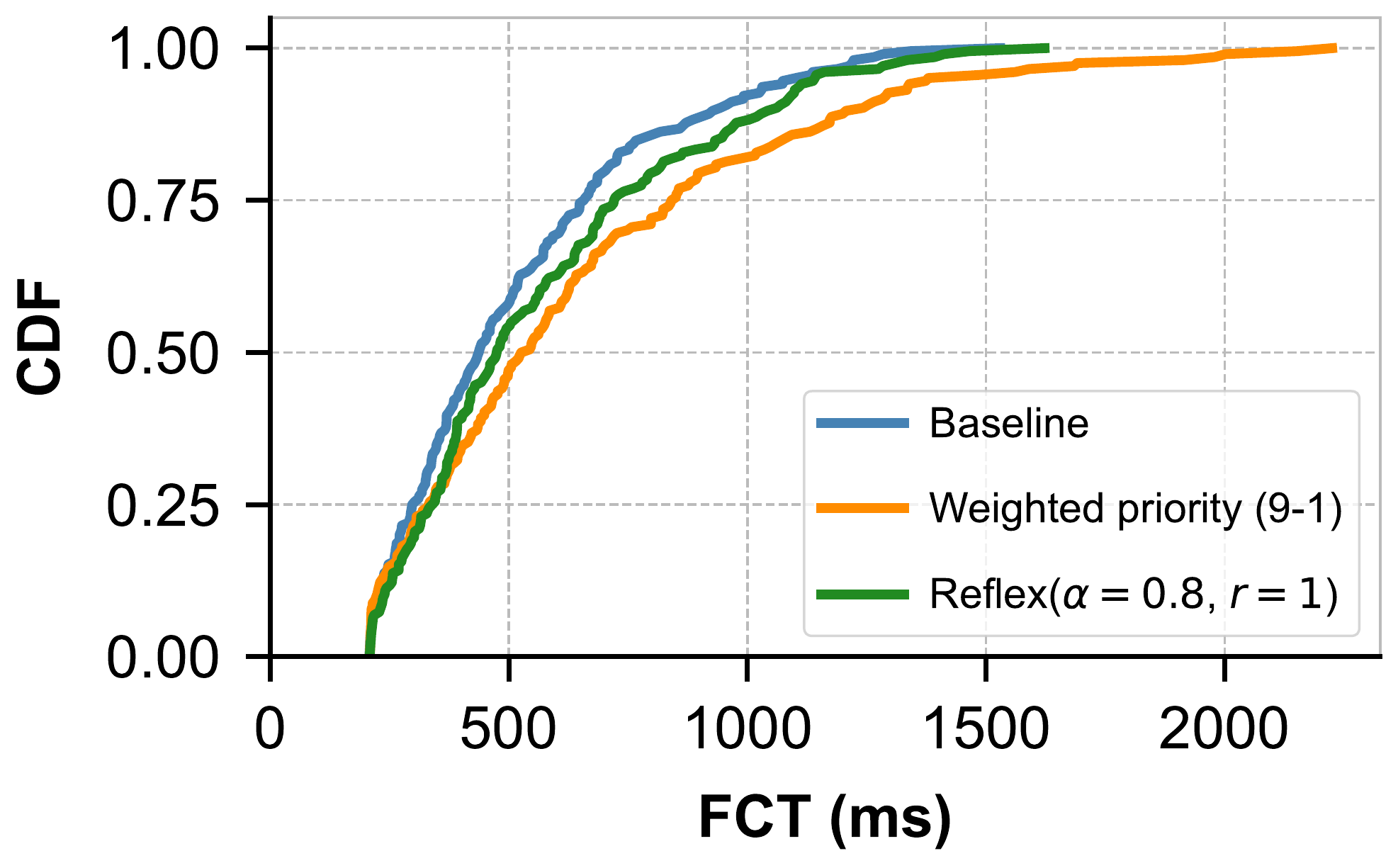}%
        \caption{Flexible: FCT}
        \label{fig:cdf-fct:flexible}
    \end{subfigure}
    \caption{Workload 1: large flow competition. For one repetition, the distribution of FCT and corresponding speed-up relative to baseline of equal priority. \sysName limits the slow-down of flexible flows as it strives to uphold its guarantees at the cost of achieving less speed-up for regular flows.}
    \label{fig:torcdf}
    \vspace{-6pt}
\end{figure*}

\textbf{Workload 1: large flow competition.} In this first workload, the regular flows are similarly large (250~MB) and arrive at the same Poisson arrival rate of 20 flows per second. This achieves a utilization of around 40\%. The resulting mean and 90/99th \%-tile FCT for both regular and flexible flows are shown in Fig.~\ref{fig:largeflows}. Without prioritization, no flows receive special priority, as such the mean FCT of regular flows (460~ms) and flexible flows (465~ms) as well as the 99th \%-tile (respectively 1215~ms and 1227~ms) are similar. With the 9-1 weighted prioritization, flexible flows are degraded significantly (speed-up mean: 0.80$\times$, 99th: 0.67$\times$) to accelerate the regular flows (mean: 1.35$\times$, 99th: 1.64$\times$). \sysName due to the probing phases is unable to provide as much speed-up as weighted priority. However, with both reduced reliability and reduced aggressiveness yields a speed-up for regular flows. At $\alpha$ or $r$ less than 0.5, \sysName does not yield much additional speed-up as with the network load it is able to consistently be at low priority during exploitation phases. This is in line with prior parameterization observations in \S\ref{sec:challenge-probing}. At $\alpha=0.8$, \sysName achieves a mean FCT of 513~ms for flexible flows (speed-up: 0.91$\times$) and 441~ms for regular flows (speed-up: 1.04$\times$), with 99th \%-tiles of 1390~ms and 1178~ms respectively (speed-ups: 0.88$\times$ and 1.04$\times$). In terms of achievable speed-up, the 9-1 weighted prioritization yields significantly more which comes at the cost of being worse for the flexible flows.

We next investigate the ability of \sysName to uphold its guarantees by using the FCT for each flow in the baseline without prioritization. We observe how much the effect of going at low priority either due to \sysName or due to weighted prioritization. We plot the CDF of the FCT and speed-up for both \sysName at $\alpha=0.8$ and 9-1 weighted prioritization for workload 1 in Fig.~\ref{fig:torcdf}. As noted earlier in \S\ref{sec:achieving-max-min}, in multi-hop networks a portion of flows will inevitably experience slowdown even if they remain at high priority. This is shown in Fig.~\ref{fig:cdf-speedup:regular}, where a small tail of around 24\% experience a worse FCT for the fixed weighted prioritization scheme -- a tail is observed for \sysName as well of 45\%. Weighted prioritization causes a portion of the flexible flows to be starved, which is shown by the long tail in Fig.~\ref{fig:cdf-fct:flexible}. In contrast, \sysName does not exhibit such a long tail, which however does go at the cost of significantly less speed-up of regular flows as is shown in Fig.~\ref{fig:cdf-speedup:regular}. \sysName guarantees are regarding the fair share rate, not achieving a particular FCT or rate. Nevertheless, it is interesting as a proxy to observe how many of the flexible flows experience a speed-up of less than 0.8 in Fig.~\ref{fig:cdf-speedup:flexible}: around 14.5\% has a speed-up less than 0.8. In contrast, the fixed weighted prioritization has 32.5\% with a speed-up less than 0.8. The cause why \sysName is unable to uphold a speed-up of 0.8 is (a) fundamental, as the fair share is interdependent as such a flow going high does not guarantee an $R_{actual}$ greater than the fair share, and (b) practical due to inaccuracies in measurements and the dependency on fast convergence.

\greybox{\textbf{Takeaways from workloads with large regular flows:} For large flows competing, \sysName provides a tradeoff between prioritizing the flexible flows and regular flows. Due to its efforts to maintain the guarantees of flexible flows, it provides less speed-up than the fixed weighted prioritization scheme. Its ability to enforce guarantees is influenced by the flow fair share interdependence, measurement inaccuracies and dependency on convergence.}
\vspace{0.4cm}

% Small flows
\begin{figure*}[t]
	\centering
    \begin{subfigure}[b]{0.3\textwidth}
        \centering
        \includegraphics[width=\textwidth]{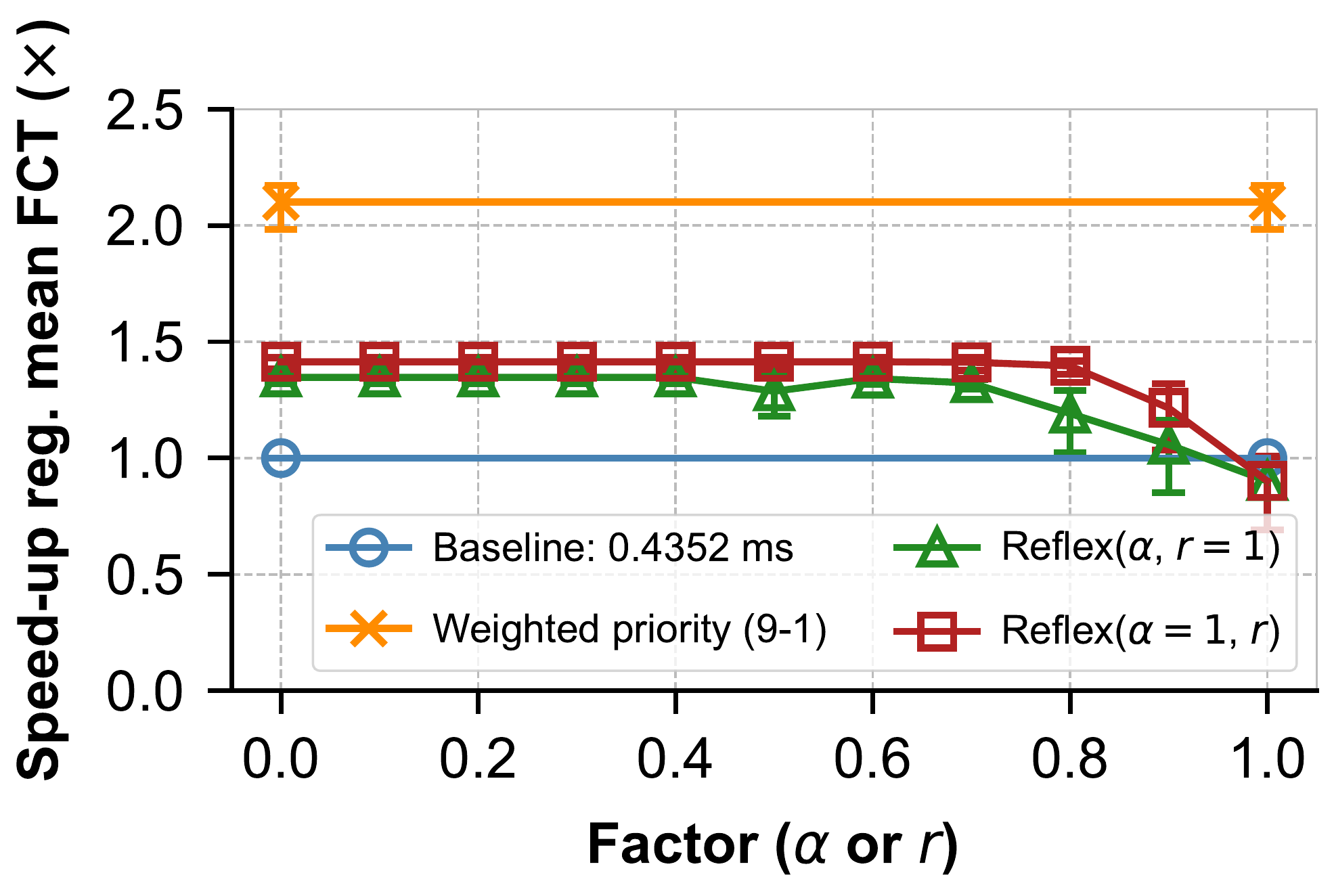}%
        \caption{Regular: mean FCT\newline\newline}
        \label{fig:shortflows:regular:mean}
    \end{subfigure}
 	\hfill
    \begin{subfigure}[b]{0.3\textwidth}
        \centering
        \includegraphics[width=\textwidth]{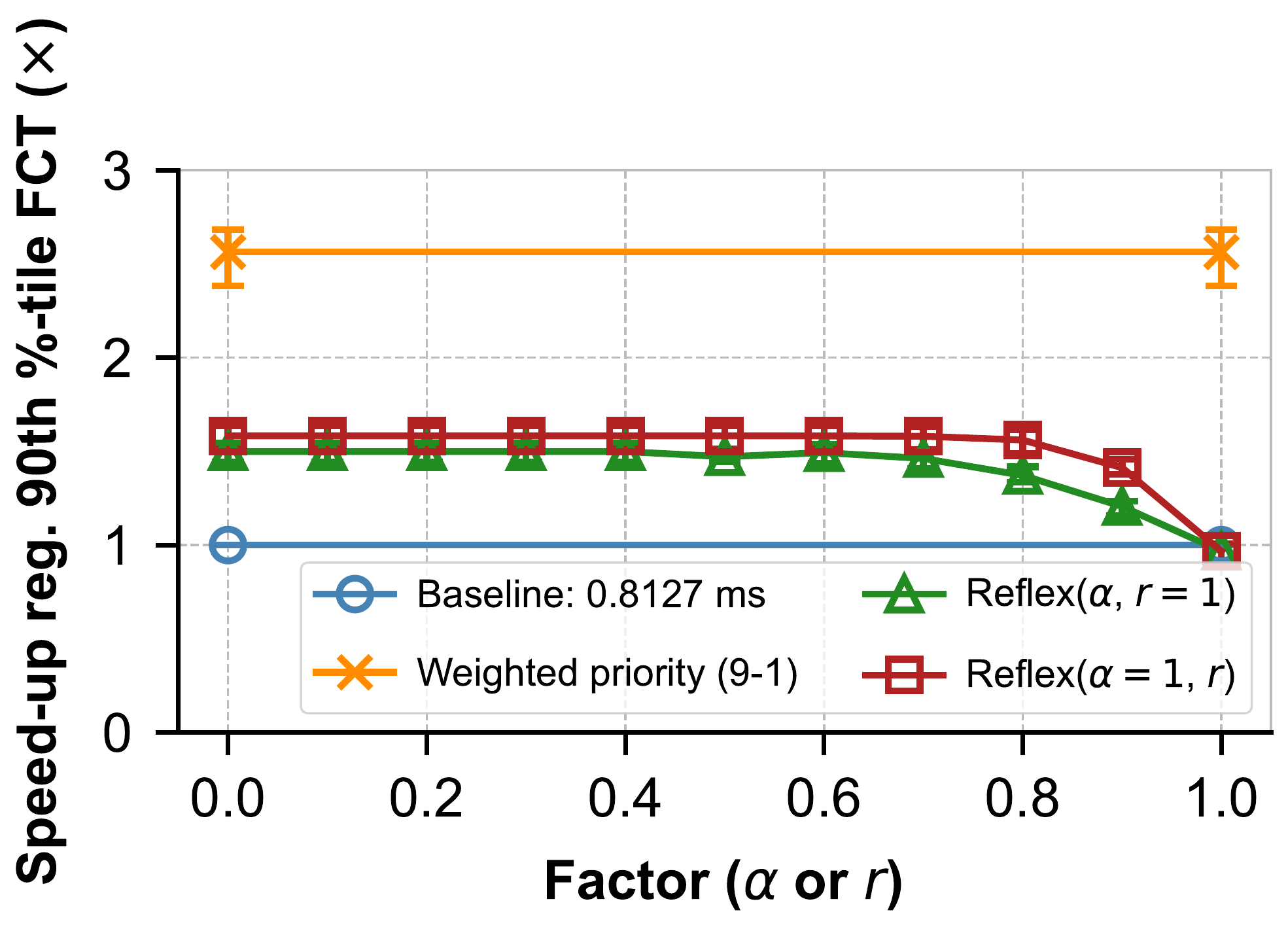}%
        \caption{Regular: 90th \%-tile FCT\newline\newline}
        \label{fig:shortflows:regular:90th}
    \end{subfigure}
 	\hfill
    \begin{subfigure}[b]{0.3\textwidth}
        \centering
        \includegraphics[width=\textwidth]{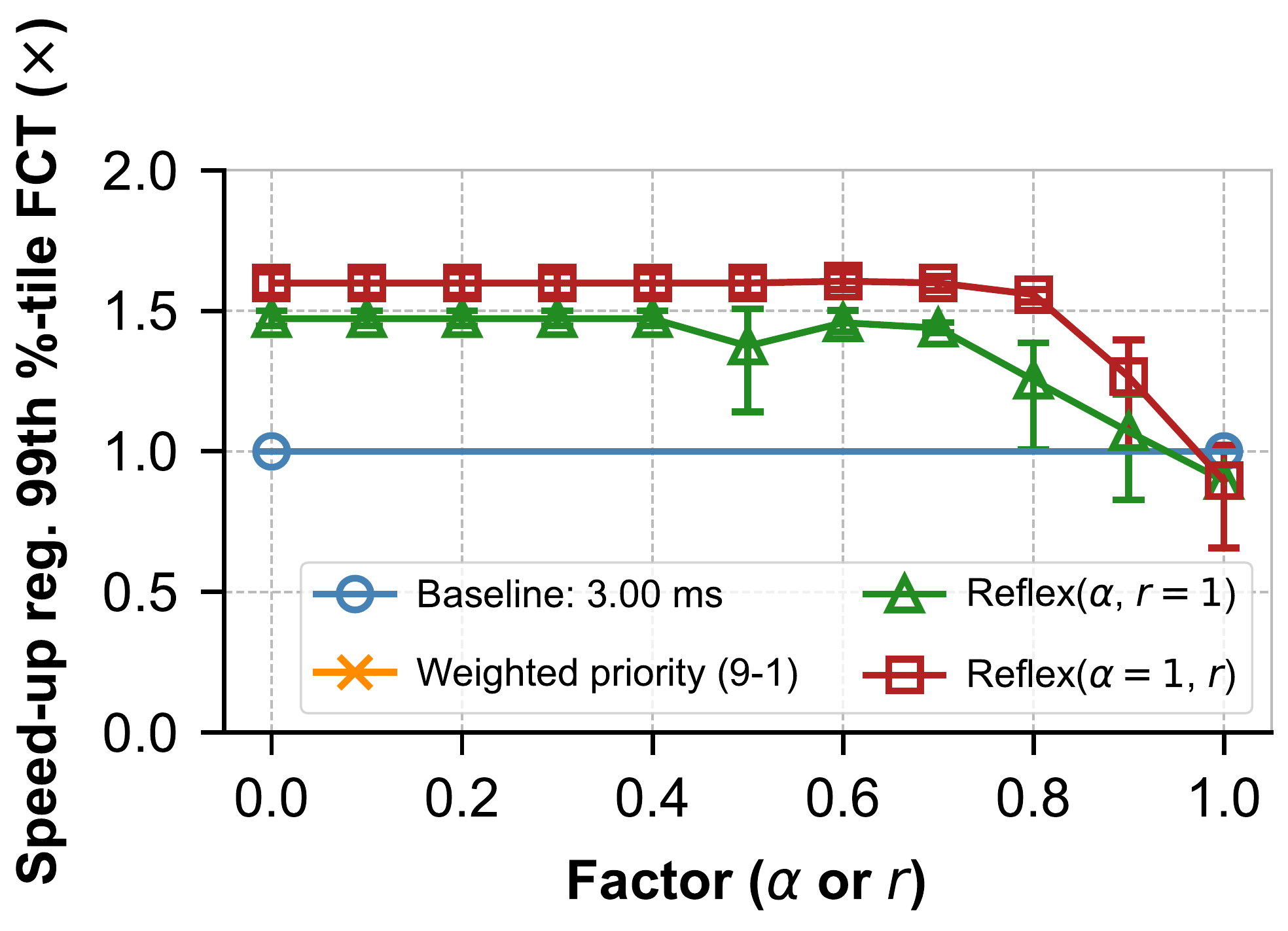}%
        \caption{Regular: 99th \%-tile FCT (weighted priority (9-1) achieved $6.8\times$ (min: $6.1\times$, max: $7.3\times$))}
        \label{fig:shortflows:regular:99th}
    \end{subfigure}
    \hfill
    \begin{subfigure}[b]{0.3\textwidth}
        \centering
        \includegraphics[width=\textwidth]{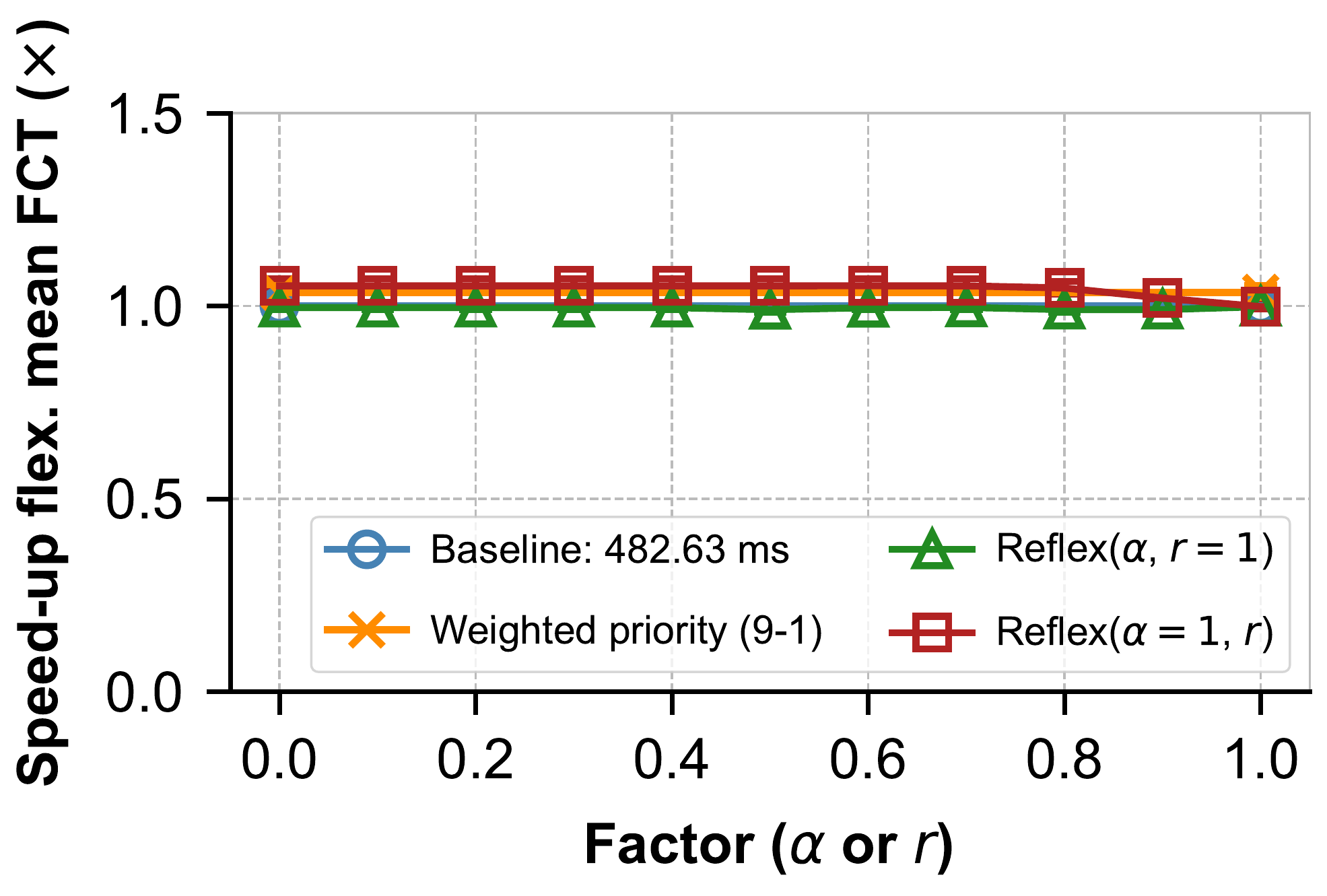}%
        \caption{Flexible: mean FCT}
        \label{fig:shortflows:flexible:mean}
    \end{subfigure}
    \hfill
    \begin{subfigure}[b]{0.3\textwidth}
        \centering
        \includegraphics[width=\textwidth]{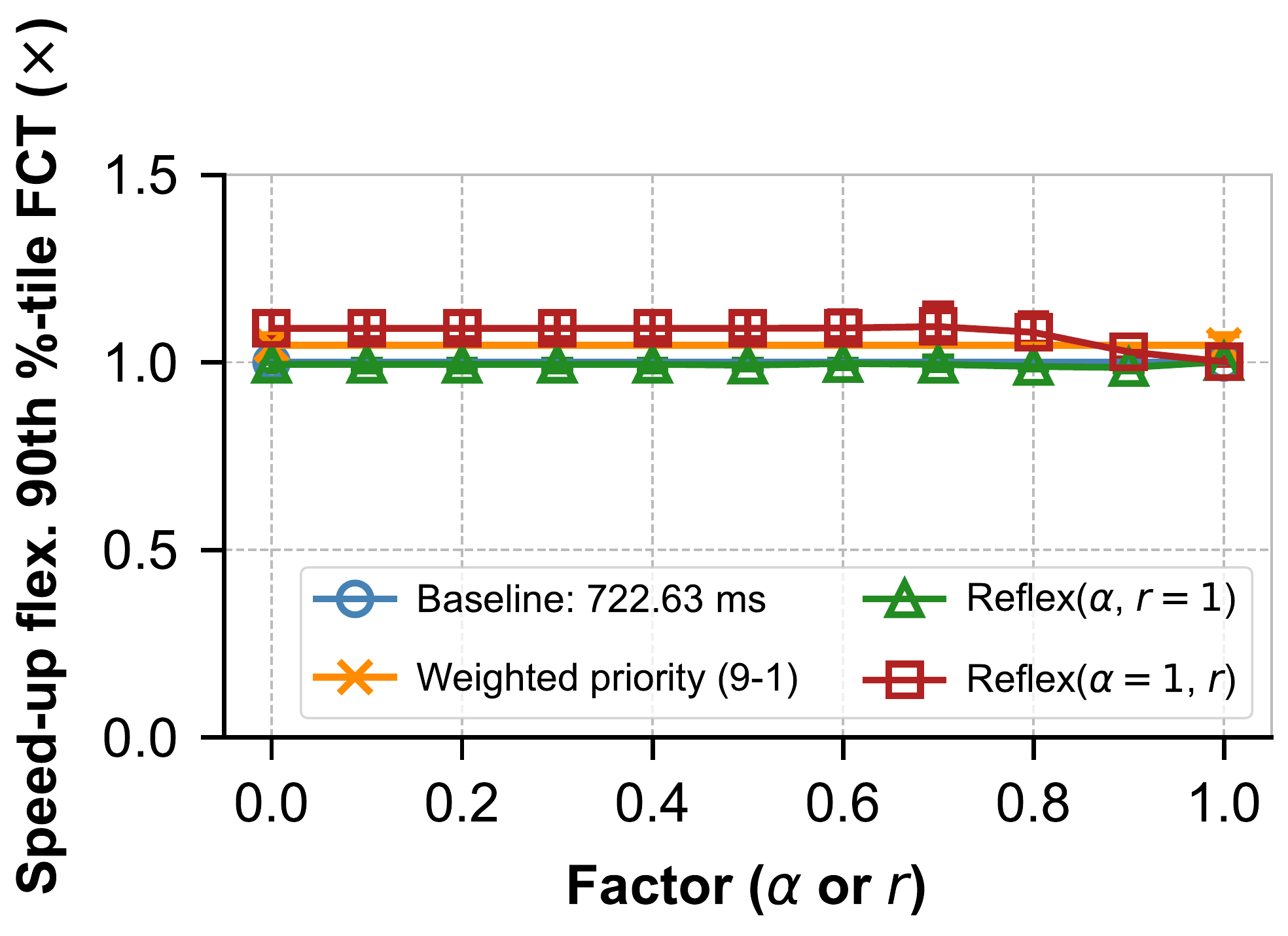}%
        \caption{Flexible: 90th \%-tile FCT}
        \label{fig:shortflows:flexible:90th}
    \end{subfigure}
    \hfill
    \begin{subfigure}[b]{0.3\textwidth}
        \centering
        \includegraphics[width=\textwidth]{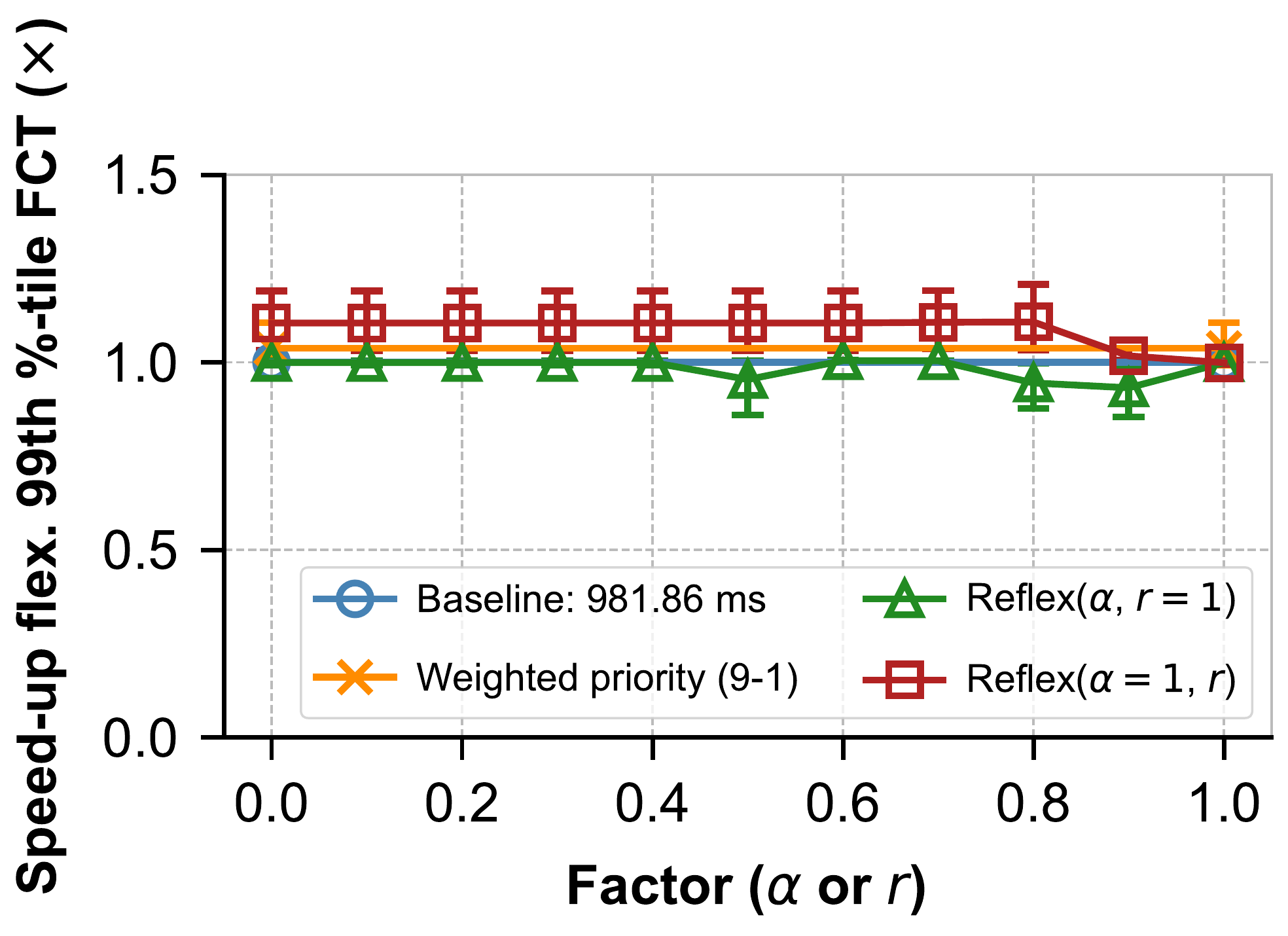}%
        \caption{Flexible: 99th \%-tile FCT}
        \label{fig:shortflows:flexible:99th}
    \end{subfigure}
    \hfill
    \caption{Workload 2: many small flows. \sysName is able to provide speed-up to the short flows, configurable by $\alpha$ and $r$, however the fixed weighted priority offers significantly more speed-up especially at higher percentiles.}
    \label{fig:shortflows}
    \vspace{-6pt}
\end{figure*}

\textbf{Workload 2: many small flows.} In this second workload, we run small regular flows of 100~kB at an Poisson arrival rate of 5000 flows/s. This similarly achieves a utilization of around 40\%. This setup mimics a scenario where an $\alpha$-flexible workload sending a small number of large messages is colocated with a latency-sensitive workload with a large number of short messages, \eg ML training colocated with Web search. The resulting mean and 99th \%-tile FCT for both regular and flexible flows are shown in Fig.~\ref{fig:shortflows}. The 9-1 fixed weighted prioritization achieves a large speed-up in both the mean FCT (2.10$\times$) and the 99th FCT (6.82$\times$). Even for the flexible flows a speed-up is achieved in both the mean and 99th of around 1.04$\times$. By separating the short flows from the large flows, queueing is significantly improved, and both performed better. For \sysName, the probing has significant impact on the speed-up, as queues are particularly slowing for small flows. As a consequence, \sysName similarly achieves a lesser speed-up, for example at $\alpha=0.8$ a speedup of 1.19$\times$ in the mean and 1.25$\times$ in the 99th \%-tile is achieved. In the 99th \%-tile of the short regular flows (Fig.~\ref{fig:shortflows:regular:99th}), \sysName experiences significant amount of variance likely caused by the probing's shifting of queues -- even in some cases performing worse than the baseline. The flexible flows are affected with a mean speed-up of 0.99$\times$ in the mean and 0.95$\times$ in the 99th percentile. For fixed prioritization, the flexible mean speed-up was 1.04$\times$ and 1.04$\times$ at the 99th -- actually providing a slight speed-up. Our algorithm had 1.2\% of flexible flows have a speed-up worse than 0.8$\times$, versus 0.2\% for fixed prioritization. Because the many short flows finish in less than the probing phase duration, they become part of the fair share estimate and as such their continuously renewed presence yields little drain of budget. Thus, the point at which reducing $\alpha$ or $r$ further is higher (around 0.7) than for the previous workload (around 0.5). In this workload where many short flows arrive, weighted prioritization yields a better speed-up in FCT for regular flows, while even resulting in a small improvement in flexible flow performance.

\greybox{\textbf{Takeaways from workload with many small regular flows:} Flexible flows must be of relatively large size, first accrue budget to be able to go at lower priority, and can only be at lower priority for limited time due to probing. Short regular flows finish in the order of 10s or 100s of microseconds, ideally completing only a few round-trips before completing. As such, they are heavily impacted when queued behind large flows or when receiving congestion signals such as reorder, loss and explicit notifications. \sysName by frequently switching queues both incurs additional reorder, as well as the instantaneous move of large queues. This particularly negatively impacts the highest percentiles, which are especially important for short flows. Fixed prioritization schemes do not have these effects as flows do not change queue over their lifetime, and as such perform better for this workload.}
\vspace{0.4cm}

% WS flows: regular
\begin{figure*}[t]
	\centering
    \begin{subfigure}[b]{0.3\textwidth}
        \centering
        \includegraphics[width=\textwidth]{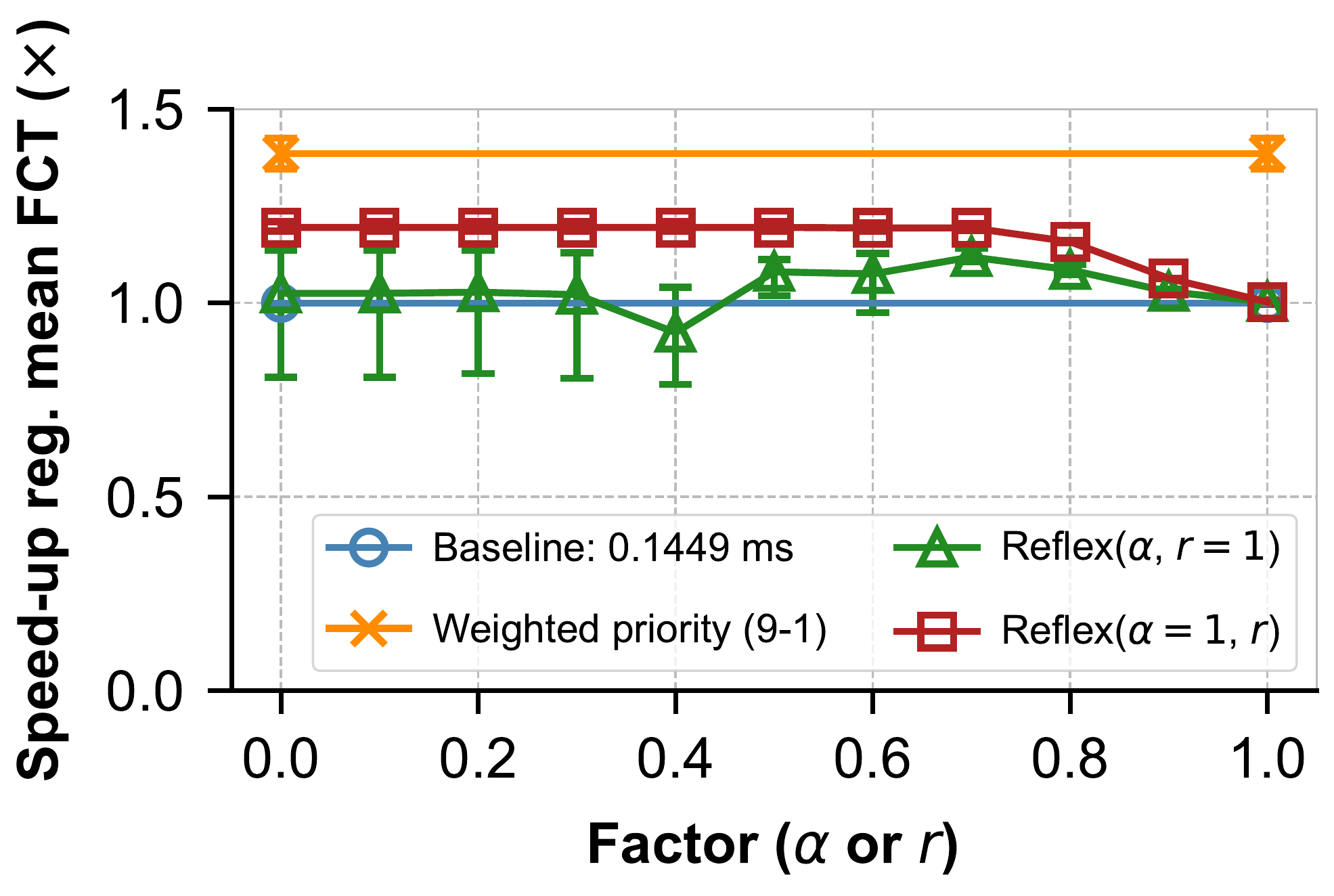}%
        \caption{Regular tiny (0, 10~kB]: mean FCT}
        \label{fig:wsflows:regular:tiny-mean}
    \end{subfigure}
    \hfill
    \begin{subfigure}[b]{0.3\textwidth}
        \centering
        \includegraphics[width=\textwidth]{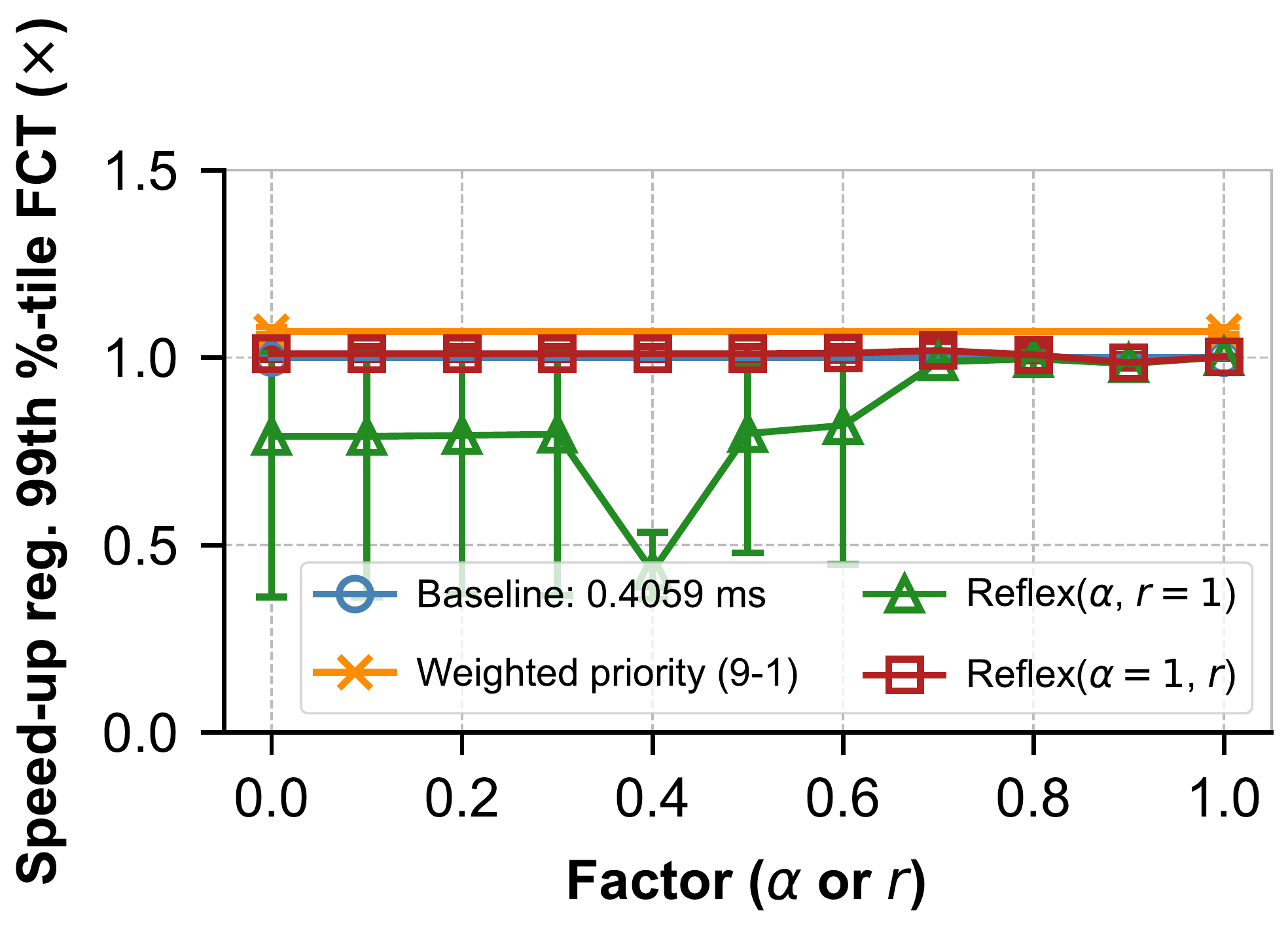}%
        \caption{Regular tiny (0, 10~kB]: 99th \%-tile FCT}
        \label{fig:wsflows:regular:tiny-99th}
    \end{subfigure}
 	\hfill
    \begin{subfigure}[b]{0.3\textwidth}
        \centering
        \includegraphics[width=\textwidth]{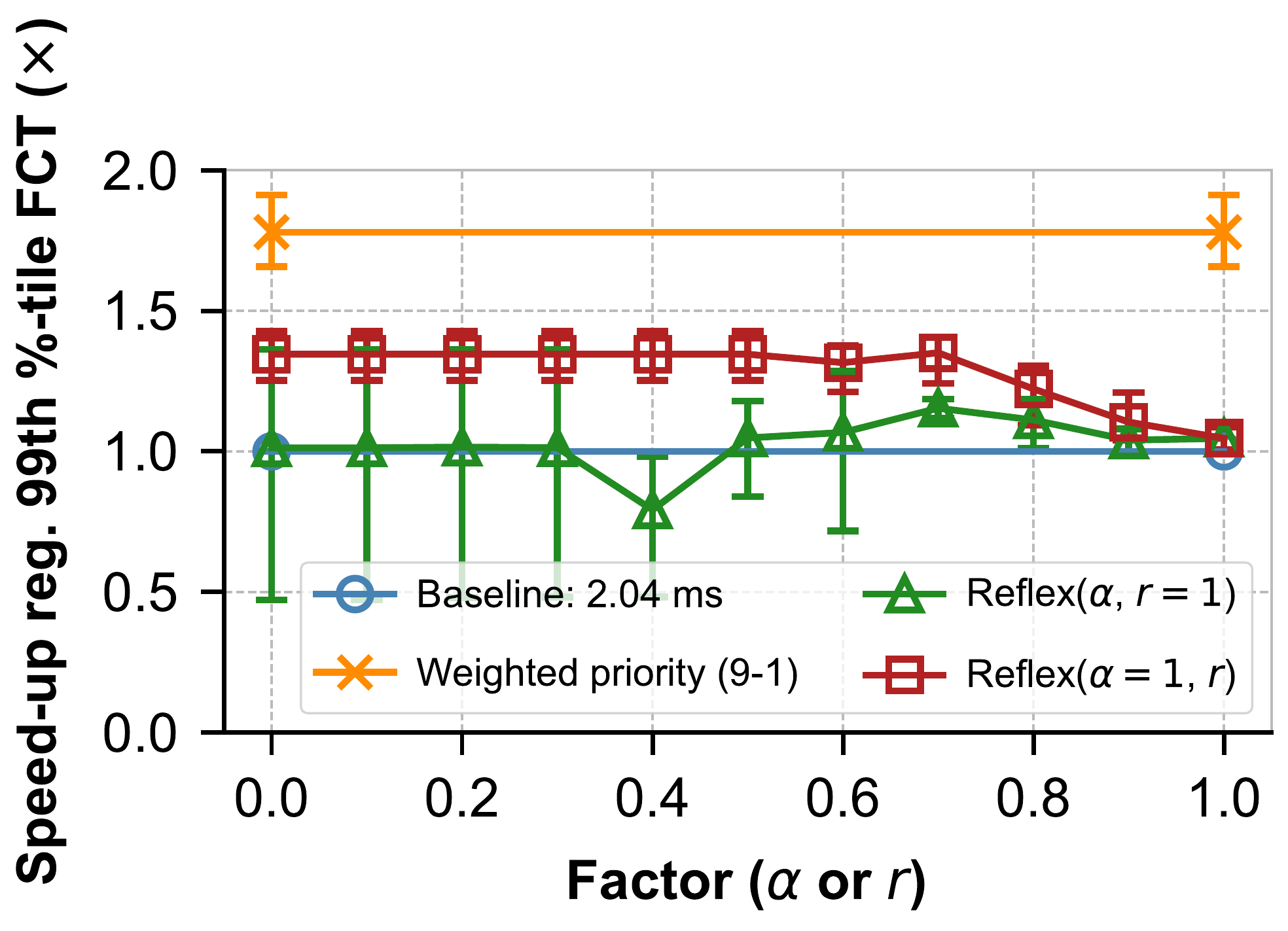}%
        \caption{Regular small (10~kB, 100~kB]: 99th \%-tile FCT}
        \label{fig:wsflows:regular:small-99th}
    \end{subfigure}
 	\hfill
    \begin{subfigure}[b]{0.3\textwidth}
        \centering
        \includegraphics[width=\textwidth]{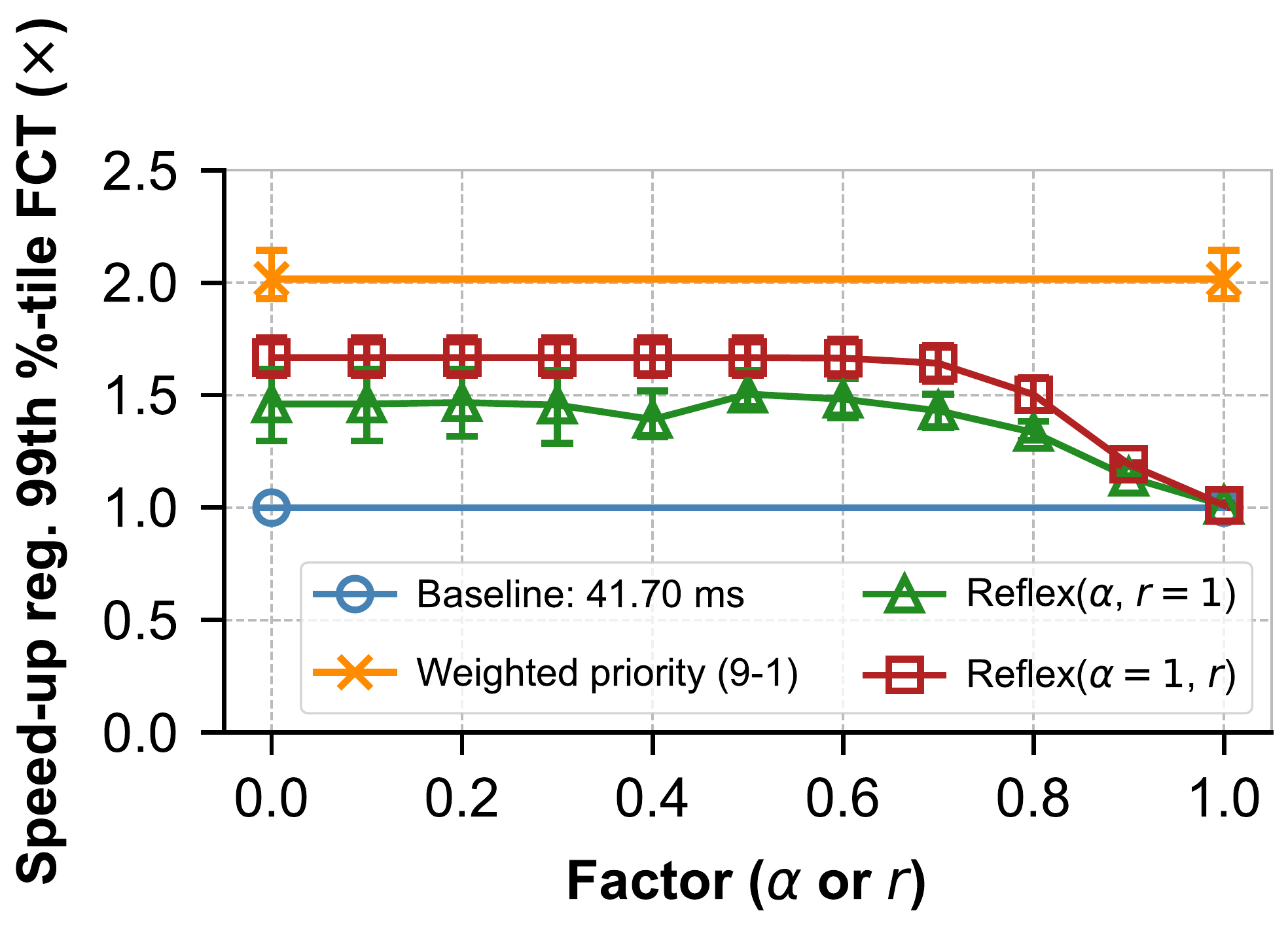}%
        \caption{Regular medium (100~kB, 10~MB]: 99th \%-tile FCT}
        \label{fig:wsflows:regular:medium-99th}
    \end{subfigure}
 	\hfill
    \begin{subfigure}[b]{0.3\textwidth}
        \centering
        \includegraphics[width=\textwidth]{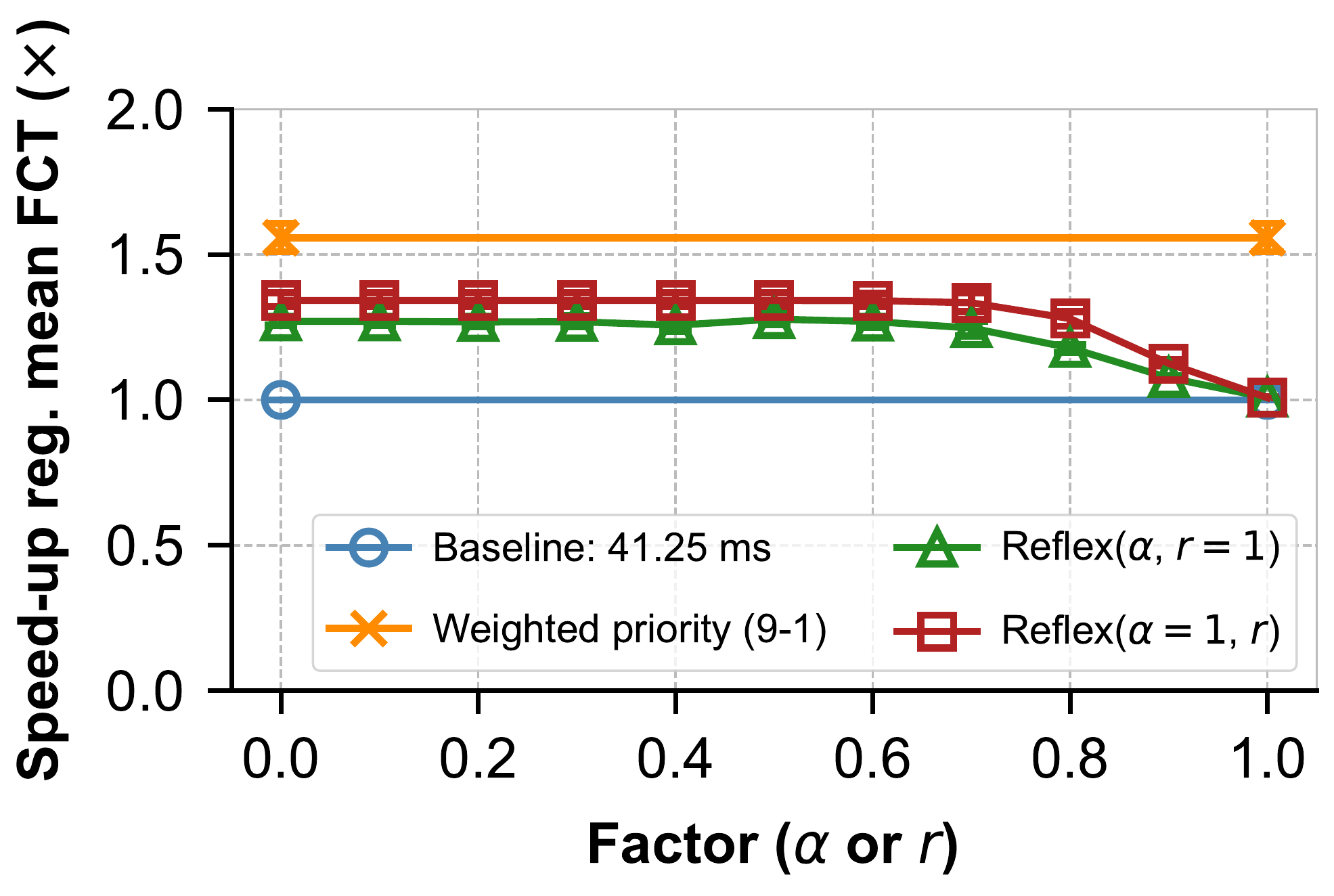}%
        \caption{Regular large (10~MB, $\infty$]: mean FCT}
        \label{fig:wsflows:regular:large-mean}
    \end{subfigure}
 	\hfill
    \begin{subfigure}[b]{0.3\textwidth}
        \centering
        \includegraphics[width=\textwidth]{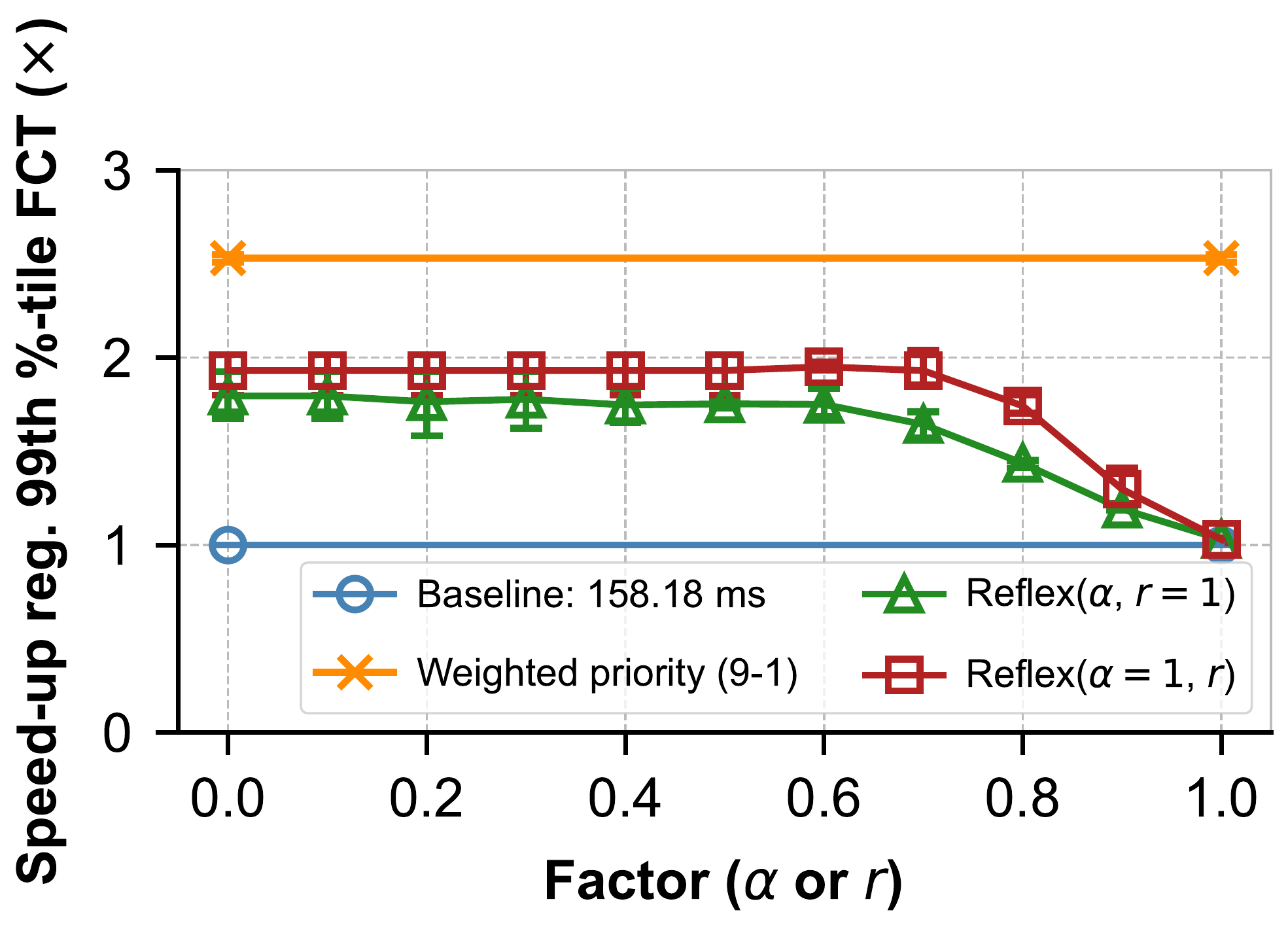}%
        \caption{Regular large (10~MB, $\infty$]: 99th \%-tile FCT}
        \label{fig:wsflows:regular:large-99th}
    \end{subfigure}
    \hfill
    \caption{Workload 3: regular flows.}
    \label{fig:wsflows:regular}
    \vspace{0pt}
\end{figure*}

% WS flows: flexible
\begin{figure*}[t]
	\centering
    \begin{subfigure}[b]{0.3\textwidth}
        \centering
        \includegraphics[width=\textwidth]{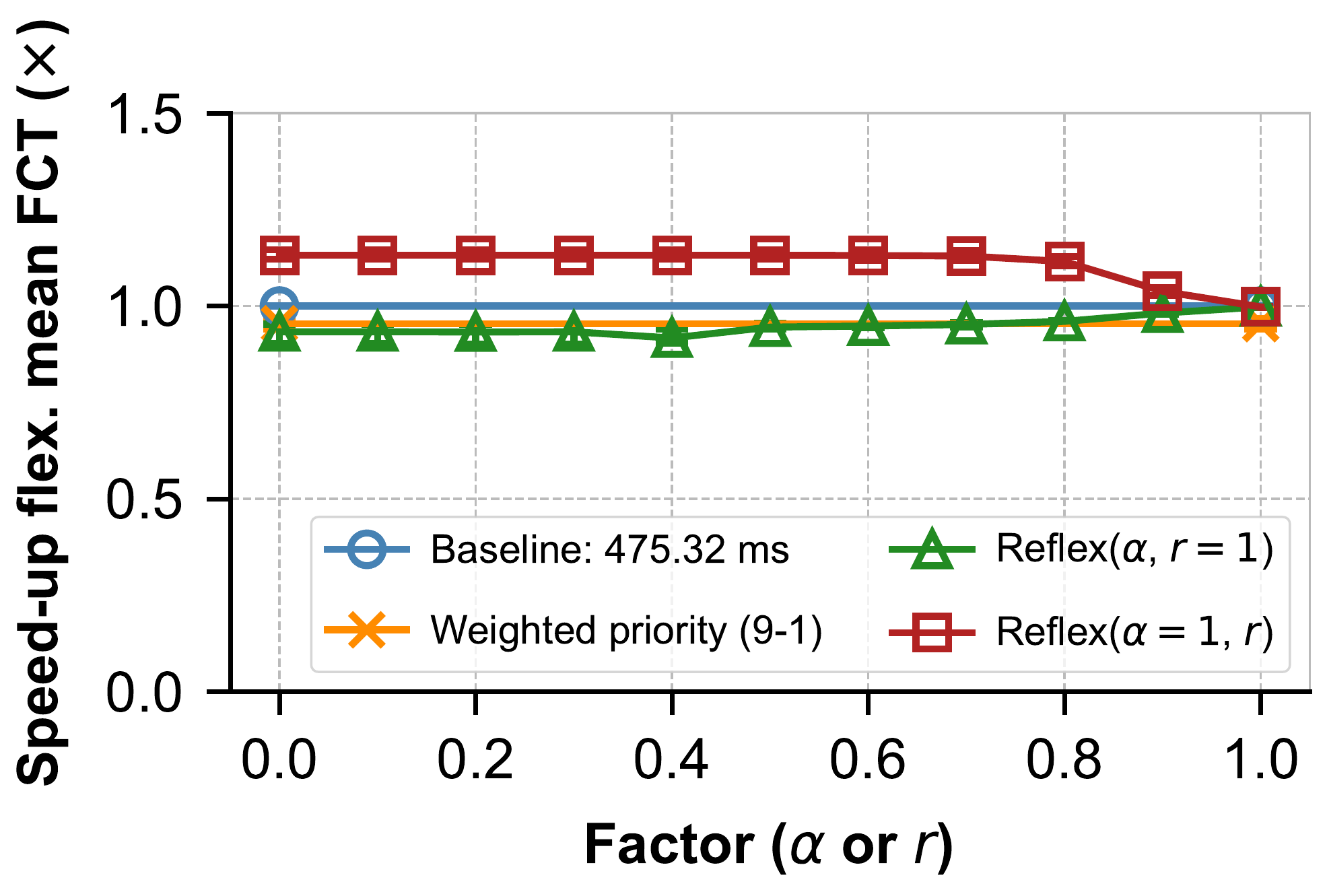}%
        \caption{Flexible: mean FCT}
        \label{fig:wsflows:flexible:mean}
    \end{subfigure}
    \hfill
    \begin{subfigure}[b]{0.3\textwidth}
        \centering
        \includegraphics[width=\textwidth]{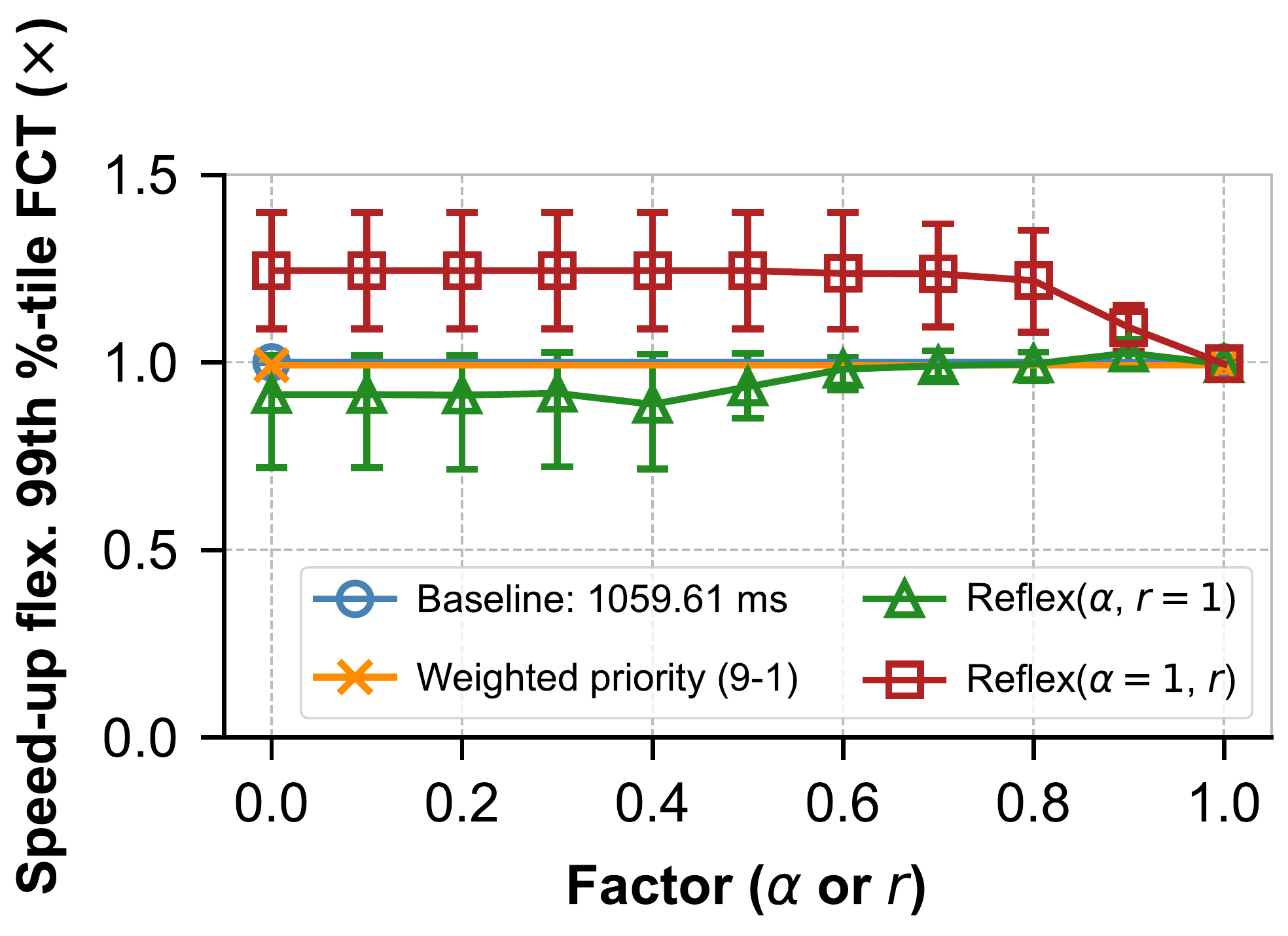}%
        \caption{Flexible: 99th \%-tile FCT}
        \label{fig:wsflows:flexible:99th}
    \end{subfigure}
    \hfill
    \begin{subfigure}[b]{0.3\textwidth}
        \centering
        \includegraphics[width=\textwidth]{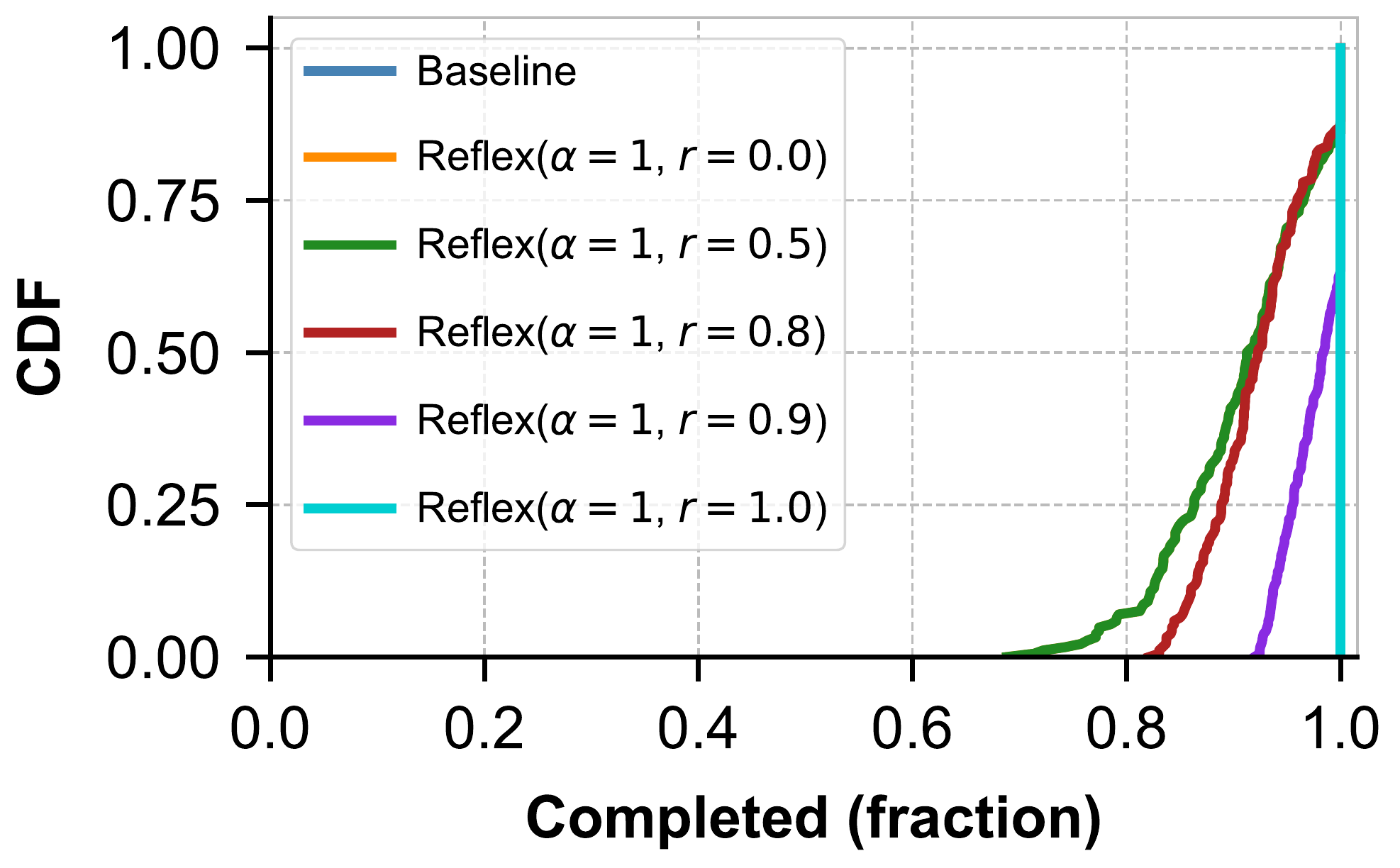}%
        \caption{For one repetition: Flexible: fraction delivered}
        \label{fig:wsflows:flexible:fraction-completed}
    \end{subfigure}
    \caption{Workload 3: flexible flows.}
    \label{fig:wsflows:flexible}
    \vspace{-6pt}
\end{figure*}

\textbf{Workload 3: WS flows.} As a continuation of the second workload, we now use the Web Search flow size distribution of pFabric~\cite{pfabric} and DCTCP~\cite{dctcp} which consists of flow sizes varying between 4~kB and 30~MB, with a majority of its flows under 100~kB. The mean flow size is 1.7~MB, as such to target 20\% utilization we configure a Poisson arrival rate of 2921~flows/s. Similar to \cite{hull}, we separate the statistics of regular flows into four categories of tiny (0, 10~kB], small (10~kB, 100~kB], medium (100~kB, 10~MB], and large (10~MB, $\infty$]. The results are shown in Fig.~\ref{fig:wsflows:regular} and Fig.~\ref{fig:wsflows:flexible}. We use the 99th percentile as key metric for the tiny and small flows. \sysName performs especially not well for the tiny and small flows (Fig.~\ref{fig:wsflows:regular:tiny-99th} and Fig.~\ref{fig:wsflows:regular:small-99th}), in which it experiences both slower FCT and large variance at lower $\alpha$. For medium and large flows it does yield speed-up although not as much as fixed weighted prioritization (Fig.~\ref{fig:wsflows:regular:medium-99th},  Fig.~\ref{fig:wsflows:regular:large-mean}, Fig.~\ref{fig:wsflows:regular:large-99th}). Flexible flows are slowed down to bring this speed-up to the regular flows (Fig.~\ref{fig:wsflows:flexible:mean} and Fig.~\ref{fig:wsflows:flexible:99th}). Across all flow size categories, reduced reliability performs well as its reduced delivery reduces general utilization of the network. The fraction of the flows actually delivered before being completed does not exceed the reliability factor $r$ (by design), and lowering the $r$ beyond 0.6 does not yield further reduced delivery as it cannot drain the budget due to the phase balance of 1:1:3 (as explained in \S\ref{sec:decision-partial-delivery}). At $\alpha=0.8$, \sysName had 1.7\% of flexible flows experience a speed-up worse than 0.8$\times$ versus 2.2\% for fixed prioritization. For the same configuration, \sysName achieved across all regular flow size categories significantly less speed-up as well, with 1.00$\times$, 1.11$\times$, 1.33$\times$ and 1.44$\times$ (\sysName) versus 1.07$\times$, 1.78$\times$, 2.02$\times$ and 2.53$\times$ (weighted prioritization) for the 99th \%-tile FCT speed-up of regular tiny, small, medium and large flows respectively. For Workload 3 as for Workload 2, the fixed prioritization scheme provided better speed-up across the regular flows at less slow-down of the flexible flows.

\greybox{\textbf{Takeaways from WS workload:} Confronted by a mix of small and large regular flows, \sysName mostly provides benefit to the relatively large regular flows. The frequent changing of queues of  \sysName combined with the presence of the other larger regular flows led to worse flow completion times of small flows with increased variance at low alpha values. \sysName performed less well than the fixed prioritization scheme for such workloads in those regular flow size categories while not providing better guarantees for the flexible flows.}

%%%%%%%%%%%%%%%%%%%%%%%%%%%%%%%%%%%%%%%%%%%%%
%%%%%%%%%%%%%%%%%%%%%%%%%%%%%%%%%%%%%%%%%%%%%
%%%%%%%%%%%%%%%%%%%%%%%%%%%%%%%%%%%%%%%%%%%%%
%%%%%%%%%%%%%%%%%%%%%%%%%%%%%%%%%%%%%%%%%%%%%
%%%%%%%%%%%%%%%%%%%%%%%%%%%%%%%%%%%%%%%%%%%%%

\section{Discussion \& future work}
\label{sec:fdt-future-work}

\sysName exposes degradation primitives to applications, thus substantially changing an application-network interface that has remained largely stable for three decades. This new approach raises several threads for further investigation.

\parab{Customizing transport:} We attempted to minimally change the current network stack to add support for bounded degradation. As a result, our approach inherits the limitations of today's transport. TCP (and its derivatives) are well known to not always converge to fair-share bandwidth, exhibiting various types of divergences from it, \eg for short flows and flows with different network round-trip times. In practice, our bounded deprioritization guarantee thus is not strict versus the actual fair-share of a flow, but only with respect to what today's transport could achieve. There would thus be value in considering what a clean-slate approach to supporting bounded degradation might look like, and then seeking a suitable pragmatic middle ground between that and our minimal changes. For instance, in a data center setting, it may even be possible to estimate fair-share bandwidth quickly~\cite{zhuo2016rack, power-of-flexible-packet-processing, jose2015high}. This would greatly enhance our bounded deprioritization's performance.

\parab{Incentivizing applications:} Clearly, unless there are appropriate incentives, applications will not ask for degraded service. The simplest incentive in a cloud setting might be cost, \ie offering slightly cheaper network data transmission for applications in proportion to their accepted service degradation. There is precedent for similar price differentiation for cloud storage offerings with different service guarantees, \eg for long-term storage~\cite{awsglacier}.

\parab{Malicious behavior:} Malicious applications could potentially degrade the performance of flexible traffic. Consider an application that splits its large regular flows into many small flows. This can distort the fair-share estimate for flexible traffic, leading it to send at a lower rate. However, this is a problem that exists in today's network stack already: it represents a fundamental limitation of flow fairness, as Bob Briscoe argued~\cite{briscoe2007flow}. Potential solutions are the same as those that apply to today's transport: instead of flow-fairness enforced at the granularity of transport flows, impose application-level fair sharing using network-level mechanisms.

\parab{Coflow support:} \sysName's primitives apply to individual flows. However, for many applications a ``coflow'' abstraction of their network traffic is more appropriate, whereby their performance depends on a collection of flows finishing~\cite{chowdhury2012coflow}. Extending our primitives to such coflows could potentially increase their benefit, as the degradations could be used more flexibly across large flow collectives instead of in a more restricted manner at a flow granularity. However, extending \sysName in this manner will require substantial effort, as it implies that the coflow traffic coflow-wide performance rather than local tracking at each flow. Exploring this is thus left to future work.

\parab{Other dimensions of service degradation:} While we have only explored bounded loss and bounded deprioritization as yet, there may be additional dimensions of degradation worth investigation. One such example is nearly-in-order packet delivery. Instead of TCP's fully ordered delivery, are there ways of benefiting from most data being delivered in order? What's the right framing for a bound on such nearly ordered delivery, and how might be exploit it to benefit other traffic? There is likely some potential in this due to the multipath nature of data center topologies: if some traffic can tolerate nearly-ordered delivery, this traffic could potentially be used for load balancing by suitably packet spraying it, while traffic that needs fully ordered delivery still uses traditional flow-affinity primitives like ECMP-per-flow and can incur link utilization imbalance. However, given the availability of flowlet switching~\cite{vanini2017let}, it is unclear how much additional value adding such a primitive brings.

%%%%%%%%%%%%%%%%%%%%%%%%%%%%%%%%%%%%%%%%%%%%%
%%%%%%%%%%%%%%%%%%%%%%%%%%%%%%%%%%%%%%%%%%%%%
%%%%%%%%%%%%%%%%%%%%%%%%%%%%%%%%%%%%%%%%%%%%%
%%%%%%%%%%%%%%%%%%%%%%%%%%%%%%%%%%%%%%%%%%%%%
%%%%%%%%%%%%%%%%%%%%%%%%%%%%%%%%%%%%%%%%%%%%%

\section{Related work}
\label{sec:fdt-related-work}

In our discussion on the need for new primitives (\S\ref{sec:why-new-primitives}), we already contrast our work against standard transport protocols like TCP, UDP, and LEDBAT, as well as work on prioritization and deadline-awareness. Besides these primitives, there is also rich literature on differentiated quality of service~\cite{ballani2014offering, rygielski2013network, wang2014sdn}, which also does not offer guarantees to degraded traffic.

Parallel work available as an online manuscript~\cite{atp} describes ATP, which provides partial delivery but with the goal of aggressively finishing lossy-flows \emph{even faster}, drawing on the rationale of approximate computing. Unlike ATP, our work is targeted at workloads that instead of aggressively competing for the network, cede ground to time-critical traffic. ATP has loss-tolerant flows behave aggressively which leads to partial delivery through actual packets being lost. In contrast, our approach determines partial delivery based on rate loss relative to the fair share. Further, ATP's complex design combines packet spraying, separately configured queue sizes for ATP and non-ATP traffic, and a new rate controller. In contrast, \sysName exploits existing primitives with small changes to how applications transmit data. ATP also does not address bounded deprioritization.

Recent work~\cite{rethinking-tl-dist-ml} explored speeding up ML training tasks at the tail by not retransmitting packets that take a long time to be detected as lost, \eg after a retransmission timeout. This is different from our approach of actively degrading service quality for certain workloads to benefit more critical traffic.

There is extensive work on adaptive bitrate video streaming~\cite{yin2015control, pensieve}, whereby the video player adjusts its playback quality to network conditions. Note that in this case, application-level quality degradation is a built-in primitive that is tolerated due to the real-time nature of the application: the only alternative to reducing video quality is to suffer pauses in streaming. The applications we consider do not have such real-time constraints and do not accept application-level quality degradation. Thus, the design goals of \sysName and its implementation share little in common with adaptive video.

The area of streaming analytics has also generated work studying how to best adapt the streamed data to network bandwidth fluctuations~\cite{awstream, jetstream}. This work takes available network bandwidth as a given, and attempts to find the most suitable way of degrading the streaming data such that the analytics task consuming it suffers the least possible impairment. However, our goals are different: we seek to allow certain flexible workloads to compete less aggressively for network resources by dropping or deprioritizing their traffic. These different goals lead to different design decisions: the streaming analytics work focuses on minimizing application-level degradation, given bandwidth changes, while \sysName takes application goals as given in terms of degradation bounds, and adapts network-level behavior.

There is also work on approximate network protocols in different contexts, \eg SAP~\cite{sapApprox} allows applications to accept partially damaged network data that would otherwise be thrown out by integrity checks like checksum failure. CoAP~\cite{coap} is a UDP-extension that allows tunable tolerance to network failures in high error rate environments. While philosophically similar, these efforts share little in common with \sysName in its use cases, the bounded degradations it allows, and its design to enforce guarantees.

%%%%%%%%%%%%%%%%%%%%%%%%%%%%%%%%%%%%%%%%%%%%%
%%%%%%%%%%%%%%%%%%%%%%%%%%%%%%%%%%%%%%%%%%%%%
%%%%%%%%%%%%%%%%%%%%%%%%%%%%%%%%%%%%%%%%%%%%%
%%%%%%%%%%%%%%%%%%%%%%%%%%%%%%%%%%%%%%%%%%%%%
%%%%%%%%%%%%%%%%%%%%%%%%%%%%%%%%%%%%%%%%%%%%%

\section{Conclusion}
\label{sec:fdt-conclusion}

We make the case that emerging workloads present opportunities for superior network multiplexing by being tolerant of degraded network service. However, to ensure that such applications still get satisfactory performance instead of being excessively penalized for being tolerant, we should offer primitives for \emph{bounded} degradation. We explore two dimensions of such bounded degradation: guaranteed partial delivery and bounded deprioritization. We show that unlike the no-guarantees degradation provided by traditional networking approaches like unreliable transport and various types of prioritization primitives, it is possible to implement bounded degradation to achieve both: (a) improved performance for traffic with strict network service requirements; and (b) assurances that tolerant traffic meets its specified reduced performance requirements.

However, there are practical considerations on (a) the mechanism to actually achieve benefit, and (b) how to achieve awareness of the baseline against which the deterioration is defined. These practical considerations both limit the maximum performance improvement that can be achieved, as well as the ability to fulfill meaningful guarantees. In our work, the continuous probing, frequent convergence of flows and the switching of queues by flows were those practical considerations which impacted performance and guarantees. The two primitives discussed in this work are (1) \textit{workload dependent} because they require flexible flows to be starved to outperform other prioritization schemes, and (2) \textit{objective dependent} as they are useful only if it is desirable in moments of high utilization to not fully prioritize regular flows.

By taking first steps in framing bounded degradation for networking, our work opens up several interesting directions worthy of future exploration, including (a) how faster-converging transport can improve the performance of bounded deprioritization; (b) how one might assess and exploit the stability of the workload distribution to configure probing appropriately; (c) in what other dimensions, \eg nearly in-order delivery, is bounded degradation potentially useful; and (d) how to incentivize good use of bounded degradation for applications that can tolerate it.

%%%%%%%%%%%%%%%%%%%%%%%%%%%%%%%%%%%%%%%%%%%%%
%%%%%%%%%%%%%%%%%%%%%%%%%%%%%%%%%%%%%%%%%%%%%
%%%%%%%%%%%%%%%%%%%%%%%%%%%%%%%%%%%%%%%%%%%%%
%%%%%%%%%%%%%%%%%%%%%%%%%%%%%%%%%%%%%%%%%%%%%
%%%%%%%%%%%%%%%%%%%%%%%%%%%%%%%%%%%%%%%%%%%%%

\balance

\bibliographystyle{plain}
\bibliography{99-bibliography}

\begin{thebibliography}{10}

\bibitem{abbas2020securing}
Syed~Hussain Abbas.
\newblock Securing the network against malicious programmable switches.
\newblock Master's thesis, ETH Zurich, 2020.

\bibitem{dctcp}
Mohammad Alizadeh, Albert Greenberg, David~A Maltz, Jitendra Padhye, Parveen
  Patel, Balaji Prabhakar, Sudipta Sengupta, and Murari Sridharan.
\newblock Data center {TCP} ({DCTCP}).
\newblock {\em ACM SIGCOMM CCR}, 41(4):63--74, 2011.

\bibitem{hull}
Mohammad Alizadeh, Abdul Kabbani, Tom Edsall, Balaji Prabhakar, Amin Vahdat,
  and Masato Yasuda.
\newblock Less is more: Trading a little bandwidth for ultra-low latency in the
  data center.
\newblock In {\em 9th USENIX Symposium on Networked Systems Design and
  Implementation (NSDI 12)}, pages 253--266, San Jose, CA, April 2012. USENIX
  Association.

\bibitem{pfabric}
Mohammad Alizadeh, Shuang Yang, Milad Sharif, Sachin Katti, Nick McKeown,
  Balaji Prabhakar, and Scott Shenker.
\newblock {pFabric}: Minimal near-optimal datacenter transport.
\newblock {\em ACM SIGCOMM CCR}, 43(4):435--446, 2013.

\bibitem{ballani2014offering}
Hitesh Ballani, Paolo Costa, Thomas Karagiannis, and Antony Rowstron.
\newblock Offering network performance guarantees in multi-tenant datacenters,
  March~11, 2014.
\newblock US Patent 8,671,407.

\bibitem{awsglacier}
Jeff Barr.
\newblock New {Amazon} {S3} storage class – {Glacier} {Deep} {Archive}.
\newblock
  \url{https://aws.amazon.com/blogs/aws/new-amazon-s3-storage-class-glacier-deep-archive/},
  2019.
\newblock Accessed: 15 August 2022.

\bibitem{briscoe2007flow}
Bob Briscoe.
\newblock Flow rate fairness: Dismantling a religion.
\newblock {\em ACM SIGCOMM Computer Communication Review}, 37(2):63--74, 2007.

\bibitem{karuna}
Li~Chen, Kai Chen, Wei Bai, and Mohammad Alizadeh.
\newblock Scheduling mix-flows in commodity datacenters with {Karuna}.
\newblock In {\em Proceedings of the 2016 ACM SIGCOMM Conference}, pages
  174--187. ACM, 2016.

\bibitem{chowdhury2012coflow}
Mosharaf Chowdhury and Ion Stoica.
\newblock Coflow: a networking abstraction for cluster applications.
\newblock In {\em HotNets}, pages 31--36, 2012.

\bibitem{cisco8queues}
Cisco.
\newblock Implementing quality of service policies with {DSCP}.
\newblock
  \url{https://www.cisco.com/c/en/us/support/docs/quality-of-service-qos/qos-packet-marking/10103-dscpvalues.html},
  2008.
\newblock Accessed: 15 August 2022.

\bibitem{cisco-priority}
Cisco.
\newblock {QoS}: Congestion management configuration guide, {Cisco} {IOS} {XE}
  release {3S}.
\newblock
  \url{https://www.cisco.com/c/en/us/td/docs/ios-xml/ios/qos_conmgt/configuration/xe-3s/qos-conmgt-xe-3s-book/qos-conmgt-oview.html},
  2018.
\newblock Accessed: 15 August 2022.

\bibitem{hanjing2020making}
Hanjing Gao.
\newblock Making machine learning friendlier in the cloud.
\newblock Master's thesis, ETH Zurich, 2020.

\bibitem{syncGoogle}
Yilong Geng, Shiyu Liu, Zi~Yin, Ashish Naik, Balaji Prabhakar, Mendel
  Rosenblum, and Amin Vahdat.
\newblock Exploiting a natural network effect for scalable, fine-grained clock
  synchronization.
\newblock In {\em 15th {USENIX} Symposium on Networked Systems Design and
  Implementation ({NSDI} 18)}, pages 81--94, 2018.

\bibitem{MITsmr}
Google.
\newblock Survey report: Behind the growing confidence in cloud security, 2019.

\bibitem{he2016deep}
Kaiming He, Xiangyu Zhang, Shaoqing Ren, and Jian Sun.
\newblock Deep residual learning for image recognition.
\newblock In {\em Proceedings of the IEEE conference on computer vision and
  pattern recognition}, pages 770--778, 2016.

\bibitem{jose2015high}
Lavanya Jose, Lisa Yan, Mohammad Alizadeh, George Varghese, Nick McKeown, and
  Sachin Katti.
\newblock High speed networks need proactive congestion control.
\newblock In {\em Proceedings of the 14th ACM Workshop on Hot Topics in
  Networks}, page~14. ACM, 2015.

\bibitem{basic-sim}
Simon Kassing, Hussain Abbas, and Hanjing Gao.
\newblock basic-sim: ns-3 module to make experimental simulation of networks a
  bit easier.
\newblock \url{https://github.com/snkas/basic-sim}, 2021.

\bibitem{hypatia}
Simon Kassing, Debopam Bhattacherjee, Andr\'{e}~Baptista \'{A}guas, Jens~Eirik
  Saethre, and Ankit Singla.
\newblock Exploring the "{Internet} from space" with {Hypatia}.
\newblock In {\em Proceedings of the ACM Internet Measurement Conference}, IMC
  '20, page 214–229, New York, NY, USA, 2020. Association for Computing
  Machinery.

\bibitem{ledbat}
M.~Kuehlewind, G.~Hazel, S.~Shalunov, and J.~Iyengar.
\newblock {RFC} 6817: Low extra delay background transport ({LEDBAT}).
\newblock \url{https://tools.ietf.org/html/rfc6817}, 2012.

\bibitem{syncHakim}
Ki~Suh Lee, Han Wang, Vishal Shrivastav, and Hakim Weatherspoon.
\newblock Globally synchronized time via datacenter networks.
\newblock In {\em Proceedings of the 2016 ACM SIGCOMM Conference}, pages
  454--467. ACM, 2016.

\bibitem{ptpimplementation}
{Linux contributors}.
\newblock The {Linux} {PTP} project.
\newblock \url{http://linuxptp.sourceforge.net/}, 2019.

\bibitem{atp}
Ke~Liu, Shin-Yeh Tsai, and Yiying Zhang.
\newblock {ATP}: a datacenter approximate transmission protocol.
\newblock {\em arXiv preprint arXiv:1901.01632}, 2019.

\bibitem{pensieve}
Hongzi Mao, Ravi Netravali, and Mohammad Alizadeh.
\newblock Neural adaptive video streaming with {Pensieve}.
\newblock In {\em Proceedings of the Conference of the ACM Special Interest
  Group on Data Communication}, pages 197--210. ACM, 2017.

\bibitem{ns3}
{ns-3 contributors}.
\newblock ns-3 network simulator.
\newblock \url{https://www.nsnam.org/}, 2021.

\bibitem{fastpass}
Jonathan Perry, Amy Ousterhout, Hari Balakrishnan, Devavrat Shah, and Hans
  Fugal.
\newblock {Fastpass}: A centralized zero-queue datacenter network.
\newblock {\em ACM SIGCOMM Computer Communication Review}, 44(4):307--318,
  2015.

\bibitem{jetstream}
Ariel Rabkin, Matvey Arye, Siddhartha Sen, Vivek~S Pai, and Michael~J Freedman.
\newblock Aggregation and degradation in {JetStream}: Streaming analytics in
  the wide area.
\newblock In {\em 11th {USENIX} Symposium on Networked Systems Design and
  Implementation ({NSDI} 14)}, pages 275--288, 2014.

\bibitem{sapApprox}
Benjamin Ransford and Luis Ceze.
\newblock {SAP}: an architecture for selectively approximate wireless
  communication, 2015.

\bibitem{rygielski2013network}
Piotr Rygielski and Samuel Kounev.
\newblock Network virtualization for {QoS}-aware resource management in cloud
  data centers: A survey.
\newblock {\em PIK-Praxis der Informationsverarbeitung und Kommunikation},
  36(1):55--64, 2013.

\bibitem{power-of-flexible-packet-processing}
Naveen~Kr Sharma, Antoine Kaufmann, Thomas Anderson, Arvind Krishnamurthy,
  Jacob Nelson, and Simon Peter.
\newblock Evaluating the power of flexible packet processing for network
  resource allocation.
\newblock In {\em 14th {USENIX} Symposium on Networked Systems Design and
  Implementation ({NSDI} 17)}, pages 67--82, 2017.

\bibitem{coap}
Z.~Shelby, C.~Bormann, and K.~Hartke.
\newblock {RFC} 7252: The constrained application protocol ({CoAP}).
\newblock \url{https://tools.ietf.org/html/rfc7252}, 2014.

\bibitem{jupiter-rising}
Arjun Singh, Joon Ong, Amit Agarwal, Glen Anderson, Ashby Armistead, Roy
  Bannon, Seb Boving, Gaurav Desai, Bob Felderman, Paulie Germano, Anand
  Kanagala, Jeff Provost, Jason Simmons, Eiichi Tanda, Jim Wanderer, Urs
  Hölzle, Stephen Stuart, and Amin Vahdat.
\newblock {Jupiter} rising: A decade of {Clos} topologies and centralized
  control in {Google}’s datacenter network.
\newblock {\em ACM SIGCOMM}, 2015.

\bibitem{flow-size-in-advance}
Vojislav {\DH}uki{\'c}, Sangeetha~Abdu Jyothi, Bojan Karla{\v{s}}, Muhsen
  Owaida, Ce~Zhang, and Ankit Singla.
\newblock Is advance knowledge of flow sizes a plausible assumption?
\newblock In {\em 16th {USENIX} Symposium on Networked Systems Design and
  Implementation ({NSDI} 19)}, pages 565--580, 2019.

\bibitem{d2tcp}
Balajee Vamanan, Jahangir Hasan, and TN~Vijaykumar.
\newblock Deadline-aware datacenter {TCP} ({D2TCP}).
\newblock {\em ACM SIGCOMM Computer Communication Review}, 42(4):115--126,
  2012.

\bibitem{vanini2017let}
Erico Vanini, Rong Pan, Mohammad Alizadeh, Parvin Taheri, and Tom Edsall.
\newblock Let it flow: Resilient asymmetric load balancing with flowlet
  switching.
\newblock In {\em 14th {USENIX} Symposium on Networked Systems Design and
  Implementation ({NSDI} 17)}, pages 407--420, 2017.

\bibitem{wang2014sdn}
Jason~Min Wang, Ying Wang, Xiangming Dai, and Brahim Bensaou.
\newblock {SDN}-based multi-class {QoS}-guaranteed inter-data center traffic
  management.
\newblock In {\em 2014 IEEE 3rd International Conference on Cloud Networking
  (CloudNet)}, pages 401--406. IEEE, 2014.

\bibitem{d3}
Christo Wilson, Hitesh Ballani, Thomas Karagiannis, and Ant Rowtron.
\newblock Better never than late: Meeting deadlines in datacenter networks.
\newblock {\em SIGCOMM Comput. Commun. Rev.}, 41(4):50–61, aug 2011.

\bibitem{rethinking-tl-dist-ml}
Jiacheng Xia, Gaoxiong Zeng, Junxue Zhang, Weiyan Wang, Wei Bai, Junchen Jiang,
  and Kai Chen.
\newblock Rethinking transport layer design for distributed machine learning.
\newblock In {\em Proceedings of the 3rd Asia-Pacific Workshop on Networking
  2019}, pages 22--28. ACM, 2019.

\bibitem{yin2015control}
Xiaoqi Yin, Abhishek Jindal, Vyas Sekar, and Bruno Sinopoli.
\newblock A control-theoretic approach for dynamic adaptive video streaming
  over {HTTP}.
\newblock {\em SIGCOMM Comput. Commun. Rev.}, 45(4):325–338, aug 2015.

\bibitem{icml-unreliable}
Chen Yu, Hanlin Tang, Cedric Renggli, Simon Kassing, Ankit Singla, Dan
  Alistarh, Ce~Zhang, and Ji~Liu.
\newblock Distributed learning over unreliable networks.
\newblock In {\em International Conference on Machine Learning}, pages
  7202--7212, 2019.

\bibitem{zagoruyko2016wide}
Sergey Zagoruyko and Nikos Komodakis.
\newblock Wide residual networks.
\newblock {\em arXiv preprint arXiv:1605.07146}, 2016.

\bibitem{awstream}
Ben Zhang, Xin Jin, Sylvia Ratnasamy, John Wawrzynek, and Edward~A Lee.
\newblock {AWStream}: Adaptive wide-area streaming analytics.
\newblock In {\em Proceedings of the 2018 Conference of the ACM Special
  Interest Group on Data Communication}, pages 236--252. ACM, 2018.

\bibitem{microbursts}
Qiao Zhang, Vincent Liu, Hongyi Zeng, and Arvind Krishnamurthy.
\newblock High-resolution measurement of data center microbursts.
\newblock In {\em Proceedings of the 2017 Internet Measurement Conference},
  pages 78--85. ACM, 2017.

\bibitem{zhuo2016rack}
Danyang Zhuo, Qiao Zhang, Vincent Liu, Arvind Krishnamurthy, and Thomas
  Anderson.
\newblock Rack-level congestion control.
\newblock In {\em Proceedings of the 15th ACM Workshop on Hot Topics in
  Networks}, pages 148--154. ACM, 2016.

\end{thebibliography}

\end{document}